\documentclass[english]{iopart}

\usepackage{graphicx}
\usepackage{url}
\usepackage{amssymb}

\newcounter{fig}
\begin{document}

\title[Schwarzian conditions]
{\Large Schwarzian conditions for linear differential operators with selected differential Galois groups (unabridged version)}

\vskip .3cm 
 

\author{Y. Abdelaziz, J.-M. Maillard$^\dag$\footnote[2]{Recherche Publique Fran\c{c}aise.}}
\address{$\dag$ LPTMC, UMR 7600 CNRS, 
Universit\'e Pierre et Marie Curie, Paris 6, Tour 23-13,
 5\`eme \'etage, case 121, 
 4 Place Jussieu, 75252 Paris Cedex 05, France} 

\vskip .2cm 

\ead{maillard@lptmc.jussieu.fr}

\begin{abstract}
We show that non-linear Schwarzian differential equations 
emerging from covariance symmetry conditions imposed on 
linear differential operators with hypergeometric 
function solutions can be generalized to arbitrary order
linear differential operators with polynomial 
coefficients having selected differential Galois groups.
For order three and order four linear differential operators 
we show that this pullback invariance up to conjugation eventually 
reduces to symmetric powers of an underlying order-two 
operator. We give, precisely, the conditions to have 
modular correspondences solutions for such Schwarzian 
differential equations, which was an open question in a previous paper.  
We analyze in detail a pullbacked hypergeometric 
example generalizing modular forms, that ushers a pullback invariance 
up to operator homomorphisms.
We expect this new concept to be well-suited in physics 
and enumerative combinatorics. We finally consider 
the more general problem of the equivalence of  
two different order-four linear differential 
 Calabi-Yau operators up to pullbacks and conjugation, 
and clarify the cases where they have the same Yukawa 
couplings. 

\end{abstract}

\vskip .5cm


\noindent {\bf AMS Classification scheme numbers}: 34M55, 
47E05, 81Qxx, 32G34, 34Lxx, 34Mxx, 14Kxx 

\vskip .5cm


\vskip .1cm
\vskip .1cm

\section{Introduction} 
\label{Intro}

\vskip .1cm

In a previous paper~\cite{Aziz} we focused on identities relating 
the {\em same} $\, _2F_1$ hypergeometric function with 
{\em two different\footnote[2]{Beyond the 
$\, x \, \rightarrow \, 1-x, \, 1/x, ...$ known pullback symmetries of 
hypergeometric functions. The correspondence between the two pullbacks 
must be an {\em infinite order} rational or algebraic 
transformation~\cite{Aziz,Hindawi}.} algebraic pullback transformations}. 
These identities correspond to {\em modular forms}, 
the algebraic transformations being solutions of a  
(non-linear) {\em differentially algebraic}~\cite{Selected,IsTheFull} 
{\em Schwarzian equation}, that also emerged in a paper 
by Casale on Galoisian envelopes~\cite{Casale,Paul}. This covariance 
symmetry of $\, _2F_1$ hypergeometric functions could be regarded as
one of the simplest illustrations of the concept of symmetries 
 (of the renormalization group type~\cite{Hindawi,broglie}) in physics
or enumerative combinatorics, 
a {\em univariate function being covariant (automorphic) with respect 
to an infinite set of rational or algebraic transformations}.
This paper~\cite{Aziz} was essentially focused on $\, _nF_{n-1}$ 
{\em hypergeometric functions} and {\em modular forms} 
actually represented as $\, _2F_1$ hypergeometric 
function with two different algebraic pullback transformations
({\em modular correspondences}~\cite{Aziz,Eichler}). 

The applications of this Schwarzian equation~\cite{Aziz} 
known to be associated to a quite large mathematical 
framework\footnote[1]{In Casale's paper~\cite{Casale,Paul}
the Schwarzian equation is associated with meromorphic functions 
instead of the rational functions of 
our paper~\cite{Aziz}. See also~\cite{Casale2,Casale3,Casale4}.} 
(Malgrange's pseudogroup, Galois 
groupoid~\cite{Casale2,Casale3,Casale4,Casale5,Casale6,Casale7,Malgrange}), 
extend well beyond hypergeometric functions in physics. We have 
seen, for instance in~\cite{Aziz}, an example of identity relating 
the same Heun function with two different 
pullbacks\footnote[9]{This Heun function being not, in general, 
reducible to a $\, _2F_1$ pullbacked hypergeometric function~\cite{maier-05}.}. This 
Heun example~\cite{Aziz} could suggest that 
such Schwarzian differential equations emerge in physics with holonomic 
functions having a narrow set of singularities (three for hypergeometric 
functions, four for Heun functions, ...) like the Heun 
example in~\cite{Aziz}. 
Going further we show, in this paper, that such 
{\em differentially algebraic}~\cite{Selected,IsTheFull} 
Schwarzian equations do emerge in a much more general holonomic framework. 

We will show in section \ref{order-two} 
that the  covariance symmetry condition 
of {\em general} order-two linear 
differential operators with polynomial coefficients automatically 
yields this Schwarzian differential equation. We will then show 
in sections \ref{order-three} and  \ref{order-four}
that the  covariance symmetry condition imposed on linear differential 
operators having  order three and order four 
with respective orthogonal and symplectic  differential Galois groups,
yield Schwarzian differential equations like the one examined in~\cite{Aziz}.
When their respective symmetric and exterior powers are of order {\em five}
(instead of six),  
one finds that these  order-three and order-four operators 
 reduce to symmetric square and symmetric cube of an underlying 
order-two operator.
In section \ref{order-N} we show that the Schwarzian condition 
can be derived for linear differential operators 
of arbitrary order $\, N$. The reduction of the solutions of this
Schwarzian differential equation to only 
{\em modular correspondences}~\cite{Eichler} was an open question in~\cite{Aziz}:
in section \ref{CalabiSchwResults} a necessary condition to have such
modular correspondences~\cite{Eichler} is derived. 
In section \ref{pullbackhomomorphism} generalizations 
of modular forms provide  examples of   
{\em pullback invariance of an operator, up to operator homomorphism}.
 This invariance  should be important to describing the symmetries of 
linear differential operators and thus, is of relevance to physics. 
Finally  in section \ref{CalabiSchwatext}, we consider the more general 
problem already addressed in~\cite{Duco} where Schwarzian 
differential equations also occurred, of the 
equivalence of {\em two different} order-four linear differential 
{\em Calabi-Yau operators}~\cite{Tables}
up to pullbacks and conjugation, possibly yielding the 
{\em same Yukawa couplings}~\cite{Duco}, and 
we will generalize it to linear differential operators of arbitrary orders. 

\vskip .2cm 

\section{Beyond hypergeometric and Heun functions: order-two linear differential operators} 
\label{order-two}

We will show here that non-linear ODEs involving Schwarzian derivatives 
(cf. equation (\ref{condition1}) below), that we will 
call ``Schwarzian ODEs''\footnote[2]{See~\cite{Aziz,What} for 
a definition. See also~\cite{Schwarzian2,Schwarzian}.}, obtained 
in~\cite{Aziz} for hypergeometric and Heun functions~\cite{Vidunas,Belyi3}, 
can be generalized to arbitrary globally nilpotent~\cite{bo-bo-ha-ma-we-ze-09} 
linear differential operators having an arbitrary numbers of 
singularities (as opposed to three and four singularities
for hypergeometric and Heun functions).

Let us consider a linear differential operator of order two
\begin{eqnarray}
\hspace{-0.95in}&& \quad \quad  
\label{L_2}
L_2 \, \, = \, \, \, \, 
D_x^2  \, \,\,  + \, p(x) \cdot \, D_x \,\,  + \, \, q(x), 
\quad \quad \quad \hbox{where:} \quad  \quad \quad  \quad
D_x \, = \, \, {{ d} \over {dx}}, 
\end{eqnarray}
and let us also introduce two other linear differential operators 
of order two: the operator 
$\, \, L_2^{(c)} \, = \, \, 1/v(x) \cdot \, L_2 \cdot \, v(x)\, $
being the  conjugate of (\ref{L_2}) by a function $\, v(x)$, and 
the pullbacked operator $\, L_2^{(p)}$ which amounts to changing 
$\, x \, \rightarrow \, \, y(x) \, $ in (\ref{L_2}), 
the head coefficient being normalized\footnote[1]{Throughout 
the paper we consider, for clarity and simplicity, this 
normalized form for the linear differential operators. The
 ``true'' pullbacked operator which amounts to  changing 
$\, x \, \rightarrow \, \, y(x)$ (see the command ``dchange''
 in PDEtools in Maple) is in fact $\, 1/y'(x)^2 \cdot \,L_2^{(p)}$
where $\, L_2^{(p)}$ is given by (\ref{L2p}).} to $\,1$. These two 
linear differential operators read respectively:
\begin{eqnarray}
\hspace{-0.95in}&&  
\label{L2c} \, 
L_2^{(c)} \,  \, \, = \, \, \, \, \, D_x^2  \,\, \,  + \, 
\Bigl(p(x) \, + \, 2 \cdot \, {{v'(x)} \over {v(x)}} \Bigr) \cdot \, D_x 
\, \, \, + \, \, q(x) 
 \, \, + \, p(x) \cdot \,  {{v'(x)} \over {v(x)}} 
\, \,  + \, {{v"(x)} \over {v(x)}},  
\end{eqnarray}
where 
\begin{eqnarray}
\label{L2cwhere}
\hspace{-0.95in}&& \quad \quad \quad \quad  \quad \quad  \quad 
v'(x) \, \, = \, \, \, {{d v(x)} \over {dx}},
 \quad  \quad \quad \quad 
 v"(x) \, \, = \, \, \, {{d^2 v(x)} \over {dx^2}},
\end{eqnarray}
and
\begin{eqnarray}
\hspace{-0.95in}&&  \quad 
\label{L2p} \, \, \, \, 
L_2^{(p)} \, \, = \, \, \, \, \, \, D_x^2  \, \, \, 
+ \,  \Bigl( p(y(x)) \, \cdot \, y'(x) \, - \, {{y"(x)} \over {y'(x)}}  \Bigr)
  \cdot \, D_x 
\,\, \,  + \, \, q(y(x)) \cdot \, y'(x)^2, 
\end{eqnarray}
where:
\begin{eqnarray}
\hspace{-0.95in}&& \quad  \quad \quad \quad \quad  \quad 
\label{L2pwhere}
y'(x) \, \, = \, \, \, {{d y(x)} \over {dx}},
 \quad  \quad \quad \quad 
 y"(x) \, \, = \, \, \, {{d^2 y(x)} \over {dx^2}}.
\end{eqnarray}
The identification of these two  linear differential operators  
$\, L_2^{(c)} \, = \, \, L_2^{(p)}$ gives two conditions:
\begin{eqnarray}
\hspace{-0.95in}&& \quad \quad \quad \quad \quad \quad 
\label{firstcond}
p(x) \, \, + \, 2 \cdot \, {{v'(x)} \over {v(x)}} 
 \, \, \, = \, \, \,  \,\,  p(y(x)) \, \cdot \, y'(x)
 \, \,\, - \, {{y"(x)} \over {y'(x)}} , 
\\
\hspace{-0.95in}&& \quad \quad \quad \quad \quad \quad 
\label{seccond}
 q(x) \, \, + \, p(x) \cdot \,  {{v'(x)} \over {v(x)}} 
\,  \,+ \, {{v"(x)} \over {v(x)}} \, \,  \,= \, \,\,  \,  \,
 q(y(x)) \cdot \, y'(x)^2.
\end{eqnarray}
Since 
\begin{eqnarray}
\hspace{-0.95in}&& \quad \quad \quad \quad \quad \quad  \quad  \quad 
\label{since}
 {{v"(x)} \over {v(x)}} \, \,  \, = \, \,\,  \, 
{{d} \over {dx}} \Bigl({{v'(x)} \over {v(x)}}\Bigr)
  \, \,  + \,  \Bigl({{v'(x)} \over {v(x)}}\Bigr)^2,  
\end{eqnarray}
one can eliminate the log-derivative $\, v'(x)/v(x)\, $ between  (\ref{firstcond}) 
and (\ref{seccond}), and obtain the {\em Schwarzian condition}
previously given in~\cite{Aziz}
\begin{eqnarray}
\label{condition1}  
\hspace{-0.95in}&& \quad   \quad \quad   \quad \quad  \quad \quad  \quad 
 W(x) 
\, \, \,\,  \,-W(y(x)) \cdot  \, y'(x)^2
\, \, \, \, \,+ \,  \{ y(x), \, x\} 
\, \,\, \, = \,\, \, \,  \, 0, 
\end{eqnarray}
where 
\begin{eqnarray}
\label{wherecond}
\hspace{-0.95in}&& \quad   \quad \quad  \quad \quad  \quad \quad  \quad 
W(x)  \, \, = \, \,  \, \, \, {{ d p(x)} \over { dx}} 
 \, \,\,  \,  + \, \, 
   {{p(x)^2} \over {2 }} \, \,  \, \,  -2 \cdot \, q(x),
\end{eqnarray}
and where $\,\{ y(x), \, x\}\, $ denotes the 
{\em Schwarzian derivative}~\cite{What}:
\begin{eqnarray}
\label{Schwa}
\hspace{-0.95in}&& \quad \, \,     \, 
\{ y(x), \, x\}    \, \, = \, \,  \,  \,  \,
{{y'''(x) } \over{ y'(x)}} 
 \,  \,  - \, \, {{3} \over {2}}
 \cdot \, \Bigl({{y''(x)} \over{y'(x)}}\Bigr)^2
 \, \, = \, \,  \,  \,  \,
{{ d } \over { dx  }} \Bigl( {{y''(x) } \over{ y'(x)}}   \Bigr) 
\, \, - {{1} \over {2}} \cdot \, \Bigl( {{y''(x) } \over{ y'(x)}}   \Bigr)^2, 
\nonumber \\
\hspace{-0.95in}&& \quad \, \,     \,  \hbox{ where:}  
\, \,   \quad  \quad \quad 
y'''(x)   \, \, = \, \,  \, {{d^3 y(x)} \over{ dx^3}},  \quad  \,  \,  \,   \,
y''(x)   \, \, = \, \,  \, {{d^2 y(x)} \over{ dx^2}}, \quad  \,  \,  \,    \,
   y'(x)  \, \, = \, \,  \, {{d y(x)} \over{ dx}}.
\nonumber 
\end{eqnarray}
Unlike in~\cite{Aziz}, the 
{\em number of singularities of the second order operator} (\ref{L_2}) 
{\em  is arbitrary}: it does not need to be three or four like in the hypergeometric 
or Heun examples in~\cite{Aziz}. The second order linear differential 
operator $\, L_2$ is a {\em general} order-two linear differential 
operator with polynomial coefficients. Introducing $\, w(x)$ the wronskian of $\, L_2$ 
\begin{eqnarray}
\hspace{-0.95in}&& \quad \quad \quad \quad  \quad 
\label{globnilp}
p(x) \, \,  = \, \,  \,  \,  -\,{{w'(x)} \over {w(x)}} 
\quad \quad  \quad  \,  \, \hbox{where:}  \quad \quad \quad \quad 
 w'(x) \, \, = \, \,  \, {{ d w(x)} \over {dx}}, 
\end{eqnarray}
we see that the LHS and  RHS of the first condition (\ref{firstcond})
are both log-derivatives. Thus one can immediately integrate  
the first condition (\ref{firstcond}) and get (up to a multiplicative 
factor $\, \mu$) the conjugation function
$\, v(x)$ in terms of
the wronskian $\, w(x)$ and the pullback function $\, y(x)$: 
\begin{eqnarray}
\hspace{-0.95in}&& \quad \quad \quad \quad   \quad \quad  \quad \quad 
\label{integrate}
v(x) \, \,  = \, \,  \,  \,  \, 
\mu \cdot \, \Bigl({{w(x)} \over {w(y(x)) \cdot y'(x)}}\Bigr)^{1/2}.
\end{eqnarray}

\vskip .1cm

{\bf Remark 1:} When the wronskian $\, w(x)$ is an $\, N$-th root 
of a rational function, the exact expression (\ref{integrate}) 
for the conjugation function $\, v(x)$, becomes an algebraic function 
when $\, y(x)$ {\em is an algebraic function}. This is actually 
the case when the order-two linear 
differential operator  $\, L_2$ is 
{\em globally nilpotent}~\cite{bo-bo-ha-ma-we-ze-09}.
In this case the linear differential operator is simply conjugated to its adjoint 
through its wronskian $\, w(x)$ which is a $\, N$-th root 
of a rational function:
\begin{eqnarray}
\label{simplyconj}
\hspace{-0.95in}&& \quad  \quad \quad 
 \quad  \quad  \quad  \quad \, \, \,   \,  \,    \, 
L_2 \cdot \, w(x) \, \, \, = \,  \, \,\,  \,   w(x) \cdot \, adjoint(L_2).
\end{eqnarray}

\vskip .1cm

{\bf Remark 2:} If the linear differential operator is {\em not} 
globally nilpotent~\cite{bo-bo-ha-ma-we-ze-09}
the wronskian is {\em not} necessarily an algebraic function. Introducing 
$\, L_v(x)$, the log-derivative of the 
conjugation function $\, v(x)$, one can rewrite the two 
conditions (\ref{firstcond}) and (\ref{seccond}) as:
\begin{eqnarray}
\hspace{-0.95in}&& \quad \quad \quad
\label{firstcondbis}
p(x) \, \, + \, 2 \cdot \, L_v(x)
 \, \, \, = \, \, \,  \,\,  p(y(x)) \, \cdot \, y'(x)
 \, \,\, - \, {{y"(x)} \over {y'(x)}} , 
\\
\hspace{-0.95in}&& \quad \quad \quad
\label{seccondbis}
 q(x) \, \, + \, p(x) \cdot \,  L_v(x)
\,  \,+ \, {{d  L_v(x)} \over {dx}} \, + L_v(x)^2 
\,  \,= \, \,\,  \,  \,
 q(y(x)) \cdot \, y'(x)^2.
\end{eqnarray}
The elimination of  $\, L_v(x)$ in (\ref{firstcondbis}) and (\ref{seccondbis})
gives the Schwarzian condition (\ref{condition1}) 
with (\ref{wherecond}), however the conjugation function $\, v(x)$ 
{\em is no longer an algebraic function when $\, y(x)$ is an algebraic function}
(see (\ref{integrate})): it is a transcendental function, and we certainly 
move away from a modular correspondence~\cite{Aziz,Eichler} 
framework\footnote[1]{For modular correspondences  see also the concept 
of modular equations~\cite{Morain,Yi,Chan,Yi2}.}.  

\vskip .2cm 
\vskip .2cm 

\section{Order-three linear differential operators} 
\label{order-three}

\subsection{General order-three linear differential operators.\\} 
\label{order-threegener}

\vskip .1cm 

Considering an  {\em irreducible} order-three linear differential operator
\begin{eqnarray}
\hspace{-0.95in}&& \quad \quad  \quad \quad  \quad  \quad 
\label{L_3}
L_3 \, \, \,  = \, \, \,  \, 
 D_x^3  \,  \,\,  + \, p(x) \cdot \, D_x^2 
\, \,  \,  + \, \, q(x)\cdot \, D_x \, \,  + \,  \, r(x),
\end{eqnarray}
we introduce two other  linear differential operators of order three 
defined as previously in section \ref{order-two}:  
the operator $\, L_3^{(c)}$ conjugated of (\ref{L_3}) by a function $\, v(x)$,
namely $\, L_3^{(c)} \, = \, \, 1/v(x) \cdot \, L_3 \cdot \, v(x)$, and the 
pullbacked\footnote[1]{The $\, D_x^3$ coefficient is normalized to $\, 1$.  } 
operator $\, L_3^{(p)}$ which amounts to changing 
$\, x \, \rightarrow \, \, y(x)\, $ in $\, L_3$. These two linear differential 
operators read respectively
\begin{eqnarray}
\hspace{-0.95in}&&  
\label{L3c} \, \,  \quad  \quad  \quad  \quad  \quad 
L_3^{(c)} \, \, = \, \, \, \, \, 
D_x^3  \, \, + \, 
\Bigl(p(x) \, + \, 3 \cdot \, {{v'(x)} \over {v(x)}} \Bigr) \cdot \, D_x^2 
\nonumber \\
\hspace{-0.95in}&&  \quad  \quad  \quad  \quad  \quad  \quad  \quad 
\, \, \, \, \, + \, \, \Bigl(q(x) \, + \, 2 \cdot \,  p(x) \cdot \,  {{v'(x)} \over {v(x)}} 
\, + \, 3 \cdot \, {{v"(x)} \over {v(x)}}\Bigr) \cdot  \, D_x
 \\
\hspace{-0.95in}&&  \quad  \quad  \quad  \quad  \quad  \quad  \quad  \quad  \quad 
\, \, \,  + \, \,  r(x) \, \, + \, q(x) \cdot \,  {{v'(x)} \over {v(x)}} 
  \, \, + \,  p(x) \cdot \,  {{v"(x)} \over {v(x)}} \,  \, + \,  {{v^{(3)}(x)} \over {v(x)}},
  \nonumber
\end{eqnarray}
and: 
\begin{eqnarray}
\hspace{-0.95in}&&  
\label{L3P} \, \, \quad  \quad  \,  \, 
L_3^{(p)} \, \, = \, \, \, \, \, D_x^3  \,\, \,   + \, 
\Bigl(p(y(x)) \cdot \, y'(x) \, -\,  3  \, {{y"(x)} \over {y'(x)}} \Bigr) \cdot \, D_x^2 
\nonumber \\
\hspace{-0.95in}&&  \quad  \quad  \quad  \quad   
\, \, \, + \, \, \Bigl(  q(y(x)) \cdot \, y'(x)^2\,  \, -p(y(x)) \cdot \, y"(x)
\, \,\,  -\, {{y^{(3)}(x) } \over {y'(x) }} 
 \,\,  +3 \cdot \, \Bigl({{y"(x) } \over {y'(x) }}  \Bigr)^2  \Bigr) \cdot \, D_x
\nonumber \\
\hspace{-0.95in}&&  \quad  \quad  \quad  \quad  \quad \quad \quad  \quad  \quad   
\, + \, r(y(x)) \cdot \, y'(x)^3.
\end{eqnarray}

The equality of these two order-three linear differential operators gives 
three conditions $\, {\cal C}_n$, 
with $\, n= \, 0, \, 1, \, 2$, corresponding, respectively, to the identification 
of the $\, D_x^n$ coefficients of $\, L_3^{(p)}$ and $\, L_3^{(c)}$. Introducing 
the wronskian $\, w(x)$ of $\, L_3$,  the LHS and RHS of condition 
$\, {\cal C}_2$ being, again, log-derivatives,
one can easily integrate  condition $\, {\cal C}_2$ and get the exact expression 
of the conjugation function $\, v(x)$ in terms of the wronskian of $ \, L_3$
and of the pullback $\, y(x)$:
\begin{eqnarray}
\label{v3} 
\hspace{-0.95in}&&  
 \, \, \quad  \quad  \quad  \quad  \quad  \quad  \quad 
v(x) \, \,  = \, \,  \,  \,  \, 
\mu \cdot \, \Bigl( {{w(x)} \over {w(y(x)) \cdot \, y'(x)^3 }} \Bigr)^{1/3}.
\end{eqnarray}
Similarly the elimination of the log-derivative $\, v'(x)/v(x)$ between condition 
$\, {\cal C}_2$ and condition $\, {\cal C}_1$ yields the Schwarzian condition
\begin{eqnarray}
\label{condition3}
\hspace{-0.95in}&& \quad   \quad \quad  \quad \quad  \quad  \quad  \quad 
 W(x) 
\, \, \,  \,-W(y(x)) \cdot  \, y'(x)^2
\, \, \,  \,+ \,  \{ y(x), \, x\} 
\, \,\, \, = \,\, \, \,  \, 0, 
\end{eqnarray}
where this time $\, W(x)$ reads:
 \begin{eqnarray}
 \label{wherecond3bis}
 \hspace{-0.95in}&& \quad   \quad \quad  \quad \quad  \quad  \quad  \,\, \, \, 
 W(x)  \, \, \, = \, \,  \, \, \,
 {{1} \over {2}} \cdot \, {{d p(x)} \over {dx }}
 \,\,   \,+{{p(x)^2} \over {6}} \, \,\,  -{{q(x)} \over {2}}.
 \end{eqnarray} 

\subsection{Symmetric Calabi-Yau condition.\\} 
\label{SymCalab}

\vskip .1cm 

Let us consider the condition corresponding to imposing  
the {\em symmetric square} of $\, L_3$ to be of order {\em five} 
instead of the generic order six. This 
(``symmetric'' Calabi-Yau~\cite{Diffalg}) 
condition reads:
\begin{eqnarray}
\label{CalabiSymm}
\hspace{-0.95in}&& \quad \quad \quad  \quad  
r(x) \, \, = \, \, \,  \, -{{2} \over {27}} \cdot \, p(x)^3 \, 
\, \, + \, {{1} \over {3}} \cdot \, p(x) \cdot \, q(x) \,  \, \, \,
 -{{1} \over {3}} \cdot \,  p(x) \cdot \, {{ d p(x)} \over { dx}}
\nonumber \\
\hspace{-0.95in}&&  \quad  \quad \quad   \quad \quad \quad \quad   \quad  \quad 
\, \, + {{1} \over {2}} \cdot \, {{ d q(x)} \over { dx}}
 \, \,\,  -  {{1} \over {6}} \cdot \,  {{ d^2 p(x)} \over { dx^2}}.
\end{eqnarray}
For a  globally nilpotent~\cite{bo-bo-ha-ma-we-ze-09} linear differential operator, 
this  (symmetric Calabi-Yau) condition (\ref{CalabiSymm}) together 
with (\ref{globnilp}) yields an order-three linear differential operator
 (\ref{L_3}) simply conjugated to its adjoint:
\begin{eqnarray}
\label{simplyconj3}
\hspace{-0.95in}&& \quad  \quad \quad  \quad \quad 
 \quad \, \, \, \, \,    \,  \,    \, 
L_3 \cdot \, w(x)^{2/3} \, \, \, = \,  \, \,\,  \,   w(x)^{2/3} \cdot \, adjoint(L_3),
\end{eqnarray}
where the wronskian $\, w(x)$ is a $\, N$-th root of a rational function.

Again for a  globally nilpotent~\cite{bo-bo-ha-ma-we-ze-09}
 linear differential operator, the 
exact expression (\ref{v3}) for the conjugation function $\, v(x)$, 
becomes an algebraic function {\em when $\, y(x)$ is an algebraic function}.

The symmetric square of an order-two linear differential operator 
$\, L_2 \, = \, D_x^2 \, + \, A(x) \cdot \, D_x \, + \, B(x) \, $ is an order-three
 linear differential operator (\ref{L_3}) with the following coefficients:
\begin{eqnarray}
\hspace{-0.95in}&&  
\label{param3} \quad  \quad   \quad  
p(x) \, \, = \, \,\,  
3  \cdot \, A(x), \quad  \quad  \, \,   \,   
q(x) \, \, = \, \,\, 
2 \cdot \, A(x)^2  \, \, + 4 \cdot \,B(x)  \, \, + {{ d A(x)} \over { dx}}, 
 \\
\label{param32}
\hspace{-0.95in}&&  \quad  \quad \quad  \quad \quad \quad  \quad  \quad  
r(x) \, \, = \, \,\, \,  
4\cdot \, B(x) \cdot \, A(x) 
\, \, \,  + 2  \cdot \, {{ d B(x)} \over { dx}}.
\end{eqnarray}
These coefficients (\ref{param3}), (\ref{param32})
{\em automatically verify the (symmetric Calabi-Yau) condition}
(\ref{CalabiSymm}): the symmetric square of a symmetric square 
of  an order-two  linear differential operator is of order {\em five}
instead of the generic order six.  Conversely, the (symmetric Calabi-Yau) 
condition (\ref{CalabiSymm}) can be parametrized\footnote[1]{Note that 
rewriting the exact expression of $\, W(x)$ given by (\ref{wherecond3bis})
in terms of $\, A(x)$ and $\, B(x)$ using  (\ref{param3})
one recovers (\ref{wherecond}), 
$\, p(x)$ and $\, q(x)$ in (\ref{wherecond}) being now 
$\, A(x)$ and $\, B(x)$.} 
by (\ref{param3}) and 
(\ref{param32}) and amounts to imposing the order-three linear differential 
operator  (\ref{L_3}) to be exactly the symmetric square of an order-two operator.

\vskip .2cm 

\vskip .2cm 

Thus our  calculations show that the pullback-compatibility 
of an order-three linear differential operator is equivalent to saying that this  
order-three operator {\em reduces to} (the symmetric square of) 
 an underlying {\em order-two linear differential operator}. The 
Schwarzian condition (\ref{condition3}) 
with $\, W(x)$ given by (\ref{wherecond3bis}), 
is {\em thus inherited from the Schwarzian condition} 
(\ref{condition1}) {\em of the underlying order-two linear differential operator}.

\vskip .2cm 

\section{Order-four linear differential operators} 
\label{order-four}

Consider the {\em irreducible} order-four linear differential 
operator
\begin{eqnarray}
\label{L_4}
\hspace{-0.95in}&& \quad \quad  \quad \,  \,
L_4 \, \, \, = \, \, \, \,  \,
 D_x^4  \, \,  \, + \, p(x) \cdot \, D_x^3 \, \, \, + \, \, q(x)\cdot \, D_x^2
 \, \,  \, + \,  \, r(x) \cdot \, D_x \,  \, + \, \, s(x),
\end{eqnarray}
and  introduce two other  linear differential operators of order four
defined as previously 
in sections \ref{order-two} and \ref{order-threegener}: 
the linear differential operator $\, L_4^{(c)}$ conjugated of (\ref{L_4}) 
by a function $\, v(x)$ and the 
(normalized) pullbacked operator $\, L_4^{(p)}$. These two 
linear differential operators read respectively
\begin{eqnarray}
\label{L2c} 
\hspace{-0.95in}&&  \quad \quad \quad 
\, \, 
L_4^{(c)} \, \, = \, \, \, \, \, D_x^4  \, \, \, + \, 
\Bigl(p(x) \, + \, 4 \cdot \, {{v'(x)} \over {v(x)}} \Bigr) \cdot \, D_x^3 
 \\
\hspace{-0.95in}&&  \quad  \quad \quad \quad \,
\, \, \,  + \, \, \Bigl(q(x) \, + \, 3 \cdot \, p(x) \cdot \,  {{v'(x)} \over {v(x)}} 
\, + \,  6 \cdot \,  {{v"(x)} \over {v(x)}} \Bigr) \cdot \, D_x^2
\nonumber \\
\hspace{-0.95in}&&  \quad  \quad \quad\quad  \quad  \, \,  \, 
+ \, \, \Bigl(r(x) \, + \, 2 \cdot \, q(x) \cdot \,  {{v'(x)} \over {v(x)}}
 \, + \, 3 \cdot \, p(x) \cdot \,  {{v"(x)} \over {v(x)}}
 \, + \, 4 \cdot \,  {{v^{(3)}(x)} \over {v(x)}}     \Bigr) \cdot \, D_x  
\nonumber \\
\hspace{-0.95in}&&  \quad   \quad \quad \quad \quad 
\,\, \,  +  \, \, s(x) \, \, + \,  \, r(x) \cdot \,  {{v'(x)} \over {v(x)}}
 \, \, + \,  q(x) \cdot \,  {{v"(x)} \over {v(x)}}
 \,\,  + \,   p(x) \cdot \,  {{v^{(3)}(x)} \over {v(x)}} 
 \, \, + \,   {{v^{(4)}(x)} \over {v(x)}}, 
 \nonumber
\end{eqnarray}
and:
\begin{eqnarray}
\label{L2c}
\hspace{-0.95in}&&    
L_4^{(p)} \, \, = \, \, \, \, \, 
D_x^4  \,  \,  \, + \,  \Bigl(p(y(x)) \cdot \, y'(x)  \,
 - \, 6 \cdot \,  {{y"(x)} \over {y'(x)}} \Bigr) \cdot \, D_x^3
 \nonumber \\
\hspace{-0.95in}&&  \quad  
\, +  \Bigl( q(y(x)) \cdot \, y'(x)^2 \, - \, 3 \cdot \, p(y(x)) \cdot \, y"(x) \, 
-4 \cdot \, {{y^{(3)}(x)} \over {y'(x)}} 
 \, + 15 \cdot \, \Bigl( {{y"(x)} \over {y'(x)}} \Bigr)^2 \Bigr) \cdot \, D_x^2 
\nonumber \\
\hspace{-0.95in}&&  \quad   
\, +  \Bigl(r(y(x)) \cdot \,y'(x)^3 \,
 -q(y(x)) \cdot \, y'(x)\cdot \, y"(x)  \, -p(y(x)) \cdot \, y^{(3)}(x) 
\nonumber \\
\hspace{-0.95in}&&  \quad   \quad   \quad \quad   
 + 3 \cdot \, p(y(x))  \cdot \, {{y"(x)^2} \over {y'(x)}} 
 \, \,  - {{y^{(4)} } \over {y'(x)}} \, \, 
+10  \cdot \, {{y"(x) \cdot \, y^{(3)} } \over { y'(x)^2 }} \, \,
  -15 \cdot  \, \Bigl( {{y"(x)  } \over { y'(x) }} \Bigr)^3
 \Bigr) \cdot \, D_x
\nonumber \\
\hspace{-0.95in}&&  \quad   \,\, 
\, \, + \, s(y(x)) \cdot y'(x)^4.
\end{eqnarray}
The identification of these two order-four linear differential 
operators $\,L_4^{(p)}$ and $\,L_4^{(c)}$ gives this time four conditions 
$\, {\cal C}_n$, $\, n= \, 0,  \, 1, \, 2, \, 3$, corresponding, 
respectively, to the identification  of the $\, D_x^n$ coefficients 
of $\, L_4^{(p)}$ and $\, L_4^{(c)}$. 

Eliminating once again the log-derivative $\, v'(x)/v(x)$ between  
$\, {\cal C}_3$  and  $\, {\cal C}_2$ one deduces a Schwarzian condition
\begin{eqnarray}
\label{condition4}
\hspace{-0.95in}&& \quad   \quad \quad  \quad \quad  \quad 
 W(x) 
\, \, \,  \,-W(y(x)) \cdot  \, y'(x)^2
\, \, \,  \,+ \,  \{ y(x), \, x\} 
\, \,\, \, = \,\, \, \,  \, 0, 
\end{eqnarray}
where this time:
\begin{eqnarray}
\label{wherecond4}
\hspace{-0.95in}&& \quad   \quad \quad  \quad \quad  \quad 
W(x)  \, \, = \, \,  \, \, \,
{{3} \over {10}} \cdot \,  {{ d p(x)} \over { dx}}  \, \, \,  + \, \, 
 {{3} \over {40}} \cdot \,   p(x)^2  \, \,  \, \,  - \, {{q(x)} \over {5}}.
\end{eqnarray}

Introducing the wronskian $\, w(x)$ of the order-four linear differential
operator $\, L_4$ with (\ref{globnilp}), the condition  $\, {\cal C}_3$ 
just corresponds to log-derivatives and can  be easily  integrated
giving the exact expression of the conjugation function 
$\, v(x)$ as:
\begin{eqnarray}
\label{v4}
\hspace{-0.95in}&& \quad   \quad \quad  \quad \quad  \quad \quad  \quad 
v(x) \, \, = \, \, \,  
\Bigl({{ w(x) } \over { w(y(x)) \cdot \, y'(x)^6 }}\Bigr)^{1/4}.
\end{eqnarray}
The next conditions  $\, {\cal C}_1$ and  $\, {\cal C}_0$ yield extremely 
involved non-linear differential conditions on the miscellaneous 
derivatives of the various coefficients. It turned out to be
 very difficult to proceed 
with such huge expressions. Yet when the linear differential
operator $\, L_4$ has a selected (symplectic) differential 
Galois group one can go much further in the calculations, 
as we will see in the coming subsection.

\vskip .2cm  

\subsection{Calabi-Yau condition (exterior square).} 
\label{CalabiYaucondi}

Imposing the 
{\em Calabi-Yau condition}~\cite{IsingCalabi,IsingCalabi2}
in the case of an order-four  linear differential operator gives:
\begin{eqnarray}
\hspace{-0.95in}&&  
\label{Calabi} 
\, r(x) \, \, = \, \, \,   
 {{p(x) \cdot \, q(x)} \over {2}} \, \,  \, 
 -\,  {{p(x)^3 } \over {8}} \, \, \,   + \, {{d q(x)} \over {dx}} \,  \,  \, 
 - {{3} \over {4}} \cdot \,  p(x) \cdot \,  {{d p(x)} \over {dx}} 
\,  \,  \,  - {{1} \over {2}} \cdot \,  {{d^2 p(x)} \over {dx^2}}.
\end{eqnarray}
In this case the exterior square of the order-four operator $\, L_4$ 
has order {\em five} instead of order six. 

When condition (\ref{Calabi}) is verified,
the order-four linear differential operator $\, L_4$ 
has a {\em symplectic} differential Galois group 
$\, Sp(4, \,  \mathbb{C})$. Note that if condition (\ref{Calabi}) is verified, 
the Calabi-Yau conditions for the pullbacked and conjugated 
operators $\, L_4^{(p)}$ and  $\, L_4^{(c)}$ 
{\em are automatically verified}: this is a consequence of the fact 
that the Calabi-Yau condition  (\ref{Calabi}) is left invariant by 
conjugation and pullback\footnote[5]{To see that the  Calabi-Yau condition 
is preserved by conjugation is straightforward. However, 
as remarked in~\cite{Duco}, to see that the
Calabi-Yau condition is preserved by pullback transformations 
is very hard to see by direct computation, since one gets 
an enormous fourth-order nonlinear differential equation.
}. In other words the {\em following identification of the} $\, D_x$ 
{\em coefficients of} $\, L_4^{(p)}$ and $\, L_4^{(c)}$ 
{\em is automatically verified when the Calabi-Yau condition} 
(\ref{Calabi})  {\em is verified}. 

Recall that the Calabi-Yau condition (\ref{Calabi}) together 
with the globally nilpotent condition~\cite{bo-bo-ha-ma-we-ze-09} 
automatically  yields $\, L_4$ 
to be conjugated to its adjoint
\begin{eqnarray}
\label{simplyconj4}
\hspace{-0.95in}&& \quad  \quad \quad  
\quad \quad  \quad \quad  \quad \, \,   \,  \,    \, 
L_4 \cdot \, w(x)^{1/2} \, \, \, = \,  \, \,\,  \, 
  w(x)^{1/2} \cdot \, adjoint(L_4),
\end{eqnarray}
where $\, w(x)$ is a $\, N$-root of a rational function.

At the last step we consider the identification of the constant 
terms in $\, D_x$ in $\, L_4^{(p)}$ and $\, L_4^{(c)}$. After injecting 
in this ``large'' non-linear differential equation, equation 
(\ref{globnilp}), the Schwarzian condition (\ref{condition4})
with $\, W(x)$ given by (\ref{wherecond4}), and the 
Calabi-Yau condition (\ref{Calabi}), we eventually find that 
this last ``large''  equation {\em becomes independent of the pullback} 
$\, y(x)$ and reduces to a quite simple condition {\em giving} $\, s(x)$ 
{\em as a polynomial expression in the two coefficients 
$\, p(x)$ and  $\, q(x)$ and their derivatives}:
\begin{eqnarray}
\hspace{-0.95in}&&  
\label{s(x)} 
s(x) \, \, = \, \, \, \,
 {{9} \over {100}} \cdot \, q(x)^2 
\, \,\,   - \, {{1} \over {200}} \cdot \, q(x) \cdot \, p(x)^2 \, \,  
+\, {{1} \over {4}} \cdot \, p(x) \cdot \, {{ d q(x)}  \over {dx}}
 \,\,   - \, {{1} \over {50 }} \cdot \, q(x) \cdot \, {{ d p(x)}  \over {dx}}
\nonumber \\
\hspace{-0.95in}&&  \quad  \quad \quad \quad 
\, + \, {{3} \over {10 }} \cdot \, {{ d^2 q(x)}  \over {dx^2}} \, \,
 -  \, {{11} \over {1600}} \cdot \, p(x)^4 \,  \,
 -  \, {{9} \over {50}} \cdot \, p(x)^2 \cdot \,  {{ d p(x)}  \over {dx}}
 \,  \, -  \, {{21} \over {100}} \cdot \, \Bigl({{ d p(x)}  \over {dx}}\Bigr)^2 
\nonumber \\
\hspace{-0.95in}&&  \quad  \quad \quad \quad  \quad \quad 
 -  \, {{1} \over {5}} \cdot \,{{ d^3 p(x)}  \over {dx^3}} \, 
-  \, {{7} \over {20}} \cdot \, p(x) \cdot \, {{ d^2 p(x)}  \over {dx^2}}.
\end{eqnarray}
In order to understand what this new condition (\ref{s(x)}) 
coming on top of the Calabi-Yau condition (\ref{Calabi}) really means,
we calculated, for various MUM\footnote[1]{Maximal unipotent 
monodromy (MUM) linear operators~\cite{bo-bo-ha-ma-we-ze-09,Christol}. } 
order-four linear differential operators
$\, L_4$ verifying (\ref{Calabi}) and (\ref{s(x)}), the corresponding nome 
and {\em Yukawa couplings}~\cite{Christol}. The corresponding Yukawa 
couplings {\em were actually found to be trivial}: $\, K_q \, = \, 1$ !! 

This amounts to saying that combining the two conditions (\ref{Calabi}) 
and (\ref{s(x)}) corresponds to a drastic reduction: 
 the (irreducible) order-four linear differential operator $\, L_4$
is not a ``true'' order-four operator. Typically one can imagine that $\, L_4$
reduces to an order-two operator, being homomorphic to the {\em symmetric cube} 
of an underlying order-two linear differential operator. In fact it is exactly
the symmetric cube of an order-two operator.

Let us consider the {\em symmetric cube} of an {\em order-two} linear differential 
operator $\, L_2 \, = \, D_x^2 \, + \, A(x) \cdot \, D_x \, + \, B(x) \, $ which is an 
order-four linear differential (\ref{L_4}) with the following coefficients:
\begin{eqnarray}
\hspace{-0.95in}&&  
\label{param} 
p(x) \, = \, \,\,  6  \cdot \, A(x), \,\,   \quad   \quad   \quad 
q(x)  \, = \, \,\,  11\cdot \, A(x)^2 \,\, 
 +4 \cdot \,  {{ d A(x)} \over {dx}} \, \, +10 \cdot \, B(x),  
 \nonumber  \\
\hspace{-0.95in}&&  
r(x) \, \, =  \, \, \, 
6\cdot \, A(x)^3 \,\, +7 \cdot \,A(x) \cdot \,  {{ d A(x)} \over {dx}}
 \,\, +30 \cdot \,B(x) \cdot \,A(x) \,\,
 + {{ d^2 A(x)} \over {dx^2}} 
\,\, +10 \cdot \,  {{ d B(x)} \over {dx}},  
\nonumber \\
\hspace{-0.95in}&&  
s(x) \, \, = \, \, \,\,\,  18 \cdot \, A(x)^2\cdot \,B(x)\,\,\, 
  +6  \cdot \, B(x) \cdot \, {{ d A(x)} \over {dx}}\,\,\,
 +15 \cdot \,  {{ d B(x)} \over {dx}} \cdot \, A(x)
\nonumber \\
\hspace{-0.95in}&&  \quad  \quad   \quad   \quad \quad   \quad   \quad 
\,\, +9 \cdot\, B(x)^2\,\,\,  +3 \cdot\,  {{ d^2 B(x)} \over {dx^2}}.
\end{eqnarray}
One finds straightforwardly that the coefficients given by (\ref{param}) 
verify the Calabi-Yau condition (\ref{Calabi}), 
{\em as well as the new condition} 
(\ref{s(x)}). In this case the differential Galois group  is no longer 
the symplectic differential Galois group $\, Sp(4,  \,  \mathbb{C})$, 
but actually reduces\footnote[2]{When an  order-four linear differential 
operator is the symmetric cube of an underling order-two operator 
its symmetric square is no longer of order $\, 10$ 
but reduces to order $\, 7$.} to the
differential Galois group of the underlying order-two linear differential 
operator, namely $\, SL(2,  \,  \mathbb{C})$. The fact that the Calabi-Yau 
condition (\ref{Calabi}) is verified is not a surprise: the exterior 
square of a symmetric cube is naturally of order less than six. The 
fact that being the symmetric cube of an underlying order-two operator 
verifies automatically the new condition (\ref{s(x)}) emerging 
from a compatibility condition of an order-four linear differential 
operator by pullback is far less obvious. The ``parametrization'' 
(\ref{param}) necessarily yields the Calabi-Yau condition (\ref{Calabi}) 
and the new condition (\ref{s(x)}), and, conversely, (\ref{Calabi})
and (\ref{s(x)}) can be  parametrized by (\ref{param}). 

\vskip .1cm

Our large calculations thus show that the pullback-compatibility 
of an order-four linear differential operator which verifies the Calabi-Yau 
condition (\ref{Calabi}), amounts to saying that this order-four linear differential
operator {\em reduces to} (the symmetric cube of) an underlying 
{\em order-two linear differential operator}. The Schwarzian 
condition (\ref{condition4}) with $\, W(x)$ given by (\ref{wherecond4}), 
is {\em thus inherited from the Schwarzian condition} (\ref{condition1})  
{\em of the underlying order-two linear differential operator}.

\vskip .1cm   

\subsection{Reducible operators} 
\label{nonirreducible}

Throughout the paper we make the assumption that the 
linear differential operators are {\em irreducible}. The reduciblility 
of the linear differential operators is not an academic
scenario: it is  the situation {\em we encounter} (almost)  
{\em systematically with the linear differential operators emerging in physics}, 
typically in the case of the 
$\, n$-fold integral $\, \chi^{(n)}$ of the two-dimensional 
Ising model~\cite{High,bo-gu-ha-je-ma-ni-ze-08,ze-bo-ha-ma-05b}. 
When the linear differential operators are {\em reducible}, 
it is clear that all the calculations of this paper must be 
revisited, taking into account the miscellaneous factorization 
scenarios. 

Sketching the kind of situation we may encounter,
let us consider an order-four linear differential operator
$\, L_4 \, = \, D_x^4 \, + \, p_r(x) \cdot \, D_x^3$
$ \, + \,  q_r(x) \cdot \, D_x^2 \, + \, \cdots \, \, $ 
which factorizes into the product of two  
order-two linear differential operators: 
\begin{eqnarray}
\label{p4q4}
\hspace{-0.95in}&& \quad  
L_4 \, = \,  \, M_2 \cdot \, L_2, \quad \quad \quad  \quad  \quad 
 \quad \quad \quad \hbox{where:} 
\nonumber \\
 \hspace{-0.95in}&& \quad  
 L_2 \, = \, \, D_x^2 \,  \,+ p(x) \cdot \, D_x \, +q(x), \quad \quad \quad 
 M_2 \, = \,  \,D_x^2 \, \, + \tilde{p}(x) \cdot \, D_x \, +\tilde{q}(x),
 \\
 \hspace{-0.95in}&& \quad   
 p_r(x) \, = \,  \, p(x) \, \, + \tilde{p}(x),  \, \quad
 q_r(x)  \, = \,  \,   \tilde{p}(x) \cdot \, p(x) \,  \, +\tilde{q}(x) \,  \, 
+2 \cdot \, {{d p(x)} \over {dx}} \, \,  +q(x), 
 \quad  \, \, \cdots 
\nonumber
\end{eqnarray}
The simple case where the two operators $\, M_2$ and $\, L_2$ are 
identical is sketched in \ref{reducibleapp}.   
In general the exterior square of $\, L_4$ is an order-six linear differential 
operator which is the product of an order-one operator, of 
the symmetric product of $\, L_2$ and $\, M_2$, and of the order-one
linear differential operator $\, D_x \, +p(x)$. Therefore, this reducible
order-four  linear differential operator $\, L_4$ 
{\em does not verify in general}
the Calabi-Yau condition (\ref{Calabi}). 

Imposing the (normalized) pullback 
by $\, y(x)$ of this reducible order-four  linear differential operator 
$\, L_4 \, = \, M_2 \cdot \, L_2$ to be equal to a conjugation by a function
$\, v(x)$ of that operator, it is important to remember that a change 
of variable $\, x \, \rightarrow \, y(x)$ on a 
linear differential operator which is the product of two  operators, 
is the product of these two linear differential operators 
on which this change of variable has been performed. One gets a set 
of equations where one can disentangle two Schwarzian equations
\begin{eqnarray}
\label{disentang}
\hspace{-0.95in}&& \quad  \quad  \quad  \quad  \quad 
 W(x) \, \, \,\,  \,-W(y(x)) \cdot  \, y'(x)^2
\, \, \, \, \,+ \,  \{ y(x), \, x\} 
\, \,\, \, = \,\, \, \,  \, 0, 
\\
\label{disentang2}
\hspace{-0.95in}&& \quad  \quad  \quad  \quad  \quad 
 {\tilde W}(x) \, \, \,\,  \,-{\tilde W}(y(x)) \cdot  \, y'(x)^2
\, \, \, \, \,+ \,  \{ y(x), \, x\} 
\, \,\, \, = \,\, \, \,  \, 0, 
\end{eqnarray}
where $\, W(x)$ and $\, {\tilde W}(x)$ are the functions  (\ref{wherecond})
already encountered in the analysis of order-two linear differential 
operators
\begin{eqnarray}
\label{disentangwhere} 
\hspace{-0.95in}&& \quad  \quad  \quad \quad  \quad  \quad \quad 
W(x)  \, \, = \, \,  \, \, \, {{ d p(x)} \over { dx}} 
 \, \,\,  \,  + \, \, 
   {{p(x)^2} \over {2 }} \, \,  \, \,  -2 \cdot \, q(x),
\\
\hspace{-0.95in}&& \quad  \quad  \quad \quad  \quad  \quad \quad 
\label{disentangwhere2}
 {\tilde W}(x)  \, \, = \, \,  \, \, \, {{ d  {\tilde p}(x)} \over { dx}} 
 \, \,\,  \,  + \, \, 
   {{ {\tilde p}(x)^2} \over {2 }} \, \,  \, \,  -2 \cdot \, {\tilde  q}(x),
\end{eqnarray}
corresponding to  the Schwarzian conditions written separately 
on $\, L_2$ and $\, M_2$, 
together with another relation which couples  $\, L_2$ and $\, M_2$:
\begin{eqnarray}
\label{couples}
\hspace{-0.95in}&& \quad  \quad  \quad  \quad 
4 \cdot\, {{  y''(x)} \over { y'(x)}} \,\, \,+  {\tilde p}(x) \, -p(x)
 \, \, = \, \, \, \Bigl({\tilde p}(y(x)) \, -p(y(x)) \Bigr) \cdot \, y'(x).
\end{eqnarray}
Among the four solutions of the order-four operators $\, L_4 \, = \, M_2 \cdot \, L_2$, 
the two solutions of the order-two linear differential operator $\, L_2$ transform 
nicely under the pullback $\, \, x \, \rightarrow \, \, y(x)$, provided the 
Schwarzian condition (\ref{disentang}) is satisfied, but this just corresponds 
to a {\em partial symmetry}. In general the set of equations (\ref{disentang}),  
(\ref{disentang2}), (\ref{couples}) seems to be too rigid to allow solutions
other than trivial symmetries or partial symmetries.

\vskip .1cm

It is however worth mentioning a quite curious result. If one imposes 
the reducible order-four  linear differential operator 
$\, L_4 \, = \, M_2 \cdot \, L_2 \, $ to verify the Calabi-Yau condition (\ref{Calabi})
(i.e. to be such that the exterior square of that operator is order five instead of order six),
one gets a condition that becomes remarkably simple when written in terms of 
the functions  $\, W(x)$ and $\, {\tilde W}(x)$ given by (\ref{disentangwhere}) 
and  (\ref{disentangwhere2}). Introducing the difference 
 $\, \Delta W(x) \,  = \, \,   W(x)\, -   {\tilde W}(x)$, 
the Calabi-Yau condition (\ref{Calabi}) simply reads:
\begin{eqnarray}
\label{verycurious}
\hspace{-0.95in}&& \quad  \quad  \quad \quad \quad \quad
 2 \cdot \, {{ d \Delta W(x)} \over { dx}}
 \, \,\,  = \, \,  \, \,\, 
 (p(x) \, - \,  {\tilde p}(x))  \cdot  \, \Delta W(x).
\end{eqnarray}
Therefore, if one restricts oneself to $ \,W(x)\, = \, \,  {\tilde W}(x)$ 
which identifies the two Schwarzian conditions
(\ref{disentang}) and (\ref{disentang2}), one sees that condition (\ref{verycurious})
is automatically verified: condition  $ \,W(x)\, = \, \,  {\tilde W}(x)$ 
{\em is thus a sufficient condition for  the Calabi-Yau condition} (\ref{Calabi}).

\vskip .1cm

The analysis of pullback symmetry on {\em reducible} linear differential operators 
is clearly an interesting and challenging problem in physics. It will require 
many more calculations to explore the arborescence of these various 
factorization scenarios.

\vskip .1cm

\subsection{Symmetric Calabi-Yau condition} 
\label{symmCalabiYau}

The condition, we called in~\cite{Diffalg,Canonical} 
{\em symmetric Calabi-Yau condition}
for the order-four linear differential operator $\, L_4$ 
(which correspond to impose that its symmetric square is 
of order less than $\, 10$), is a huge\footnote[1]{This polynomial 
is the sum of $ \, 3548$ monomials in the coefficients of $\, L_4$ and 
their derivatives.} polynomial condition 
on the coefficients of $\, L_4$ and their derivatives.
This condition is invariant by pullback 
and conjugation. Provided the Schwarzian condition (\ref{condition4}) 
with $\, W(x)$ given by (\ref{wherecond4}) 
is satisfied, this symmetric Calabi-Yau 
condition alone is not sufficient  to have $\, L_4^{p} \, = \, \, L_4^{c}$.
Similarly to what we saw with the 
Calabi-Yau condition (\ref{Calabi}), would a supplementary 
condition to the symmetric Calabi-Yau 
condition be sufficient to have $\, L_4^{p} \, = \, \, L_4^{c}$ ?
Could one also have, in this selected
subcase, a reduction of $\, L_4$  to an underlying order-two operator? This 
scenario remains open.

\vskip .1cm 
\vskip .1cm 

Working with such huge polynomials will not get us far.  In order to advance, 
let us introduce a parametrization based on the 
ideas explained in~\cite{Canonical}, namely that an order-four linear 
differential operator $\, L_4$, with an orthogonal differential 
Galois group $\, SO(4, \,  \mathbb{C})$ and such that its symmetric square  
is of order less than $\, 10$, is necessarily of the form\footnote[2]{The differential 
Galois group $\, SO(4, \,  \mathbb{C})$ with an order-$10$ symmetric square situation
corresponds to a decomposition 
$\, L_4 \, = \, \, (U_3 \cdot U_1 \, + \, \, 1) \cdot \, d(x)$, see~\cite{Canonical}.} 
\begin{eqnarray}
\label{L4symm}
\hspace{-0.95in}&& \quad \quad \quad \quad  \quad \quad  \quad \quad  \quad 
L_4 \, \,\, = \, \, \,\, (U_1 \cdot U_3 \, + \, \, 1) \cdot \, d(x), 
\end{eqnarray}
where $\, U_1$ and $\, U_3$ are order-one and  order-three {\em self-adjoint}
linear differential operators:
\begin{eqnarray}
\label{U1U3}
\hspace{-0.95in}&& 
U_3 \,  = \, \, a(x) \cdot \, D_x^3 
\,\,  + \,  {{3} \over {2}} \cdot \, {{d a(x)} \over {dx}}  \cdot \, D_x^2
 \, \, + \,  b(x) \cdot \, D_x \,\,
 + \,  {{1} \over {2}} \cdot \,  {{d b(x)} \over {dx}} 
\,\,  - \,  {{1} \over {4}} \cdot \,  {{d^3 a(x)} \over {dx^3}}, 
\quad \quad 
\\
\label{U1U3U1}
\hspace{-0.95in}&& 
U_1 \,\, \, = \, \, \,\, c(x) \cdot \, D_x
\,\,\,  + \,  \, \, {{1} \over {2}} \cdot \,  {{d c(x)} \over {dx}}.  
\end{eqnarray}
This yields a parametrization of this huge polynomial differential 
(symmetric Calabi-Yau) condition:
\begin{eqnarray}
\label{U1U3param}
\hspace{-0.95in}&& \quad \quad 
p(x) \, \, = \, \, \, \,   {{5} \over { 2}} \cdot \, {{a'(x)} \over {a(x)}}
\,\,  + \, \,  {{1} \over { 2}} \cdot \, {{c'(x)} \over {c(x)}}
\,  \, + \, \,  4 \cdot \, {{d'(x)} \over {d(x)}},
\\
\label{U1U3param1}
\hspace{-0.95in}&& \quad \quad 
q(x) \, \, = \, \, \,\,  {{b(x)} \over {a(x)}}
 \,\,  + \,  {{3} \over { 2}} \cdot \, {{a''(x)} \over {a(x)}}
\,\,  + \,  {{3} \over { 4}} \cdot \, {{a'(x)} \over {a(x)}}
 \cdot \, {{c'(x)} \over {c(x)}} 
 \, \, + \, \,  6 \cdot \, {{d''(x)} \over {d(x)}} 
\nonumber \\
\hspace{-0.95in}&& \quad \quad \quad  \quad \quad 
 \, + \,  {{15} \over { 2}} \cdot \, {{a'(x)} \over {a(x)}}
 \cdot \, {{d'(x)} \over {d(x)}} 
\, + \,  {{3} \over { 2}} \cdot \, {{c'(x)} \over {c(x)}}
 \cdot \, {{d'(x)} \over {d(x)}}, 
\\
\label{U1U3param2}
\hspace{-0.95in}&& \quad \quad 
r(x) \, \, = \, \, \,\,   {{1} \over { 2}}
 \cdot \, {{c'(x)} \over {c(x)}} \cdot \,  {{b(x)} \over {a(x)}}
\, \, + \, \,  4 \cdot \, {{d'''(x)} \over {d(x)}}
 \,\,  + \, \,  4 \cdot \,  {{a'(x)} \over {a(x)}} 
\cdot \, {{c'(x)} \over {c(x)}} \cdot \, {{d'(x)} \over {d(x)}}
\nonumber \\
\hspace{-0.95in}&& \quad \quad \quad \quad \quad 
+{{3} \over { 2}} \cdot \, {{d''(x)} \over {d(x)}} \cdot \, {{c'(x)} \over {c(x)}} 
\,\,  - \,  {{1} \over { 4}} \cdot \, {{a'''(x)} \over {a(x)}} 
\,\,  + \,  {{3} \over { 2}} \cdot \, {{b'(x)} \over {a(x)}} 
\,\,   + \, {{15} \over { 2}} \cdot \, {{d''(x)} \over {d(x)}}
 \cdot \, {{a'(x)} \over {a(x)}}
\nonumber \\
\hspace{-0.95in}&& \quad \quad \quad \quad \quad 
+ 2 \cdot \, {{d'(x)} \over {d(x)}} \cdot \,  {{b(x)} \over {a(x)}} 
\,\,  + 3 \cdot \, {{d'(x)} \over {d(x)}} \cdot \,  {{a''(x)} \over {a(x)}}, 
\\
\label{U1U3param3}
\hspace{-0.95in}&& \quad \quad 
s(x) \, \, = \, \, \, \,  {{d^{(4)}} \over {d(x)}} \, \, \, 
+ \,  {{1} \over { 2}} \cdot \, {{c'(x)} \over {c(x)}} \cdot \, {{d'''(x)} \over {d(x)}}
 \, \, + \,  {{1} \over { 2}} \cdot \, {{b"(x)} \over {a(x)}}
 \, \, - \,  {{1} \over { 4}} \cdot \, {{a^{(4)}(x)} \over {a(x)}}
\nonumber \\
\label{36}
\hspace{-0.95in}&& \quad \quad \quad \quad \quad  \, \,
 - \,  {{1} \over { 8}} \cdot \, {{a'''(x)} \over {a(x)}} \cdot \,  {{c'(x)} \over {c(x)}} \,\, 
+ \,  {{1} \over { 4}} \cdot \,  {{b'(x)} \over {a(x)}}  \cdot \,  {{c'(x)} \over {c(x)}}
\, + {{1} \over {a(x) \, c(x)}}
\nonumber \\
\hspace{-0.95in}&& \quad \quad \quad \quad \quad  \, \,  
- \,  {{1} \over { 4}} \cdot \, {{a'''(x)} \over {a(x)}}\cdot \, {{d'(x)} \over {d(x)}} \, 
+ \,  {{3} \over { 2}} \cdot \, {{b'(x)} \over {a(x)}} \cdot \, {{d'(x)} \over {d(x)}}
\,\,  + \,\, {{b(x)} \over {a(x)}} \cdot \, {{d"(x)} \over {d(x)}}
 \\
\hspace{-0.95in}&& \quad \quad \quad \quad \quad \,\,  
+ \,  {{3} \over { 2}} \cdot \,  {{a''(x)} \over {a(x)}} \cdot \, {{d''(x)} \over {d(x)}}
 \,\,  
+ \,  {{5} \over { 2}} \cdot \,  {{a'(x)} \over {a(x)}} \cdot \, {{d'''(x)} \over {d(x)}}
\nonumber \\
\hspace{-0.95in}&& \quad \quad \quad \quad \quad  \,\,  
+ \,  {{1} \over { 2}} \cdot \,  {{c'(x)} \over {c(x)}}
 \cdot \, {{d'(x)} \over {d(x)}} \cdot \, {{b(x)} \over {a(x)}}
 \,\,  + \,  {{3} \over { 4}} \cdot \, {{a'(x)} \over {a(x)}} 
\cdot \, {{c'(x)} \over {c(x)}} \cdot \, {{d''(x)} \over {d(x)}}.
\nonumber 
\end{eqnarray}
One easily verifies that this parametrization 
(\ref{U1U3param}) ... (\ref{36}) is 
such that the polynomial encoding the symmetric Calabi-Yau condition, 
{\em  is identically equal to zero}. Moreover one verifies 
that the order-four linear differential operator (\ref{L4symm}), with 
parametrization (\ref{U1U3param}), (\ref{U1U3param1}),  (\ref{U1U3param2}),  
(\ref{U1U3param3}), is, generically, such that its symmetric 
square has order $\, 9$ (instead of $\, 10$), its exterior 
square being of order $\, 6$.

Imposing $\, L_4^{(p)} \, = \, \, L_4^{(c)}$ 
for an order-four linear differential operator, 
corresponding to this parametrization (such that it verifies 
the symmetric Calabi-Yau condition, and such that its symmetric 
square is of order nine), one naturally finds the Schwarzian 
condition (\ref{condition4}) with (\ref{wherecond4}), as well
as the exact expression  (\ref{v4}) for the conjugation 
function $\, v(x)$. Taking into account the Schwarzian 
condition (\ref{condition4}), the identification of 
the coefficients of $\, D_x$  for $\, L_4^{(p)}$ 
and  $ \, L_4^{(c)}$ yields a relation of the form 
$\, \Phi(x) \, \,\,  = \, \, \, \,  \Phi(y(x)) \cdot \, y'(x)^3$,
where $\, \Phi(x)$ is a rational function. 
Together with the last condition, this gives an invariance  of the form 
 $\, \Psi(x) \,\,  \, = \, \,  \, \Psi(y(x)) \, $ yielding only 
trivial cases\footnote[2]{See~\cite{Aziz} for similar calculations.} 
for $\, L_4^{(p)} \, = \, \, L_4^{(c)}$.

\vskip .1cm

This symmetric Calabi-Yau condition,  
{\em even if it is invariant by pullback and conjugation},
 is thus {\em not sufficient to get} $\, L_4^{(p)} \, = \, \, L_4^{(c)}$. We 
have here a situation similar to the one described in the previous 
section \ref{CalabiYaucondi}, with the emergence of the additional 
condition (\ref{s(x)}) on top of the Calabi-Yau condition 
(\ref{Calabi}). However here the calculations are way too large: 
finding the additional condition(s) together with the symmetric Calabi-Yau 
condition yielding  $\, L_4^{(p)} \, = \, \, L_4^{(c)}$, is beyond 
our reach for now. The case, described in the previous 
section \ref{CalabiYaucondi}, where the order-four operator 
(\ref{L4symm}) is the symmetric cube of an underlying order-two operator is 
also such that the symmetric square of $\, L_4$ is not of the generic 
order $\, 10$, but, in fact, of order $\, 7$: in this case the coefficients
of  $\, L_4$ {\em verify}\footnote[1]{This can be verified straightforwardly 
substituting (\ref{param}) in the   $ \, 3548$ monomials symmetric 
Calabi-Yau condition.} {\em the symmetric Calabi-Yau condition}. 
Since the calculations are way too large, it is not possible for now to tell
if the additional condition(s) to the symmetric Calabi-Yau condition, also 
gives eventually a linear differential operator that is the 
symmetric cube of an order-two 
operator, as described in the previous section \ref{CalabiYaucondi}, 
or whether it would give something else. This would mean the emergence
of the ``classic'' Calabi-Yau condition (\ref{Calabi}) 
combined with the condition (\ref{s(x)}). This remains an open question.

\vskip .2cm 

\section{Order-$N$ linear differential operators} 
\label{order-N}

The analysis of {\em irreducible} order-five operators is sketched 
in \ref{order-five}. Let us now consider an  {\em irreducible} 
order-$N$ linear differential operator
\begin{eqnarray}
\label{L_N}
\hspace{-0.95in}&& \quad \quad  \quad  \quad \quad  
L_N \, \, \, = \, \, \, \, 
 D_x^N  \, \, \, + \, p(x) \cdot \, D_x^{N-1} \, \,\,  + \, \, q(x)\cdot \, D_x^{N-2}
 \, \,\, \,   + \, \,  \, \cdots
\end{eqnarray}
and let us also introduce two other  linear differential operators of order $N$:
the operator $\, L_N^{(c)}$ conjugated of (\ref{L_N}) by a function $\, v(x)$,
namely $\, L_N^{(c)} \, = \, \, 1/v(x) \cdot \, L_N \cdot \, v(x)$, and the 
(normalized) pullbacked operator $\, L_N^{(p)}$ which amounts to changing 
$\, x \, \rightarrow \, \, y(x)$ in $\, L_N$.  
The pullbacked operator $\, L_N^{(p)}$ reads
\begin{eqnarray}
\label{L_Np}
\hspace{-0.95in}&& \quad   \quad  
L_N^{(p)} \, \, \, = \, \, \, \,\,  D_x^N  \, \,\,  \, + \, 
\Bigl(p(y(x)) \cdot \, y'(x) \,
 - {{N\cdot \, (N-1)} \over {2}} \cdot \,  {{y"(x)} \over {y'(x)}} \Bigr) 
\cdot \, D_x^{N-1}  
\nonumber \\
\hspace{-0.95in}&& \quad \quad \quad  \quad  \quad  
+ \, \Bigl(  q(y(x)) \cdot \, y'(x)^2 \, 
 - {{ (N-2) \cdot \, (N-1)} \over {2}} \cdot \, p(y(x)) \cdot \, y"(x) 
\nonumber \\
\hspace{-0.95in}&& \quad \quad \quad \quad \quad  \quad  \quad  
\,  - {{ N \cdot \, (N-1) \cdot \, (N-2)} \over {6}} 
\cdot \, {{y^{(3)} } \over { y'(x)}}
 \\
\hspace{-0.95in}&& \quad \quad \quad \quad \quad  \quad  \quad  
\,  - {{ (N+1) \cdot \, N \cdot \, (N-1) \cdot \, (N-2)} \over {8}}
 \cdot \, \Bigl({{y^{(2)} } \over { y'(x)}}\Bigr)^2
 \Bigr)  \cdot \, D_x^{N-2} 
\, \,\,  \, \, + \, \,\, \cdots
 \nonumber 
\end{eqnarray}
and the conjugate of (\ref{L_N}) reads:
\begin{eqnarray}
\label{L_Nc}
\hspace{-0.95in}&& \quad \, \, 
L_N^{(c)} \, \, \, = \, \, \, \, \, D_x^N  
\, \, \, \, + \, 
\Bigl(p(x) \, +N \cdot \, {{v'(x)} \over {v(x)}}\Bigr) \cdot \, D_x^{N-1} 
 \\
\hspace{-0.95in}&& \quad \quad 
 \, \, \,\,
 + \,  \Bigl( q(x) \,
 \, +(N-1) \cdot \,  {{v'(x) } \over {v(x) }}
  \cdot \, p(x)  
 \,  \,  + {{N\cdot \, (N-1)} \over {2}} \cdot \, {{v"(x) } \over {v(x) }} 
    \Bigr)  \cdot \, D_x^{N-2} 
\,\, \, \, \, \,  + \, \, \, \cdots
 \nonumber 
\end{eqnarray}
We impose the identification of these two order-$N$ linear differential operators:
\begin{eqnarray}
\label{vLNgenerggfirst}
\hspace{-0.95in}&&  \quad \quad  \quad   \quad   \quad   \quad  
 {{1} \over {v(x)}} \cdot \,  L_N  \cdot \, v(x) 
\, \,\,  = \, \, \,  \, pullback\Bigl(L_N, \,  y(x)  \Bigr).
\end{eqnarray}
The identification of the $\, D_x^{N-1} $ coefficients gives the 
exact expression of $\, v(x)$ in terms of the wronskian $\, w(x)$ 
and of the pullback $\, y(x)$:
\begin{eqnarray}
\label{vN}
\hspace{-0.95in}&& \quad  
v(x) \, \, = \, \, \,  
 y'(x)^{ -(N-1)/2} \cdot \,   \Bigl({{w(x) } \over {w(y(x)) }}\Bigr)^{1/N}
 \quad  \,   \, \hbox{where:}   \,  \,   \,   \,   \,   \quad 
p(x) \, \, = \, \, \, - \, {{w'(x)} \over {w(x)}}.
\end{eqnarray}
Injecting this exact expression in (\ref{L_Nc}), or eliminating the 
log-derivative $\, v'(x)/v(x)$, the identification of the
$\, D_x^{N-2} $ coefficients gives the following Schwarzian equation
\begin{eqnarray}
\label{SchwarN}
\hspace{-0.95in}&& \quad \quad   \quad \quad  \quad \quad  
W(x) 
\, \, \,  \,-W(y(x)) \cdot  \, y'(x)^2
\, \, \,  \,+ \,  \{ y(x), \, x\} 
\, \,\, \, = \,\, \, \,  \, 0, 
\end{eqnarray}
where 
\begin{eqnarray}
\label{wherecondN}
\hspace{-0.95in}&&  
W(x)  \, \, = \, \,  \,  \,
{{6} \over { (N+1) \cdot \, N }} \cdot \,  {{ d p(x)} \over { dx}} 
 \, \,  \,   + \, \,  {{6 \cdot \,   p(x)^2 } \over {(N+1) \cdot \, N^2}} 
 \,  \,  \, 
 - \, {{12 \cdot \,  q(x)} \over {(N+1) \cdot \, N \cdot \, (N-1) }},
\end{eqnarray}
i.e.
\begin{eqnarray}
\label{wherecondNmore}
\hspace{-0.95in}&&  \quad \quad 
W(x)  \, \, = \, \,  \,  \, 
 {{6} \over { (N+1) \cdot \, N }} \cdot \, {\cal W}(x)
 \quad \quad \quad\quad \quad \quad
\hbox{where:} 
 \\
\hspace{-0.95in}&&  \quad \quad 
{\cal W}(x)  \, \, = \, \,  \,  \,  
 {{ d p(x)} \over { dx}} 
 \, \,   \,  + \, \,  {{  p(x)^2 } \over { N}} 
 \,  \,  \,  - \, {{2 \cdot \,  q(x)} \over {N-1 }}
 \, \, \,  = \, \,  \,  \,  \,  
  N \cdot \,  {{z"(x)} \over  {z(x)}} 
 \, \,  \,  - \, {{2 \cdot \,  q(x)} \over {N-1 }},  
\end{eqnarray}
where: 
\begin{eqnarray}
\label{wherecondNbis}
\hspace{-0.95in}&&  \quad \quad \quad  \, \, \, \,  \quad \quad \quad 
 z(x) \, \,  = \, \,  \, w(x)^{-1/N}, \quad \quad \quad \,\, 
 p(x) \, \, = \, \, \, - {{w'(x)} \over {w(x)}}.
\end{eqnarray}

\vskip .1cm

This is in agreement with the fact that the symmetric $\, (N-1)$-th 
power of an order-two linear differential operator 
$\, L_2 \, = \, D_x^2 \, +A(x) \cdot \, D_x \, + \, B(x)\, $ 
gives an order-$N$ linear differential operator 
$\, L_N\, = \,\, D_x^N \,\, +p(x) \cdot \, D_x^{N-1} 
\,\, + \, q(x)\cdot \, D_x^{N-2} \,\, + \, \, \cdots \, \,\,\, $ 
such that
\begin{eqnarray}
\label{suchthat}
\hspace{-0.95in}&&  
p(x) \, \, = \, \, \,
 {{ N \cdot \, (N-1)} \over {2}} \cdot A(x), \quad 
\nonumber \\ 
\hspace{-0.95in}&&  
q(x) \, \, = \, \, \, \,
 {{ (3\,N-1) \cdot \, N \cdot \, (N-1)  \cdot \, (N-2) } \over {24}} \cdot \,  A(x)^{2}
\, \, +\,  {{ N \cdot \, (N-1)  \cdot \, (N+1)} \over {6}} \cdot \,  B(x) 
\nonumber \\ 
\hspace{-0.95in}&&  \quad  \quad  \quad \quad  \quad \quad  \quad 
+ \, {{ N \cdot \, (N-1)  \cdot \, (N-2)} \over  {6}} 
 \cdot \,  {{ d A(x)} \over {dx}}, 
\end{eqnarray}
and thus conversely:
\begin{eqnarray}
\label{suchthatconverse}
\hspace{-0.95in}&&  \quad \quad \quad \quad 
A(x) \,\, = \, \,\, {{2 } \over { N \cdot \, (N-1)}} \cdot \, p(x), \quad 
\nonumber \\
\hspace{-0.95in}&&  \quad \quad \quad \quad 
B(x) \,\, = \, \,\, \,
 {{6 \cdot \, q(x)} \over { (N+1) \cdot \, N \cdot \, (N-1)}}
 \,  \, \,  \, - {{(3\, N \, -1) \cdot \, (N -2)  \cdot \, p(x)^2 } \over {
   (N+1) \cdot \, N^2 \cdot \, (N-1)^2 }} 
\nonumber  \\ 
\hspace{-0.95in}&& \quad  \quad  \quad \quad \quad  \quad \quad \quad \quad 
\,  -\, {{2 \cdot \, (N-2) } \over {(N+1) \cdot \, N \cdot \, (N-1) }} 
 \cdot \,  {{ d p(x)} \over { dx}}. 
\end{eqnarray}
Injecting (\ref{suchthatconverse}) in the expression of $\, W(x)$ for 
an order-two linear differential operator $\, L_2$ (see (\ref{wherecond}))
\begin{eqnarray}
\label{WL2}
\hspace{-0.95in}&&   \quad \quad  \quad \quad  \quad \quad 
W(x)  \, \, = \, \,  \,  \, {{ d A(x)} \over { dx}} 
 \, \, \,   + \, \,  {{ A(x)^2 } \over {2}} 
 \,  \, \,  - \, 2 \cdot \, B(x),
\end{eqnarray}
one gets again the expression (\ref{wherecondN}) 
for $\, W(x)$ for an order-$N$ linear differential operator 
$\,\,\, L_N\, = \,\, $
$\,\,D_x^N \, \, +p(x) \cdot \, D_x^{N-1} $
$\,\,  + \, q(x)\cdot \, D_x^{N-2} \,\,  + \, \, \cdots $ 

\vskip .1cm 
\vskip .1cm 

{\bf Remark:} the Schwarzian condition (\ref{SchwarN}) and the 
associated function $\, W(x)$ given by (\ref{wherecondN}), 
correspond to an elimination of the conjugation function 
$\, v(x)$ in (\ref{vLNgenerggfirst}). If one changes 
the order-$N$ linear differential operator $\, L_N$ by conjugation, 
$\,  L_N \, \rightarrow \, \, $
$\tilde{L}_N \, = \, \, 1/\rho(x) \cdot \,  L_N   \cdot \, \rho(x)$,
one gets again (\ref{vLNgenerggfirst}),  $\, L_N$ being replaced 
by $ \, \tilde{L}_N$ and $\, v(x)$ being replaced by  $ \, \tilde{v}(x)$:
\begin{eqnarray}
\label{new}
\hspace{-0.95in}&&  \,   \,   \quad \quad  \quad \quad  \quad \, \,    
v(x) \quad \quad  \longrightarrow \quad \, \,   \quad 
 \tilde{v}(x) \, \, = \, \,  {{ v(x) \cdot \,  \rho(y(x)) } \over {  \rho(x)  }}.  
\end{eqnarray}
Consequently one gets again the same Schwarzian condition (\ref{SchwarN}) 
with the function $\, W(x)$ given by (\ref{wherecondN}), 
since they are obtained by elimination of the conjugation functions 
$\, v(x)$ or $ \, \tilde{v}(x)$. Therefore $\, W(L_N, \,x)$ given 
by (\ref{wherecondN}), which is invariant by the conjugation 
$\,  L_N \, \rightarrow \, \, 1/\rho(x) \cdot \,  L_N   \cdot \, \rho(x)$, 
is left invariant by:
\begin{eqnarray}
\label{accordingxtext}
\hspace{-0.95in}&&  \,   \,   
 p(L_N, \, x) \, \, \,   \,  \longrightarrow \, \, 
 \quad   \, \,  \, \,   p(L_N, \, x)
\, \, \,\, + \, N \cdot \, {{\rho'(x)} \over {\rho(x)}},
\\
\label{accordingx2text}
\hspace{-0.95in}&&  \,  \,   
  q(L_N, \, x) \, \,  \,   \, \longrightarrow \, \,  \, \, 
 \nonumber \\
\hspace{-0.95in}&&  \quad   \quad  \quad    \quad  
 q(L_N, \, x) \,\,  \, \,
 + \, (N-1) \cdot \, {{\rho'(x)} \over {\rho(x)}} \cdot \,  p(L_N, \, x)
\, \, \, \, 
+ \, {{N \cdot \, (N-1)} \over {2}} \cdot \, {{\rho"(x)} \over {\rho(x)}}. 
\end{eqnarray}
Conversely imposing this invariance by conjugation (\ref{accordingxtext}),  
(\ref{accordingx2text}), on a function of the form 
$\,\,  W(x)  =   \alpha_N \cdot \, p'(x) 
 + \beta_N \cdot \, p(x)^2  + \gamma \cdot \, q(x)\, $
gives  (\ref{wherecondN}) {\em up to an overall constant factor}.

\vskip .2cm 

\section{Solutions of the Schwarzian conditions} 
\label{CalabiSchwResults}

Let us study the solutions $\, y(x)$ of the  Schwarzian equation (\ref{SchwarN})
that  emerge for any pullback-symmetry condition of linear differential 
operators of arbitrary order $\, N$. This should provide valuable
information on the pullbacks that are symmetries of linear 
differential operators.

\vskip .2cm

\subsection{Solutions of the Schwarzian equation that are 
diffeomorphisms of the identity: a condition on $\, W(x)$} 
\label{around}

The Schwarzian condition (\ref{condition1}) has been shown in~\cite{Aziz} 
to be compatible under the {\em composition of the pullback-functions} 
$\, y(x)$ verifying (\ref{condition1}). 
The fact that the composition of two solutions  $\, y(x)$ 
of the  Schwarzian condition (\ref{condition1})
is also a solution\footnote[2]{See Appendix D in~\cite{Aziz}.} of the  
Schwarzian condition (\ref{condition1}), is crucial to describe the set 
of solutions $\, y(x)$ of (\ref{condition1}). Once a solution 
$\, y(x)$ of the  Schwarzian condition (\ref{condition1}) is known, 
the $\, n$-th composition 
$\, y^{(n)}(x) \, = \, \, y(y( \, \cdots \, y(x) \, \cdots \, )))$, 
yields automatically a commuting set of solutions\footnote[5]{Cum grano salis:
when the pullbacks $\, y(x)$ are algebraic functions, they are 
{\em multivalued functions}. The composition 
of multivalued functions is limited to their analytic series expansions 
(setting aside Puiseux series).} 
of (\ref{condition1}). 
By obtaining the series expansions of these solutions, one 
can extend to non integer complex values 
of $\, n$, and in order to build a one-parameter family of {\em commuting} solution 
series, consider the infinitesimal composition~\cite{Hindawi}:
\begin{eqnarray}
\label{yinfini}
\hspace{-0.95in}&& \quad  \quad \quad 
 \quad  \quad  \quad \, \,   \,  \,    \, 
y_{\epsilon}(x) \, \, \, = \,  \, \,\,  \,
  x \, \, \, \,  + \epsilon \cdot \, F(x) \,  \,\, +  \, \, \, \cdots 
\end{eqnarray}
The  one-parameter family of commuting solution series $\, y^{(n)}(x)$ 
commutes with (\ref{yinfini}) yielding the functional 
equations~\cite{Hindawi}:
\begin{eqnarray}
\label{functional}
\hspace{-0.95in}&& \, \,\, \quad  \quad   \, \, 
F(x)  \cdot \,  {{d y^{(n)}(x)} \over {dx}}  \, \, = \, \, \,  F(y^{(n)}(x)),
\quad \quad  \, \,  
 F(x)  \cdot \,  {{d y_{\epsilon}(x)} \over {dx}}  \, \, = \, \, \,  F(y_{\epsilon}(x)).
\end{eqnarray}
Inserting (\ref{yinfini}) in the Schwarzian condition (\ref{condition1}), 
one sees that $\, F(x)$ is actually {\em holonomic} being solution of 
the linear differential equation of {\em order-three}:
\begin{eqnarray}
\label{equaF}
\hspace{-0.95in}&& \quad  \quad \quad  \quad \quad  \,\,
{{d^3 F(x)} \over { dx^3}}  \, \,  \,
 - \, 2 \cdot \,  W(x) \cdot \,  {{d F(x)} \over { dx}}
 \, \, \,  - \, {{d W(x)} \over { dx}}     \cdot \,   F(x) 
\,\, \, \,  = \, \,   \, \,\, \, 0,
\end{eqnarray}
whose corresponding order-three linear differential operator 
$\, {\cal L}_3$ is {\em exactly the symmetric square of an 
underlying order-two linear differential operator}\footnote[3]{The 
reduction of $\, {\cal L}_3$ to a symmetric square  (\ref{equaFop}) 
{\em does not mean that} $\, F(x)$ is solution of a second order 
linear differential (Liouvillian) equation 
$\, F"(x)/F(x) \, \, \, = \,  \, \,\, \,  W(x)/2$.} 
$\, {\cal L}_2$:
\begin{eqnarray}
\label{equaFop}
\hspace{-0.95in}&& \quad  \quad  
 {\cal L}_3  \,  \, = \,  \, 
 D_x^3  \, \, - \, 2 \cdot \,  W(x) \cdot \, D_x \, \, - \, {{d W(x)} \over { dx}} 
 \, \,  \,  \, = \,  \,  \, \,\,  Sym^2\Bigl( D_x^2 \, -\, {{W(x)} \over {2}}  \Bigr).
\end{eqnarray}
Conversely $\, W(x)$ can be expressed in terms of $\, F(x)$ as follows:
\begin{eqnarray}
\label{equaFthus}
\hspace{-0.95in}&& \quad  \quad \quad \quad \quad 
W(x)\, \, \, = \,  \, \,\, \, \,   
{{F"(x) } \over { F(x)}} \,  \,
 -{{1} \over {2}} \cdot \,\Bigl( {{F'(x) } \over { F(x)}} \Bigr)^2 
\, \, \,  + \, \, {{ \lambda} \over {F(x)^2}}
\\
\label{equaFthusricatti}
\hspace{-0.95in}&& \quad  \quad \quad \quad \quad \quad \quad 
\, \, \, = \,  \, \,\, \, \,  
{{d } \over {dx}} \Bigl(  {{F'(x) } \over { F(x)}} \Bigr)  \, 
+ \, \, {{1} \over {2}} \cdot \,\Bigl( {{F'(x) } \over { F(x)}} \Bigr)^2 
\, \, \,  + \, \, {{ \lambda} \over {F(x)^2}}.
\end{eqnarray}
This last result (\ref{equaFthus}) is easily obtained 
by multiplying the LHS of (\ref{equaF}) by $\, F(x)$ and integrating 
the result. One gets this way\footnote[1]{This ``gauge'' 
$\,  W(x) \, \rightarrow \, W(x) \,+ \, \lambda/F(x)^2 \,  $ 
in (\ref{equaFthus}) corresponds to the fact that because 
of (\ref{functional}) one has 
$\, \lambda/F(x)^2 \, - \lambda/F(y(x))^2 \cdot \, y'(x)^2 \, = \, 0$ 
which allows to change  
$\,  W(x) \, \rightarrow \, \, W(x) \,+ \, \lambda/F(x)^2\, $ 
in the Schwarzian equation  (\ref{condition1}), as well as in the 
third order linear differential ODE (\ref{equaF}). One easily verifies 
that inserting (\ref{equaFthus}) in (\ref{equaF}) {\em gives an identity}. 
}:
\begin{eqnarray}
\label{Casimir}
\hspace{-0.95in}&& \quad   \quad \quad 
F(x) \cdot \, {{d^2 F(x)} \over {dx^2}} \, \, \, 
 -{{1} \over {2}} \cdot \,  \Bigl({{d F(x)} \over {dx}}\Bigr)^2
 \, \,  + \, \lambda \, \,  \,  -\, F(x)^2 \cdot \, W(x) \,  \,\, = \, \, \, 0,
\end{eqnarray}
which is (\ref{equaFthus}). 
Thus, for a given pullback $\, y(x)$, or for
a given {\em one-parameter} family of commuting solution series (\ref{yinfini}), 
or for a given $\, F(x)$, one has a one-parameter family (\ref{equaFthus})
of $\, W(x)$ in the Schwarzian equation  (\ref{condition1}). Conversely, 
for a given  $\, W(x)$, one has {\em at least} a one-parameter family of 
commuting solution series (\ref{yinfini}). 

\vskip .1cm

\subsubsection{Selected subcase of the Schwarzian equation. \\} 
\label{subcase}

\vskip .1cm

Let us consider an order-two linear differential operator
 $\, L_2\, = \, \, D_x^2 \, + \, A(x) \cdot \, D_x  \, + \, B(x)$ 
(where $A(x)$ and $\, B(x)$ are rational functions), 
such that its corresponding function
$\, W(x) \, = \, \,\, A'(x)\, +A(x)^2/2 \, -2 \, B(x) \, \, $ 
(see (\ref{wherecond})) in the  
Schwarzian equation (\ref{condition1}), is of the form 
(see subsection 6.2 of~\cite{Aziz}) 
\begin{eqnarray}
\label{WAR}
\hspace{-0.95in}&&  \quad \quad \quad \quad \quad \quad \quad 
 W(x) \,  \, = \, \,  \, \,  
 {{d A_R(x)} \over {dx}} \,\, + \, {{ A_R(x)^2 }  \over {2}}, 
\end{eqnarray}
where $\, A_R(x)$ is a {\em rational function}.
Introducing the rational function $\, C(x) \, = \, (A(x) \, - A_R(x))/2$,  
the identification of the expression of $\, W(x)$, namely 
$\, W(x) \, = \, \, A'(x)\, +A(x)^2/2 \, -2 \, B(x)$
with (\ref{WAR}), gives $\, B(x)$ in terms of $\, A_R(x)$ and $\, C(x)$
\begin{eqnarray}
\label{Bx}
\hspace{-0.95in}&&  \quad \quad \quad \quad \quad \quad \quad 
 B(x) \,  \, = \, \,  \, \,  
  \, {{d C(x)} \over {dx}} 
\,\,   +  C(x) \cdot \, ( C(x) \,  \, + \, A_R(x)),    
\end{eqnarray}
which is the condition for the order-two linear differential operator 
$\, L_2$ {\em to factorize into two order-one linear differential operators}:
\begin{eqnarray}
\label{factoL2}
\hspace{-0.95in}&&  \quad \quad \quad \quad \quad \, \, 
 L_2\,\, = \, \, \,
 \Bigl(D_x \, + \, A_R(x) \, + \,C(x)\Bigr) \cdot \, \Bigl(D_x \, \, + \,C(x)\Bigr).
\end{eqnarray}
In other words, condition (\ref{WAR}) with $\, A_R(x)$ a rational function, is 
the condition of factorization of the order-two linear differential operator 
$\, L_2$. In this case, the Schwarzian equation (\ref{condition1}) 
reduces to a simpler second order {\em non-linear} differential 
equation (that was studied extensively in~\cite{Aziz,Hindawi}):
\begin{eqnarray}
\label{WAR2}
\hspace{-0.95in}&&  \quad \quad \quad \quad \quad 
 {{d^2 y(x)} \over {dx^2}}  \,  \, = \, \,  \, \,  
 A_R(y(x)) \cdot \, \Bigl({{d y(x)} \over {dx}}\Bigr)^2 
\,\, - \,  A_R(x) \cdot \, {{d y(x)} \over {dx}}. 
\end{eqnarray}
Seeking the following one-parameter solutions (\ref{yinfini}), 
$\, y_{\epsilon}(x) \, = \,   
x \, + \,  \epsilon \cdot \, F(x) \, +  \, \cdots$,  
one finds that $\, F(x)$ verifies a linear 
differential equation of order two~\cite{Hindawi} 
\begin{eqnarray}
\label{FepsAR}
\hspace{-0.95in}&& \quad  \,  
 \quad  \quad  \quad \, \,   \,  \,    \, 
{{d^2 F(x)} \over { dx^2}}  \, \,  \,
 -  \,  A_R(x) \cdot \,  {{d F(x)} \over { dx}}
 \, \, \,  - \, {{d A_R(x)} \over { dx}}     \cdot \,   F(x) 
\,\, \, \,  = \, \,   \, \,\, \, 0,
\end{eqnarray}
corresponding to the linear differential 
operator of order two\footnote[1]{In fact the order-two operator 
$\, {\cal L}_F$  is the adjoint of the operator 
$\, \, \Omega = \,(D_x \, +A_R(x)) \cdot \, D_x$ 
(see~\cite{Hindawi}). When  $\, A_R(x) \, = \, \, -w'(x)/w(x)$ 
the linear differential operator $\, {\cal L}_F$ is conjugated by the 
wronskian $\, w(x)$ to the linear differential operator $\, \Omega$, 
namely $\, \Omega \cdot \, w(x) = \, w(x) \cdot \,{\cal L}_F$.}:
\begin{eqnarray}
\label{FepsARop}
\hspace{-0.95in}&& \quad  \,  
 \quad  \, \,  \,  \,  \, 
{\cal L}_F  \,  \,  \, = \, \,  \,  \, 
D_x^2  \, \, 
 -  \,  A_R(x) \cdot \, D_x
 \, \, \,  - \, {{d A_R(x)} \over { dx}}
  \,  \, \, = \,  \, \,  \, D_x \cdot \, 
\Bigl(D_x\, -A_R(x)\Bigr).
\end{eqnarray}
Introducing the wronskian $\, w(x)$, $\, A_R(x)$ reads
 $\, A_R(x) \, = \, \, -w'(x)/w(x)$. Thus  the linear differential 
operator (\ref{FepsARop}) has two solutions:  $\, 1/w(x)$ which is 
the solution of the right factor $\, D_x\, -A_R(x)$, and another (transcendental) 
solution that we denote $\, {\cal S}_F$. The function $\, F(x)$ 
corresponds to this last (transcendental) solution, and 
{\em not the} $\, 1/w(x)$ 
{\em solution}. Conversely $\, A_R(x)$ can be 
expressed\footnote[2]{Just integrate the LHS of (\ref{FepsAR}).} 
in terms of $\, F(x)$ as follows:
\begin{eqnarray}
\label{conversFepsAR}
\hspace{-0.95in}&& \quad \quad \quad  \quad  \quad  \quad \quad  \,  
A_R(x) \,  \,  = \, \, \,\, {{ F'(x)} \over { F(x)}} 
\, \, + \, \, {{\mu} \over { F(x)}}.
\end{eqnarray}
One easily verifies that by inserting 
(\ref{conversFepsAR}) in (\ref{FepsAR}) ones gets
an identity, and that by inserting (\ref{conversFepsAR}) 
in (\ref{WAR}) one recovers (\ref{equaFthusricatti}) with
$\, \lambda \, = \, \mu^2/2$. Here the $\, \mu/F(x)$ 
{\em term is crucial}, because when
$\,  \mu \, = \, 0$ condition (\ref{conversFepsAR})
 with $\, A_R(x) \, = \, \, -w'(x)/w(x)$  yield 
the trivial result,  $\,\, F(x) \, = \, 1/w(x)$ which is
different from the transcendental (holonomic) function 
we are looking for. For instance in the example detailed 
in~\cite{Hindawi}, we 
had the condition (\ref{conversFepsAR})  verified 
with $\, \mu \, \ne \, 0$, namely $\,\mu \, = 1/4$:
 \begin{eqnarray}
\label{Hindwaiexample}
\hspace{-0.95in}&&  \quad  \quad 
F(x) \, = \,   \, 
 x \cdot \, (1-x)^{1/2} \cdot \, 
_2F_1\Bigl([{{1} \over {2}}, \, {{1} \over {4}}],[{{5} \over {4}}], \, x\Bigr), 
\, \, \quad  \,\,\, A_R(x) \, = \, \, {{ 3 \,-5\,x } \over {4 \, x \, (1\, -x) }}.
\end{eqnarray}

\vskip .2cm  
  
At first sight one expects the order-two linear differential 
equation (\ref{FepsAR}) on $\, F(x)$ to be a simple limit 
of the order-three linear differential 
equation (\ref{equaF}) when the condition 
(\ref{WAR}) is imposed.
This reduction is not obvious however and the interested reader can
find it explained in \ref{Reduction32}. 

\vskip .2cm    

{\bf Remark:} the global nilpotence of the linear differential 
operators gives an $\, A_R(x)$ of the form  
$\, A_R(x) \, = \, \, -w'(x)/w(x)$, 
where the wronskian $\, w(x)$ is an $\, N$-th root of a 
rational function~\cite{bo-bo-ha-ma-we-ze-09}. Using 
$\, A_R(x) \, = \, \, -w'(x)/w(x)$, 
condition (\ref{WAR2}) can be  easily integrated into
\begin{eqnarray}
\label{WAR21}
\hspace{-0.95in}&& \quad \quad \quad \quad \quad \quad \quad 
{{ d y(x)} \over { dx}} \, \, = \, \, \,  \, 
c_1 \cdot \, {{ w(x)} \over {w(y(x))}}
 \quad   \quad  \quad   \quad   \quad  \hbox{or:}  
\\
\label{WAR22}
\hspace{-0.95in}&& \quad \quad  
\Theta(y(x)) \, \, = \, \, \,  \, c_1 \cdot \,  \Theta(x) \, + \, c_2
 \quad  \quad   \quad\hbox{with:}  \quad  \quad  \quad 
\Theta(x) \, = \, \, \int^x \, w(x) \, dx 
\end{eqnarray}
where $\, c_1$ and  $\, c_2$ are constants of integration.

\vskip .1cm

\vskip .1cm

\vskip .1cm

 Now let us describe 
this one-parameter family of commuting solution series (\ref{yinfini}) 
of the  Schwarzian equation  (\ref{condition1}).

\vskip .1cm

\subsection{Solutions of the Schwarzian equation that are diffeomorphisms 
of the identity: the general formal solution} 
\label{formal}

Let us consider (\ref{yinfini}) as a series in $\, \epsilon$:
\begin{eqnarray}
\label{expansion}
\hspace{-0.95in}&&  \quad \quad \quad \, \,  \quad \quad 
y_{\epsilon}(x)  \, \, = \, \, \, \,  \,
 x \, \, \, \, +  \, \epsilon \cdot \, F(x) \, \, \, 
+ \, \sum_{n=2}^{\infty} \,  {{\epsilon^n} \over {n!}} 
 \cdot \,  F(x)  \cdot \, Q_n(x),  
\end{eqnarray}
 solution of the functional equation (\ref{functional}). This 
 is sufficient to find, order by order in $\, \epsilon$, 
the solution (\ref{expansion})  of (\ref{functional}) 
where the $\, Q_n(x)$ are given by
\begin{eqnarray}
\label{recursion}
\hspace{-0.95in}&&  \quad \quad  \quad 
Q_1(x) \, = \, \, F(x), \quad \quad \quad  \quad  \, \,  \, \,  \,
Q_2(x)  \, = \,\, \,  F(x) \cdot \, {{ d Q_1(x) } \over {dx}} 
\, = \, \, \,  F(x) \cdot \, {{ d F(x)  } \over {dx}} , 
\nonumber \\
\hspace{-0.95in}&&  \quad \quad   \quad 
Q_3(x)  \, = \, \,  F(x)  \cdot \, {{ d } \over {dx}} \, Q_2(x) 
 \,\,  = \, \, \,   F(x) \cdot \, \Bigl(F(x) \cdot \, F"(x)
 \,\,  + \, F'(x)^2\Bigr),
\nonumber \\
\hspace{-0.95in}&&  \quad  \quad  \quad  
Q_4(x)  \, = \, \,  F(x)  \cdot \,  {{ d } \over {dx}}  \, Q_3(x), 
\quad \quad  \quad  \, \,  
Q_5(x)  \, = \, \,  F(x)  \cdot \, {{ d } \over {dx}} \, Q_4(x), 
\nonumber \\
\hspace{-0.95in}&&  \quad \quad  \quad   \quad \quad 
\, \,   \, \, \,  \cdots
 \quad \quad  \quad  \quad 
Q_{n+1}(x)  \, = \, \, \,  F(x)  \cdot \, {{ d } \over {dx}}  \, Q_n(x),
\end{eqnarray}
the most general solution (\ref{expansion}) of (\ref{functional}) 
corresponding to linear combinations of the $\, Q_n$'s 
which amounts to changing 
$\, \epsilon$ in (\ref{expansion}) into:
\begin{eqnarray}
\label{change}
\hspace{-0.95in}&&  \quad  \quad  \quad \quad  \quad \quad 
\epsilon \, \, \,\, \, \, \longrightarrow \, \, \,  \, \,  \quad 
 \epsilon \cdot \, 
(1 \, + \, \lambda_1 \cdot \epsilon \, 
 +\,  \lambda_2 \cdot \epsilon^2
 \, +\,  \lambda_3 \cdot \epsilon^3 \, +\, \, \cdots ).
\end{eqnarray}
Note that all the $\, Q_n$'s are {\em polynomial expressions of} 
$\, F(x)$ {\em and its derivatives}.

The functional equation (\ref{functional}) corresponds to the one-form 
 $\,d \Theta   \,  = \, \, dx/F(x)  \,  = \, \, dy/F(y)$ giving:
\begin{eqnarray}
\label{deriv}
\hspace{-0.95in}&&  \quad \quad   \quad \quad  \quad    \quad \quad
\Theta(x)  \, \, = \, \, \, \int^{x} \, {{ dx} \over {F(x)}}, 
 \quad \quad \quad \, \,  \,  
 {{d} \over { d\Theta}} \, \,  = \, \, \,\,  F(x) \cdot \, {{ d  } \over {dx}}.   
\end{eqnarray}
Seeing $\, x$ as a function 
of $\, \Theta$, one finds that the series (\ref{expansion}) 
together with the recursion (\ref{recursion}),
gives the well-known Taylor expansion
\begin{eqnarray}
\label{Taylor}
\hspace{-0.95in}&&  \quad \quad  \quad  \quad  \quad 
 y_{\epsilon}(x(\Theta)) \, \, = \, \, \,\, \, 
 x(\Theta) \,   \,  \, 
+ \, \sum_{n=1}^{\infty} \,  {{\epsilon^n} \over {n!}}
  \cdot \,   \,  {{d^n  \, x(\Theta)} \over { d\Theta^n}}
\, \, = \, \, \,\, x(\Theta \, + \, \epsilon),
\end{eqnarray}
meaning that  $\, x \, \rightarrow \, \, y_{\epsilon}(x)$ 
{\em is just a shift in} $\, \Theta$
\begin{eqnarray}
\label{justashift}
\hspace{-0.95in}&&  \quad \quad \quad  \quad  \quad  \quad \quad  \quad    
\, \Theta_x   \,\quad   \longrightarrow \, \quad   \, 
 \Theta_y   \,  \, = \, \,  \, \,  \Theta_x \,  + \, \, \epsilon, 
\end{eqnarray}
corresponding to the integration of the one-form
$\,d \Theta   \,  = \, \, dx/F(x)  \,  = \, \, dy/F(y)$.
The two transformations 
$\, y_{\epsilon_1}(x)$ and  $\, y_{\epsilon_2}(x)$ 
of the one-parameter family clearly commute\footnote[2]{This can 
also be checked directly using (\ref{expansion}) with 
(\ref{recursion}) for any rational function $\, F(x)$.}:
\begin{eqnarray}
\label{deriv2}
\hspace{-0.95in}&&  \quad \quad  \quad  \quad  \quad 
y_{\epsilon_1}(y_{\epsilon_2}(x(\Theta))) 
\, \, = \, \, \,\, y_{\epsilon_1}(x(\Theta\, + \, \epsilon_2))
 \, \, = \, \, \,\, x(\Theta\, + \, \epsilon_1 \, + \, \epsilon_2).
\end{eqnarray}

One verifies order by order in $\, \epsilon$,  that the one-parameter 
family of commuting series (\ref{expansion}) with (\ref{recursion})
{\em is solution of the Schwarzian equation}
\begin{eqnarray}
\label{condition1ter}  
\hspace{-0.95in}&& \quad   \quad \quad   \quad  \quad   \quad 
 W(x) 
\, \, \,\,  \,-W(y_{\epsilon}(x)) \cdot  \, y_{\epsilon}'(x)^2
\, \, \, \, \,+ \,  \{ y_{\epsilon}(x), \, x\} 
\, \,\, \, = \,\, \, \,  \, 0, 
\end{eqnarray}
where $\, W(x)$ is given by (\ref{equaFthus}). In terms of $\, \Theta$, 
the expression (\ref{equaFthus}) for $\, W(x)$ 
can be  written using the Schwarzian derivative:
\begin{eqnarray}
\label{identityQbis}
\hspace{-0.95in}&&  \quad \quad  \quad  \quad   \quad  \quad  
W(x) \,\,  + \,  \{\Theta(x), \, x\}  \,\,\, 
 -\, \lambda \cdot \, \Bigl( {{d \Theta(x)} \over {dx}}\Bigr)^2
\, \, = \, \, \, 0. 
\end{eqnarray}
Recalling the chain rule for the Schwarzian derivative of a 
composition of functions\footnote[9]{Namely 
$ \, \, \, \{\Theta(y(x)), \, x\} \, \,\,  = \,\,  \,$ 
$\,\{\Theta(y(x)), \, y(x)\}  \cdot  \, y'(x)^2 \,\,\, + \,  \{ y(x), \, x\}$. } 
and the fact that $\, d \Theta(y(x))/dx \, = \, \, d \Theta(x)/dx$, 
one finds that the Schwarzian condition (\ref{condition1ter}) 
corresponds to the equality of the two Schwarzian derivatives:
\begin{eqnarray}
\label{identityQbis}
\hspace{-0.95in}&&  \quad  \quad   \quad 
 \quad    \quad  \quad    \quad   \quad    \quad  
\{\Theta(y(x)), \, x\}  \, \, \, = \, \, \, \,   \{\Theta(x), \, x\},  
\nonumber 
\end{eqnarray}
which is verified since 
$\, d \Theta(y(x))/dx \, = \, \, d \Theta(x)/dx$.
This is another way
to see that the one-parameter 
family of commuting series (\ref{expansion}) 
(with the $\, Q_n$'s given by (\ref{recursion}))
 is  solution of the Schwarzian equation.

\vskip .2cm 

\subsection{A simple modular form example. } 
\label{example}

 We have considered in~\cite{Aziz,IsingCalabi,IsingCalabi2,Christol,perimeter} 
many examples of {\em modular forms}  
represented as pullbacked $\, _2F_1$ hypergeometric functions. Each time the 
one-parameter commuting series combined with the 
modular correspondences~\cite{Eichler} series yields one-parameter series
of the form $\, y_n(x) \, = \, \, a_n \cdot \, x^n \, + \, \, \cdots, \, $
 $\, n \, = \, 2, \, 3, \, 4, \,  \cdots \, \, $ that are solutions of the
 Schwarzian equation (\ref{condition1ter}). 

In~\cite{Aziz} the pullback symmetry of the order-two linear 
differential operator was given as a covariance of its  
solution, namely a hypergeometric function
with {\em two different\footnote[5]{We exclude the trivial well-known changes 
of variables on hypergeometric functions 
$\, x \, \rightarrow $
$ \, \, 1\, -x, \, 1/x, ...$ The transformation 
$\, x \, \rightarrow \, \, y(x)$ must be an {\em infinite order} 
transformation symmetry.} pullbacks} related by modular 
equations\footnote[3]{
The emergence of a {\em modular form}~\cite{IsingCalabi,IsingCalabi2,RatMaier} 
corresponds to the emergence of a selected hypergeometric function 
having an exact covariance property~\cite{Stiller,Zudilin} 
with respect to an {\em infinite order algebraic transformation} 
(the modular correspondences).
}
\begin{eqnarray}
\label{modularform2explicit}
\hspace{-0.95in}&& \quad \quad  \quad  \quad  \quad 
 _2F_1\Bigl([{{1} \over {12}}, \, {{5} \over {12}}], \, [1], \, y(x)  \Bigr)
\,\, = \, \, \, 
 {\cal A}(x) \cdot \,
 _2F_1\Bigl([{{1} \over {12}}, \, {{5} \over {12}}], \, [1], \,  x  \Bigr),
\end{eqnarray}
the pullback $\, y(x)$ being solution of the Schwarzian 
condition (\ref{condition1ter}).

In this example, the pullback $\, y_{\epsilon}(x)$ is solution of 
the Schwarzian solution (\ref{condition1ter}) with $\, w(x)$ and $\, F(x)$ 
given by\footnote[1]{ One can easily check that these 
expressions (\ref{firstbister}) 
for $\, W(x)$ and $\, F(x)$ verify  (\ref{equaF}).}:
\begin{eqnarray}
\label{firstbister}
\hspace{-0.96in}&&  
W(x)  = \, \,
 -{\frac {32\,{x}^{2}-41\,x+36}{72\,{x}^{2} \cdot \, (x-1)^{2}}}, 
\, \,  \, \,  \,   \, 
F(x) = \, 
 x  \cdot \, (1-x)^{1/2} \cdot \,
 _2F_1\Bigl([{{1} \over {12}}, \, {{5} \over {12}}], \, [1], \, x\Bigr)^2.
\end{eqnarray}
One can also check that these expressions (\ref{firstbister})  
 verify  (\ref{equaFthus}) with\footnote[2]{This selected value of $\, \lambda$ 
has to be compared with the value $\, \mu = \, 1/4$ in (\ref{Hindwaiexample}).} 
$\, \lambda \, = \, \, 0$,
{\em thus providing a quite non-trivial (non-linear differential)
identity between the rational function} $\, W(x)$ {\em and the holonomic 
function} $\, F(x)$. 

The one-parameter commuting 
family (\ref{yinfini}) solution of the Schwarzian equation 
(\ref{condition1ter}) can be expressed using the two (mirror maps) 
{\em differentially algebraic}~\cite{Selected,IsTheFull}
functions $\, P(x)$ and $\, Q(x)$
described in~\cite{Aziz} and in \ref{Mirror},  
as $\,\,\, y_1(a_1, \, x)  =  P(a_1 \cdot \, Q(x))$:
\begin{eqnarray}
\label{Qseries}
\hspace{-0.95in}&&   
y_1(a_1, \, x) \,  = \, \, a_1 \cdot \, x \, \, 
-{\frac {31\,a_1 \cdot \, (a_1 \,-1) }{72}} \cdot \, x^2  \, 
 +{\frac {a_1 \cdot \, (9907\,{a_1}^{2} -30752\,a_1 +20845) }{82944}} \cdot \, x^3 
\nonumber  \\
\hspace{-0.95in}&&  \,     \quad    \quad      \quad  
\, -{\frac {a_1 \cdot \, (a_1 -1) \cdot \,
 (4386286\,{a_1}^{2} -20490191\,a_1 +27274051) }{161243136}} \cdot \, x^4
\, \, \,\, + \, \cdots 
\end{eqnarray}
where $\,\,\, a_1 \, = \, \, \exp(\epsilon)$.  

Besides this one-parameter commuting family (\ref{yinfini}), the Schwarzian 
equation  (\ref{condition1ter}) has a remarkable (infinite) set of algebraic 
functions solutions~\cite{Aziz}  $\, y(x)$ defined by the corresponding 
{\em modular equations}~\cite{Morain,Andrews,Atkin,Hermite,Hanna,Weisstein}. Their 
series expansions near 
$\, x \, = \, \, 0$ read:
\begin{eqnarray}
\label{Qseriesn}
\hspace{-0.95in}&&    \quad \quad       \quad \quad  \quad  
y_n(x) \, \,\, = \, \, \, P(Q^n(x)) \,  \, = \, \,\, \, \,
  1728 \cdot \, \Bigl({{ x} \over {1728}}\Bigr)^n  \, \, + \,   \, \cdots 
\end{eqnarray}
where $\, n$ is an integer $\, n \, = \, 2, \, 3, \, 4, \cdots \, \, $ These 
series expansions {\em commute for different values of the integer} 
$\, n$. This is a consequence of the fact that, up to the 
previous change of variables $\, P(x)$, $\, Q(x)$, these 
{\em modular correspondences} (\ref{Qseriesn}) correspond to taking 
the $\, n$-th power of the nome: 
$\, q \, \rightarrow \, q^n$ (see~\cite{Aziz} for more details).

\vskip .2cm 

\subsubsection{A pre-modular concept. \\} 
\label{premodul}

The composition of the one-parameter series (\ref{yinfini}) 
(which corresponds to 
$\, q \, \rightarrow \, a_1 \cdot \, q$) and of the modular correspondences 
(\ref{Qseriesn}), yields an {\em infinite set of one-parameter series} 
$\, y_n(x) \, = \, \, a_n \cdot \, x^n \, + \, \, \cdots, \, \, $
 $ n \, = \, 2, \, 3, \, 4, \,\, \cdots \, \,  \, $
for instance~\cite{Aziz}:
\begin{eqnarray}
\label{seriesmodcurve3a}
\hspace{-0.95in}&& 
y_3  = \,  
a_3 \cdot \, {x}^{3} \, 
+{\frac {31\, a_3  }{24}} \cdot \, {x}^{4}\, 
+{\frac {36221\, a_3  }{27648}} \cdot \, {x}^{5}  \,  
-{\frac {a_3   \cdot \, (23141376\, a_3 -66458485)
}{53747712}} \cdot \,  {x}^{6}  + \, \, \cdots 
\nonumber 
\end{eqnarray}
These  one-parameter series do not commute but verify~\cite{Aziz} 
the simple composition 
formulae\footnote[5]{Consequence of the fact, in the nome, they 
correspond to the composition of transformations like 
$\,  q \, \rightarrow \, a_n \cdot \, q^n$.}:
\begin{eqnarray}
\label{curious5}
\hspace{-0.95in}&& \quad   \quad  \quad   \,  \,  \, 
y_n(a_n, \, \, y_m(a_m,  \,\, x)) \, \, = \, \,  \,\, y_{n m}(a_n a_m^n, \, \, x),
 \quad \, \,   \, \, \,\, \,\, \,
n, \, m \, \, = \, \, \, 1, \, 2, \, 3, \,  \, \,  \, \cdots 
\end{eqnarray}
When the $\, a_n$ are arbitrary rational numbers the corresponding series 
$\, y_n(a_n, \, x)$ are {\em not globally bounded series}~\cite{Christol} in general. 
Therefore, they {\em cannot be the series expansion of an algebraic function}: 
they are {\em differentially algebraic}~\cite{Selected,IsTheFull} since 
they are solutions of the Schwarzian equation (\ref{condition1ter}).

\vskip .2cm 

In general, finding the Schwarzian equation (\ref{condition1ter}) is easy, and
getting solutions order by order as series expansions is also easy. 
However finding the selected values of the rational numbers $\, a_n$ 
such that the 
{\em differentially algebraic}~\cite{Selected,IsTheFull} series
$\, y_n(a_n, \, x)$ are globally bounded and 
{\em thus can be algebraic functions}, and, possibly, modular 
correspondences, is a quite difficult 
task\footnote[2]{Similar to finding the selected values
of the parameters so that a quantum Hamiltonian becomes  integrable,
or finding modular forms among Beukers' second order differential equations
depending on three parameters~\cite{Zagier} (36 cases emerging
of a numerical exploration of 10 millions triples). }.

We will call ``pre-modular\footnote[8]{Of course, this ``pre-modular'' term
should not be confused with the term premodular in premodular 
categories, (ribbon fusion categories). Here we mean prerequisites
for the emergence of modular forms.}'' the existence of an infinite set of 
one-parameter differentially algebraic series (solution of the 
Schwarzian equation) of the form 
$\, y_n(x) \, = \, \, a_n \cdot \, x^n \, + \, \, \cdots \, $ which 
verify (\ref{curious5}), but for which 
{\em one does not know if there exist some selected values of the parameter} 
$\, a_n$ such that these {\em differentially algebraic} 
series~\cite{Selected,IsTheFull} become {\em algebraic functions}.

\vskip .1cm

In the next section, we will characterize the  Schwarzian 
equations corresponding to these ``pre-modular'' structure, thus finding 
{\em conditions that are necessary} for the emergence of modular forms.

\vskip .2cm 

\subsection{Schwarzian equation: conditions for modular correspondence } 
\label{minus1over2}

In the previous sections it was shown that the pullback symmetry condition 
of {\em arbitrary} order-two linear differential operators
yields Schwarzian equation (\ref{condition1ter}). The solutions of these
order-two linear differential operators {\em are much more general than 
hypergeometric functions and Heun functions}~\cite{Aziz}: they can 
have an {\em arbitrary number of singularities}. Let us see which  
Schwarzian equation (\ref{condition1ter}), or equivalently, which 
function $\, W(x)$ gives relations (\ref{curious5}) corresponding to
{\em  rigid constraints necessary to have modular correspondences}~\cite{Aziz}.

Series calculations give the conditions on $\, W(x)$ 
such that series solutions of the form 
$\,  \, y_n(x) \, = \, \, a_n \cdot \, x^n \, + \, \, \cdots \, $ are 
solutions of the Schwarzian equation with these $\,  \, y_n(x)$'s verifying 
relations (\ref{curious5}). These constraints are  conditions on the  
{\em Laurent series} of $\, W(x)$. 
For the solution series of the Schwarzian equation to have the 
pre-modular structure (\ref{curious5}), i.e. the same structure 
as modular correspondences, 
the  Laurent series of $\, W(x)$ must be of the form:
\begin{eqnarray}
\label{Laurent}
\hspace{-0.95in}&& \quad \quad \quad \quad  \quad \quad 
W(x)  \, \, \, = \,\, \,\,
  \, -\, {{1} \over {2 \, x^2}}\, \, \,  + \,  {{b_1} \over {x}}
 \,\,\, + \, \sum_{m=0}^{\infty} \, a_m \cdot \, x^m. 
\end{eqnarray}
One easily verifies that this is the case for the previous modular form 
example where $\, W(x)$ reads (\ref{firstbister}), as well as for all the other 
modular forms emerging in physics or enumerative combinatorics we mentioned 
in previous papers~\cite{IsingCalabi,IsingCalabi2,Christol,Diffalg,perimeter}.

\vskip .2cm

Condition (\ref{Laurent}) is a result whose scope transcends the
hypergeometric functions framework. 
In order to show this, let us apply this result on the open
problem of finding Heun functions\footnote[1]{Finding the selected values 
of the parameters of a Heun function~\cite{Heun1} (in particular the 
{\em accessory parameter}~\cite{Katz}) such that its series
expansion is a series with {\em integer coefficients} (or more generally is globally 
bounded~\cite{Christol}), or such that the corresponding order-two linear differential 
operator is {\em globally nilpotent}~\cite{bo-bo-ha-ma-we-ze-09} is a difficult problem. 
These classification problems are closely related to finding the Heun functions 
reducible to pullbacked hypergeometric functions~\cite{VidunasHoeij}, 
and to modular forms~\cite{Zagier}.} that  
could be modular forms~\cite{RatMaier}, or pullbacked $\, _2F_1$ 
functions~\cite{maier-05,BelyiMaier}.
 The Heun function 
$\, {\it HeunG} \left( a, \,q, \, \alpha,\, \beta,\, \gamma,\, \delta, \, x \right)$ 
is solution of a linear differential operator of order two 
$\, L_2 \, = \, \, D_x^2 \, +A(x) \cdot \, D_x \, +B(x)\,$ 
where $\, \,A(x)$ and  $\, B(x)$
read:
\begin{eqnarray}
\label{Heun2F1with}
\hspace{-0.96in}&&   \quad    
A(x)  \,\, = \, \, \,  \,  \, 
{\frac { (\alpha+\beta \, +1) \cdot \,  {x}^{2} \, 
- \, ( ( \delta +\gamma) \cdot \,  a 
\, +\alpha \, -\delta\,  +\beta\, +1) \cdot \,  x 
\,\, +\gamma \cdot \,a}{ x \cdot \, (x-1)  \cdot \, (x \, -a) }},
\\
\label{Heun2F1withB}
\hspace{-0.96in}&& 
\quad \quad \quad \quad \quad \quad \quad 
B(x) \, \, = \, \, \, \,  \, 
 {\frac {\alpha \, \,\beta \cdot \,x \, \, -q}{
x \cdot \, (x-1)  \cdot \, (x \, -a) }}.
\end{eqnarray}
The corresponding function $\, W(x)$ is easily deduced
from the formula (\ref{wherecond}) given by 
$\, W(x) \, = \, A'(x) \, A^2(x)/2 \, -2 \, B(x)$. It
has the following {\em Laurent series} expansion:
 \begin{eqnarray}
\label{Heunexpansion}
\hspace{-0.96in}&&    \quad     \,    
W(x)\, \, = \, \, \, \,  \, 
\,{\frac {\gamma \cdot \, (\gamma  -2) }{{2 \, \, x}^{2}}}  \, \, \, \,   
-{\frac {a \, \delta\,\gamma \, +\alpha\,\gamma+\beta\,\gamma \,
 -\delta\,\gamma-{\gamma}^{2}+\gamma-2\,q}{ a \, \, x}}
 \,\,  \,  \, \, + \, \, \, \cdots,  
\end{eqnarray}
and has the form (\ref{Laurent}) given by
$\, -1/2/x^2 \, + \, \, \cdots \, \,\, $
 {\em only when} $\, \gamma \, = \, \, 1$. 
Thus a general analytical constraint like (\ref{Laurent}) yields a
simple exact constraint on the intriguing problem of the classification 
of the Heun functions that can be modular forms, and more specifically 
on the necessary conditions for the Heun functions  to have a ``pre-modular'' structure.

\vskip .2cm

\subsubsection{Schwarzian equation for $\, \, \, W(x) \, = \, -1/2/x^2$. \\} 
\label{limit}

In order to understand the Laurent series condition (\ref{Laurent}), let 
us try to see what is so ``special'' in the case where
 $\, W(x) \, = \, -1/2/x^2$. For
\begin{eqnarray}
\label{first}
\hspace{-0.95in}&& \quad \quad   \quad \quad \quad \quad   \, 
W(x)  \, \, \, = \,\, \,  
-\, {{1} \over {2 \, x^2}} \, \, = \,\, \,  \,  \, - \,  \{ \ln(x), \, x\},
\end{eqnarray}
the most general solutions of corresponding Schwarzian equation read:
\begin{eqnarray}
\label{solSchwarrec}
\hspace{-0.95in}&& \quad \quad  \quad \quad  \quad  \quad \quad  \quad \quad
y(x) \, \, \,  = \, \,  \, \,
 \exp\Bigl({{a \cdot \ln(x) \, + \, b } \over { 
c \cdot \ln(x) \, + \, d }}   \Bigr),
\end{eqnarray}
which just amounts to a simple transformation on $\, \ln(x)$:
\begin{eqnarray}
\label{justamounts}
\hspace{-0.95in}&& \quad \quad    \quad \quad  \quad \quad
\ln(x) \quad \longrightarrow \quad \quad 
\ln(y(x))  \, \, \,  = \, \,  \, \, \,
 {{a \cdot \ln(x) \, + \, b } \over { c \cdot \ln(x) \, + \, d }}.
\end{eqnarray}
The solutions of the form
$\, y_n(x) \, = \, \, a_n \cdot \, x^n \, + \, \, \cdots \, \, $ 
are given by $\, \, \, y_n(x) \, = \, \, \, a_n \cdot \, x^n$ and are thus ``trivial'': 
this is the case because the nome\footnote[2]{Such that the transformations 
$ x \, \rightarrow \,  y_n(x) \, = \, \, a_n \cdot \, x^n \, + \, \, \cdots \, \, $ 
simply reduce to $\, q \, \rightarrow \,  a_n \cdot \, q^n$, see 
the concept of mirror maps~\cite{Aziz}. } $\, q$ {\em is nothing but 
the} $\, x$ {\em variable!} Similarly, the ratio of periods $\, \tau$ is 
just $\, \ln(x)$, and thus the condition $\, W(x) \, = \, -1/2/x^2$ 
is a ``trivialization'' of the mirror map.

\vskip .1cm

\subsubsection{Rank-two condition (\ref{WAR2})  and pre-modular  structures. \\} 
\label{rank}

The factorization of the order-two linear differential 
operator which corresponds to $\, W(x)$ of the form (\ref{WAR}), yields the 
rank-two {\em non-linear} differential equation (\ref{WAR2}) 
(see section \ref{subcase}).
We would like to know when the modular correspondences structures  (existence of 
solutions series $\,  \, y_n(x) \, = \, \, a_n \cdot \, x^n \, + \, \, \cdots, \, $
$n \, = \, 2, \, 3, \, 4, \cdots \, \, $ such that (\ref{curious5}), thus 
requiring $\, W(x)\, = \, -1/2/x^2 \, + \, \cdots$) 
are compatible with a factorization of the order-two linear 
differential operator and thus with condition (\ref{WAR}). Imposing 
\begin{eqnarray}
\label{WARminus1over2}
\hspace{-0.95in}&&  \quad \quad \quad  \quad \quad 
 W(x) \,  \, = \, \,  \, \,  
 {{d A_R(x)} \over {dx}} \,\, + \, {{ A_R(x)^2 }  \over {2}}
 \,  \, = \, \,  \, \,  
- \, {{1} \over {2 \, x^2}} \,  \, \, \, + \, \, \cdots  
\end{eqnarray}
where $\, A_R(x)$ is a rational function, one finds that $\, A_R(x)$  
must have the following Laurent series expansion:
\begin{eqnarray}
\label{LaurentAR}
\hspace{-0.95in}&& \quad \quad \quad \quad  \quad \quad 
 A_R(x)  \, \,\, \, = \,\, \,\,
  \, \, {{1} \over { x}} 
\,\,\, \, + \, \sum_{m=0}^{\infty} \, r_m \cdot \, x^m. 
\end{eqnarray}
This result (\ref{LaurentAR}) can be directly obtained 
by looking for the {\em Laurent series} 
for $\, A_R(x)$ with a pre-modular structure, i.e. such that the series 
$\,  \, y_n(x) \, = \, \, a_n \cdot \, x^n \, + \, \, \cdots, \, $
$\, n \, = \, 2, \, 3, \, 4, \cdots \, \, $ are 
solutions of 
 condition (\ref{WAR2}). As a byproduct, one finds that 
in the case (\ref{LaurentAR}) the 
solutions $\,  \, y_n(x) \, = \, \, a_n \cdot \, x^n \, + \, \, \cdots \, $
are such that (\ref{curious5}). In particular the 
solution $\,  \, y_1(x) \, = \, \, a_1 \cdot \, x \, + \, \, \cdots \, $
is a one-parameter family of commuting series. The case $\, W(x)\, = \, -1/2/x^2$, 
or $\, A_R(x)  \, \, \, = \,\, 1/x$,
corresponds to the simple order-two linear differential 
operator $\, \theta^2$ where $\, \theta$ is 
the homogeneous derivative $\, \theta\, = \, \, x \cdot \, D_x$.

\vskip .2cm

More specifically, if one revisits our Heun classification problems,
imposing the {\em factorization condition} (see the analysis sketched 
in \ref{subcaseHeun}) {\em together} with the condition 
(\ref{Laurent}) required for the 
{\em emergence of modular correspondence structure} 
(\ref{curious5}), one gets the following Laurent series expansion
(see (\ref{mustbe}) for the definition of the $\, u, \, v, \, w$ parameters): 
\begin{eqnarray}
\label{LaurentARHeun}
\hspace{-0.95in}&& \quad \quad \quad \quad \quad \quad \quad 
W(x) \,\, = \, \, \,  \,
{\frac {v \cdot \, (v \, -2) }{ 2 \cdot \, x^{2}}}
 \, \,\,\, -{\frac {v \cdot \, (a\,w\,+u) }{a \,\cdot  \, x}}
\, \,\,\,\, + \, \,\, \cdots 
\end{eqnarray}
This gives the condition $\, v\, = \, \, 1$ (in agreement 
with condition (\ref{LaurentAR})) and four other conditions.
Excluding the case $\, a \, =\, 0$ corresponding to the reduction 
from the four singularities of the Heun function 
to three singularities, one gets 
$\, \gamma \, = \, \, v\, = \, \, 1$. 
The Heun function
$\, {\it HeunG} \left( a, \,0, \, 0,\, \beta,\, 1,\, \delta, \, x \right)$ 
is a (Liouvillian) solution of a reducible linear differential operator 
of order two 
$\, L_2 \, = \, \, (D_x \, +A_R(x)) \cdot \, D_x$,  where $\, A_R(x)$
then reads:
\begin{eqnarray}
\label{ARgood}
\hspace{-0.96in}&&   \quad  \quad \quad 
   \quad  \quad   \quad  \quad \quad    
A_R(x)  \,\, = \, \, \,  \,  \, 
{{1} \over {x}} \,\, \, + \, {{\delta } \over {x \, -1 }}
 \,\,  \,+ \, {{ \beta \, -\delta} \over {x\, -a }}.
\end{eqnarray}
The pullbacks $\, y(x)$ are solutions of the rank-two non-linear 
differential equation (\ref{WAR2}) which can easily be 
integrated into (see (\ref{WAR21}), (\ref{WAR22})):
\begin{eqnarray}
\label{ARgood1}
\hspace{-0.96in}&&   \quad   \quad  
\quad \quad  \quad  \quad  \quad 
x \cdot \, {{ y'(x)} \over {y(x)}} 
  \, \,\, = \, \, \, \,  \,  c_1 \cdot \, 
 {{ (y(x)-1)^{\delta} \cdot \, (y(x)-a)^{\beta \, -\delta}} \over { 
(x-1)^{\delta} \cdot \, (x-a)^{\beta \, -\delta}} }, 
\end{eqnarray}
giving a functional equation on the pullbacks 
$\, y(x)$ with an Abel integral $\, \Theta(x)$:
\begin{eqnarray}
\label{WAR2good}
\hspace{-0.96in}&&  
\Theta(y(x) \, = \, \, c_1 \cdot \Theta(x) \, + \, c_2 
\quad \hbox{where:} \, \,  \quad  
\Theta(x)  \,\, = \, \, \, 
\int^x \, {{ dx } \over {
x \cdot \, (x\, -1)^{\delta} \cdot \, (x-a)^{\beta \, -\delta}  }}. 
\end{eqnarray}
One has for instance the following one-parameter series solutions 
for the pullback $\, y(x)$, which verify (\ref{curious5}):
\begin{eqnarray}
\label{WAR2goody1}
\hspace{-0.96in}&& \quad  \quad  \quad   \quad  \quad  
y_1 \,\,\, = \,\, \, \,\, a_1 \cdot \, x \,\,\, \, 
 - \, a_1 \cdot \, (a_1 \, -1)\cdot \, 
{{ a\, \delta \, + \beta \, -\delta} \over {a }} \cdot \, x^2
 \, \,\,\, + \, \, \, \cdots 
\\
\hspace{-0.96in}&& \quad  \quad  \quad   \quad  \quad  
y_2 \,\,\, = \,\, \, \,\, a_2 \cdot \, x^2 \,\, 
\, \, + \, \,  2 \cdot \,
 {{ a\, \delta \, +\beta \, -\delta } \over {a }}
 \cdot \, a_2 \cdot \, x^3
 \,\,\,\, + \, \,  \cdots 
\end{eqnarray}
The fact that solutions of the form 
$\, y(x) \, = \, \, a_n \cdot \, x^n \, + \, \, \cdots\,\,$
occur can be clearly seen on equation (\ref{ARgood1}). Even if 
the ``pre-modular'' conditions (\ref{curious5}) 
are verified for this example, this Heun function
$\, {\it HeunG} \left( a, \,0, \, 0,\, \beta,\, 1,\, \delta, \, x \right)$ 
will not be necessarily a modular form represented as a pullbacked 
$\, _2F_1$ hypergeometric function with more than one pullback
for generic parameters\footnote[1]{The exponent-differences 
at the four singularities are: 
$\,0, 1\,-\delta, \,1 \,+\delta \,-\beta, \beta$. Introducing 
$\, e_1$,$\, e_2$,$\, e_3$ the exponents difference of the three 
singular points of the $\, _2F_1$ hypergeometric function each the
previous exponent-differences must be a multiple of the $\, e_i$'s.}. 

\vskip .2cm

\section{Pullback symmetry of an operator up to equivalence of operators} 
\label{pullbackhomomorphism}

With the aim of generalizing 
 covariance  (\ref{modularform2explicit}), we introduce 
the derivative of $\, _2F_1([1/12,5/12],[1],x)$ 
\begin{eqnarray}
\label{modularform2explicitPhi}
\hspace{-0.95in}&& \quad 
 \Phi(x) \,  \, = \, \, \, 
 {{d } \over { dx}} 
\Bigl( \, _2F_1\Bigl([{{1} \over {12}}, \, {{5} \over {12}}], \, [1], \, x  \Bigr) \Bigr)
\,\, = \, \, \, 
 {{5} \over { 144}} \cdot \,
  _2F_1\Bigl([{{13} \over {12}}, \, {{17} \over {12}}], \, [2], \, x  \Bigr), 
\end{eqnarray}
which {\em does not} correspond to a modular form, 
{\em since the derivative of a modular form is not a modular form}.  A derivative 
of the simple covariance identity (\ref{modularform2explicit}) gives
\begin{eqnarray}
\label{modularform2explicitder}
\hspace{-0.95in}&& \quad  \quad  \, \,      
 \Phi(y(x)) \cdot \, y'(x) 
\, = \, \, \, \,\,  {\cal A}(x) \cdot \, \Phi(x)
\, \, \,  + \, 
  {\cal A}'(x) \cdot \,
 _2F_1\Bigl([{{1} \over {12}}, \, {{5} \over {12}}], \, [1], \,  x  \Bigr).
\end{eqnarray}
Using the order-two linear differential equation
verified by  $\, _2F_1([1/12,5/12],[1],x)$, one can rewrite the 
 $\, _2F_1([1/12,5/12],[1],x)$ in the RHS of (\ref{modularform2explicitder}),
as a linear combination of $\, \Phi(x)$
and its derivative $\, \Phi'(x)$. One then deduces from 
relation (\ref{modularform2explicitder}) a slightly 
more general relation
than the initial simple covariance (\ref{modularform2explicit})
\begin{eqnarray}
\label{moregeneralPhi}
\hspace{-0.95in}&& \quad \quad \quad  \quad  \quad 
 \Phi(y(x)) \,  \,  \, = \, \,  \, \, \Bigl(   {\cal A}_{\Phi}(x) \cdot \, {{d} \over { dx}}
\, \, + {\cal B}_{\Phi}(x)   \Bigr)
\, \cdot \, \Phi(x),
\end{eqnarray}
where $\, {\cal A}_{\Phi}(x)$ and $\,  {\cal B}_{\Phi}(x)$ read in this particular 
example\footnote[2]{If instead of the simple derivative 
(\ref{modularform2explicitPhi}) we had introduced 
 $\, \Phi(x) \, = \,\, $$L_1( \,_2F_1([1/12,5/12],[1],x))$ where  $\, L_1$ is an 
arbitrary order-one linear differential operator, we would have also obtained 
a relation of the form (\ref{moregeneralPhi}) but where  $\, {\cal A}_{\Phi}(x)$ 
and $\,  {\cal B}_{\Phi}(x)$ are much more involved expressions.   }:
\begin{eqnarray}
\label{particularmoregeneralPhi}
\hspace{-0.95in}&& 
{\cal A}_{\Phi}(x)  \,   = \, \,
  {{144  \cdot \, x \cdot \, (x-1)  \cdot \, {\cal A}(x) } \over { 5 \cdot \, y'(x)}}, 
\quad \quad 
 {\cal B}_{\Phi}(x) \,   = \, \, 
  {{ 5 \cdot \,  {\cal A}(x) \, +\, 72 \cdot \, (2 \,-3\,x) 
\cdot \,  {\cal A}'(x) } \over { 5 \cdot \,  y'(x)}}.
\nonumber 
\end{eqnarray}
Recalling two Hauptmoduls $\, p_1(x)$ and $\, p_2(x)$
\begin{eqnarray}
\label{recallHaupt}
\hspace{-0.95in}&& \quad \quad  \quad  \quad \,
p_1(x) \,\, = \, \, \,{{ 1728 \cdot \, x } \over {(x\, +16)^3 }}, 
\quad \quad  \quad \quad 
 p_2(x) \, \,= \, \,\, {{ 1728 \cdot \, x^2 } \over {(x\, +256)^3 }}, 
\end{eqnarray}
one can also write relation (\ref{moregeneralPhi}) in a more ``balanced''  form  
(see equation (7) in~\cite{Hindawi}).
Introducing the two algebraic functions $\, A_1(x)$ and  $\, A_2(x)$
\begin{eqnarray}
\label{recallHaupt2}
\hspace{-0.95in}&& \quad \quad  \quad  \quad 
A_1(x) \,\, = \, \, \,\Bigl(1 \, +\, {{x} \over {16}} \Bigr)^{-1/4}, 
 \quad \quad  \quad 
A_2(x) \, \,= \, \, \,\Bigl(1 \, +\, {{x} \over {256}} \Bigr)^{-1/4}, 
\end{eqnarray}
one has the (modular form) hypergeometric identity: 
\begin{eqnarray}
\label{modularformident}
\hspace{-0.95in}&& 
\, \, \,  \, \,  \, \, 
A_1(x) \cdot \,
 _2F_1\Bigl([{{1} \over {12}}, \, {{5} \over {12}}], \, [1], \, p_1(x) \Bigr)
\,\,  = \, \, \, 
A_1(x) \cdot \,
 _2F_1\Bigl([{{1} \over {12}}, \, {{5} \over {12}}], \, [1], \, p_2(x)  \Bigr). 
\end{eqnarray}
After performing calculations of a similar nature of the ones previously seen,
one deduces
the $\, 1 \, \leftrightarrow \, 2$ balanced relation on $\, \Phi(x)$:
\begin{eqnarray}
\label{modularformidentder}
\hspace{-0.95in}&&  \quad \quad \quad \, \, \, 
144 \cdot \, p_1(x) \cdot \, (p_1(x) \, -1) 
\cdot \, {{d A_1(x)} \over { dx}} \cdot \,  \Phi'(p_1(x))
 \nonumber \\
\hspace{-0.95in}&& \quad \quad \quad \quad \quad
\, + \, \Bigl(72 \cdot \, (3\,p_1(x) \, -2) \cdot \,  {{d A_1(x)} \over { dx}}
 \, - 5 \cdot \,  A_1(x) \cdot \,  
 {{d p_1(x)} \over { dx}} \Bigr) \cdot \, \Phi(p_1(x))
 \nonumber \\
\hspace{-0.95in}&& \quad \, \, \, 
\, = \, \,  \,  \, 144 \cdot \, p_2(x) \cdot \, (p_2(x) \, -1) 
\cdot \, {{d A_2(x)} \over { dx}} \cdot \,  \Phi'(p_2(x))
 \\
\hspace{-0.95in}&& \quad \quad \quad \quad \quad
\, + \, \Bigl(72 \cdot \, (3\,p_2(x) \, -2) \cdot \,  {{d A_2(x)} \over { dx}}
 \, - 5 \cdot \,  A_1(x) \cdot \, 
  {{d p_2(x)} \over { dx}} \Bigr) \cdot \, \Phi(p_2(x)),
\nonumber 
\end{eqnarray}
which should be viewed as a (rational) parametrization of the relation having 
the form (\ref{moregeneralPhi}).

The interested reader shall find in \ref{Asimpler} 
a detailed (and we hope pedagogical) analysis of the 
more general relation (\ref{moregeneralPhi}) given
for a selected  hypergeometric function\footnote[5]{We thank A.J. Guttmann for 
showing us this remarkable hypergeometric function emerging in a 
dual context of combinatorics and random-matrix theory, counting 
the number of avoiding permutations~\cite{Conway,Bona}.} solution 
$\, \,  _2F_1([-1/4, 3/4],[1],x)$.

\vskip .1cm

Let us provide an example of the relevance of the relation (\ref{moregeneralPhi}) 
in the context of integrable models in physics.
In the case of the two-dimensional Ising model, the covariance (\ref{moregeneralPhi})
is instantiated on  $\, \tilde{\chi}^{(2)}$, the simplest of the low-temperature
$\, n$-fold integrals
 $\, \tilde{\chi}^{(n)}$ occurring in the decomposition 
of the susceptibility of the square 
Ising model~\cite{High,bo-gu-ha-je-ma-ni-ze-08,ze-bo-ha-ma-05b} 
(see subsection 5.1 in~\cite{Automaton}). When applied to $\, \tilde{\chi}^{(2)}$,
the {\em Landen transformation}  $\, k \, \rightarrow \, \, $
$k_L \, = \, \,  {{2 \sqrt{k}} \over {1 \, + \, k}}$, which provides 
an exact representation of a generator of the renormalization 
group~\cite{Hindawi,broglie,Heegner}, gives the following covariance relation
(see equation\footnote[2]{Note a misprint in the
 expression of the Landen transformation in the unlabelled equation
 above equation (62) in~\cite{Automaton}.} (64) in~\cite{Automaton}):
\begin{eqnarray}
\label{Automat}
\hspace{-0.95in}&& \quad \quad \quad \quad \quad \,  \,  \, 
\tilde{\chi}^{(2)}\Bigl( {{2 \sqrt{k}} \over {1 \, + \, k}}  \Bigr) 
  \,\,  = \, \,  \,  \, 
 4 \cdot \, {{1\, + \, k} \over { k}} \cdot \, 
{{ d \, \tilde{\chi}^{(2)}(k)} \over { d k}}, 
 \\
\label{Automat2}
\hspace{-0.95in}&& \quad \quad \,  \,  \, 
\hbox{where:} \quad \quad \quad \quad \quad
 \tilde{\chi}^{(2)}(k)  \,\, \,  = \, \,  \,  \,  
{{k^4} \over {4^3}} \cdot \,
 _2F_1\Bigl( [{{3} \over {2}}, \,  {{5} \over {2}}], \,\, [3],\, \, k^2\Bigr). 
\end{eqnarray}
This relation (\ref{Automat}) can also be written as 
\begin{eqnarray}
\label{Automat2}
\hspace{-0.95in}&& \quad \quad \,  \,  \, 
\tilde{\chi}^{(2)}(k)  \,\,  = \, \,  \,  \, 
{{1} \over {4}} \cdot \, \Bigl( k \cdot \, (k-1)  \cdot \, {{d} \over {dk}} 
\, + \, {{k^2\, +k\, +2 } \over {k\, +1}}   \Bigr) 
\,  \,  \tilde{\chi}^{(2)}\Bigl( {{2 \sqrt{k}} \over {1 \, + \, k}}  \Bigr), 
\end{eqnarray}
or introducing the {\em inverse Landen transformation} (descending Landen transformation):
\begin{eqnarray}
\label{inverseLanden}
\hspace{-0.95in}&& \quad  \quad \quad 
{{ 1 \, -(1\, -k^2)^{1/2} } \over {  1 \, +(1\, -k^2)^{1/2} }} 
\, \,   \,   \,   = \, \,  \,  \, {\frac{k^2}{4}}  \,   \, +{\frac{k^4}{8}} \,   \,  
+{\frac{5}{64}}{k}^{6} \,   \,  +{\frac{7}{128}}{k}^{8} \,  \,   +{\frac{21}{512}}{k}^{10} \,   \,  
 \, + \, \, \cdots,  
\end{eqnarray}
\begin{eqnarray}
\label{Automat3}
\hspace{-0.95in}&& \quad  \quad   
\tilde{\chi}^{(2)}\Bigl( {{ 1 \, -(1\, -k^2)^{1/2} } \over {  1 \, +(1\, -k^2)^{1/2} }}  \Bigr)
 \,  \,  \,  \,   = \,  \,   \,  \,   
\Bigl(  {{(k^2 \, -2)  \cdot \, (1\, -k^2)^{1/2} \, + \, 2 } \over { 
4 \, k^2}} \Bigr)  \cdot \,  \tilde{\chi}^{(2)}(k)
\nonumber \\
\hspace{-0.95in}&&  \quad  \quad \quad \quad  \quad  \quad  \quad  \quad 
 \,   + \,   \,  \, 
 {{k^2 \, -1 } \over { 4 \, k}} \cdot \, 
\Bigl(1 \, - \, (1\, -k^2)^{1/2}\Bigr)  \cdot \, {{d \tilde{\chi}^{(2)}(k)} \over {d k}}. \,  \, 
\end{eqnarray}

\vskip .1cm

{\bf Remark:} Note that the premodular condition (\ref{Laurent}), 
$\, W(x) \, = -1/2/x^2 \, + \, \, \cdots, \, $
has no reason to be verified for such generalizations of modular 
forms (\ref{modularform2explicitPhi}), (\ref{moregeneralPhi}).  For instance 
for $\, \tilde{\chi}^{(2)}$ given by (\ref{Automat2}), the function 
$\, W(x) \, = \, p'(x)\, +p(x)^2/2 \, -2\, q(x)$ (see (\ref{wherecond}))  
 has the following Laurent series 
expansion (here $x\, = \, k$):
\begin{eqnarray}
\label{Laurentdonotmatch}
\hspace{-0.95in}&& \quad \quad  
W(x) \, \,  = \, \,  \,
  {{3} \over {2}} \cdot \,  {{x^2 \, -5 } \over {x^2 \cdot \, (x^2 \, -1) }} 
 \, \,  = \, \,  \,  \,
{{15} \over {2}} \cdot \, {{1} \over {x^2}}
 \, \, \, + 6 \, \,  + \, 6 \, x^2 \,  \, + \, \, 6 \, x^4 \,  \,\, + \, \, \cdots 
\end{eqnarray}

\vskip .2cm

More generally these  (hypergeometric) examples provide  simple illustrations of 
a more general pullback symmetry, where one imposes  the pullback of an  
order $\, N$ linear differential operator to be {\em homomorphic to that operator}.
In this case there exists two intertwiners (of order $\, N-1$ in general) 
$\, L_{N-1}$ and $\, \, M_{N-1}$, such that:  
\begin{eqnarray}
\label{onepullbackhomo}
\hspace{-0.95in}&&  \quad   \quad   \quad   \quad   \quad    \quad  
 M_{N-1} \cdot \,  L_N \, \,\,  = \, \, \,  \, 
pullback\Bigl(L_N, \,  y(x)  \Bigr) \cdot L_{N-1}.
\end{eqnarray}
The pullback symmetry up to conjugation studied in sections 
\ref{order-two}, \ref{order-three}, \ref{order-four}, 
\ref{order-N}, \ref{CalabiSchwResults}  is  appropriate for 
modular forms~\cite{IsingCalabi,IsingCalabi2,Christol,perimeter},
but {\em not for derivatives of modular forms} that 
{\em also occur in physics} (see for instance the previous 
relation (\ref{Automat}) on the 
square Ising model). The emergence of such generalized covariance 
(\ref{onepullbackhomo}) for the representation of the Landen transformation
(and more generally the modular correspondences providing exact 
representations of the generators of the renormalization group) on the 
other $n$-fold integrals $\, \tilde{\chi}^{(n)}$'s of the susceptibility 
of the Ising model~\cite{High,bo-gu-ha-je-ma-ni-ze-08,ze-bo-ha-ma-05b}
is a {\em challenging open problem}, that will require one to consider 
 {\em reducible operators} (see subsection \ref{nonirreducible}).

Analyzing these more general constraints (\ref{onepullbackhomo})
will require many additional assumptions (beyond the 
one of having selected differential Galois group) 
on the linear differential operator  $\, L_N$ to be able 
to perform more calculations. 

\vskip .2cm

\section{Schwarzian conditions for different Calabi-Yau operators with the same Yukawa couplings} 
\label{CalabiSchwatext}

In the previous sections we have analyzed the question of the covariance under algebraic 
pullbacks of a linear differential operator of arbitrary order $\, N$, i.e. the 
question of linear  differential operators with algebraic pullback symmetries. Let 
us consider here the more general problem  of the equivalence under 
pullbacks up to conjugations of {\em two different linear  differential operators}, 
which is an enlightening
 sieve when one tries to classify selected 
linear differential operators in theoretical physics (Calabi-Yau linear  
differential operators~\cite{Duco,Tables}). The interested reader 
will find in \ref{CalabiSchwa} an illustration of this 
important question where we revisit in detail some calculations of a paper by  Almkvist, 
van Straten and Zudilin~\cite{Duco}. This calculation reexamines the
question of pullback equivalence up to 
conjugation, of two selected order-four operators $\, L_4$ and ${\cal L}_4$ verifying 
the Calabi-Yau condition:
\begin{eqnarray}
\label{vL4text}
\hspace{-0.95in}&&   \quad \quad \quad \quad  \quad  \quad 
 v(x) \cdot \,  {\cal L}_4   \cdot \, {{1} \over {v(x)}}
\, \,   = \, \, \, 
pullback\Bigl(L_4, \,  {\frac {-4 \, x}{ (1 \, -\,x)^{2}}}  \Bigr), 
\\
\hspace{-0.95in}&&   \quad \quad \quad \quad \quad \quad \quad \quad 
\, \, \, \,  \hbox{with:} \, \, \,  \quad  \quad \quad \quad 
v(x) \, = \, \,  \Bigl({{ x \cdot (1\, +x)} \over { 1\, -x}} \Bigr)^{1/2}.
\end{eqnarray}
One finds that a  Schwarzian equation  verified 
by these two order-four linear differential operators
 $\, L_4$ and ${\cal L}_4$ reads:
\begin{eqnarray}
\label{SchwarUUhatAlmkvistext}
\hspace{-0.95in}&& \quad  \quad \quad  \quad  \quad 
\hat{U}_R(x)  \,\, \,  \, - \, U_M(y(x)) \cdot \, y'(x)^2 
\, \,\,  + \,  \{y(x), \, x\} 
 \,\,\, \, = \, \,\, \,   \, 0, 
\end{eqnarray}
where $\, U_M(x)$ and  $\, \hat{U}_R(x)$ are given by (\ref{wherecond4}),
and where $\, p(x)$ and $\, q(x)$ are the coefficients of 
$\, D_x^3$ and  $\, D_x^2$ for respectively $\, L_4$ and $\, {\cal L}_4$, 
 (see (\ref{U}) and (\ref{Uhat}) in \ref{CalabiSchwa}).

\vskip .1cm

One sees on this example that the nome and Yukawa couplings, expressed 
in terms of the $\, x$ variable, are related 
(see (\ref{nomeL4}), (\ref{YukcalL2})) by the pullback 
transformation. Yet, the Yukawa couplings of the two linear differential 
operators {\em expressed in term of the nome}, 
are related  in an even simpler and ``universal'' way:
 $\, K_q({\cal L}_4) \, \, = \, \, \, K_q(L_4)(-4 \cdot \, q)$,
as shown in Appendix E of~\cite{Christol}. For 
a pullback $\, y(x)$ with a series expansion of the form 
\begin{eqnarray}
\label{pullform}
\hspace{-0.95in}&&    \quad \quad 
\quad \quad \quad \quad \quad  \quad \quad 
y(x) \,\, \,\, = \, \, \, \,\,\lambda \cdot x^n \,\,\, \, + \, \, \cdots 
\end{eqnarray} 
the nome and Yukawa couplings expressed in terms of the $\, x$ variable,
of two order-four operators such that 
\begin{eqnarray}
\label{vL4suchthat}
\hspace{-0.95in}&&  \quad \quad  \quad \quad  \quad   \quad   \quad   \quad  
 v(x) \cdot \,  {\cal L}_4   \cdot \, {{1} \over {v(x)}}
\, \,\,  = \, \, \,  \, pullback\Bigl(L_4, \,  y(x) \Bigr),
\end{eqnarray}
are simply related through the relations
\begin{eqnarray}
\label{YukcalL2gen}
\hspace{-0.95in}&&    \quad  \quad  \quad  \quad
 q_x({\cal L}_4)^n \, \, = \, \, \,   
 \, {{1} \over {\lambda}} \cdot \, q_x(L_4)\Bigl( y(x) \Bigr), 
 \quad  \quad 
K_x({\cal L}_4) \, \, = \, \, \,K_x(L_4)\Bigl( y(x) \Bigr). 
\end{eqnarray}
The Yukawa couplings 
{\em expressed in terms of the nome}\footnote[2]{This function 
is often viewed as a function of the nome $\, q \, = \, e^{\tau}$, since 
its $ \,q$-expansion in the case of degenerating family of Calabi-Yau 3-folds is 
supposed to encode the counting of rational curves of various degrees on a mirror 
manifold.}, {\em are related in an even simpler ``universal'' way} as so: 
\begin{eqnarray}
\label{YukcalL2nomegenertext}
\hspace{-0.95in}&&    \quad  \quad  \quad 
 \quad    \quad  \quad  \quad \quad  \quad  \quad 
K_q({\cal L}_4) \, \, = \, \, \, K_q(L_4)(\lambda  \cdot \, q^n).
\end{eqnarray} 
The previous example (\ref{vL4text}) corresponds to $\, n\, = \, 1$ 
and $\, \lambda \, = -4$.
In the case  $\, n\, = \, 1$ and $\, \lambda \, = \, 1$, the pullback 
is a deformation of the identity $\, y(x) \, = \, \, x \, + \, \, \cdots \,$ 
and the  Yukawa couplings expressed in terms of the nome, of the two 
linear differential operators are equal. Thus one recovers 
Proposition (6.2) of Almkvist et al. paper~\cite{Duco}
where the Yukawa couplings coincide.
                                    
Since the Schwarzian equation (\ref{SchwarUUhatAlmkvistext}) corresponds to the 
equivalence of two linear differential operators by pullback with remarkably 
simple relations (\ref{YukcalL2nomegenertext}) on their Yukawa couplings
expressed in terms of the nome,  the Schwarzian equation (\ref{SchwarUUhatAlmkvistext})
can be seen as a condition to have simply related Yukawa couplings. In the case 
of {\em  deformation of the identity} $\, y(x) \, = \, \, x \, + \, \, \cdots \,\,$ 
pullbacks, it can be seen  as a condition of preservation of the Yukawa 
couplings ({\em seen as functions of the nome}). These results {\em are not restricted 
to order-four linear differential operators} (see Appendix E of~\cite{Christol} 
and \ref{CalabiSchwa}). For instance, one can impose that 
{\em two different pullbacks of the same order-$N$ 
linear differential operator $\, L_N$  are homomorphic}, i.e. there exist two
intertwiners (of order $\, N-1$ in general) $\, L_{N-1}$ and $\, \, M_{N-1}$
such that:    
\begin{eqnarray}
\label{twopullbacks}
\hspace{-0.95in}&&  \quad   \quad   \quad  
pullback\Bigl(L_N, \,  p_1(x)  \Bigr) \cdot L_{N-1} 
\, \,\,  = \, \, \,  \, 
 M_{N-1} \cdot \,  pullback\Bigl(L_N, \,  p_2(x)  \Bigr).
\end{eqnarray}
This last generalization turns out to be instructive for physics and 
enumerative combinatorics.

\vskip .2cm

\section{Conclusion} 
\label{Conclu}

In a previous paper~\cite{Aziz} we focused on identities relating 
the {\em same} $\, _2F_1$ hypergeometric function with 
{\em two different algebraic pullback transformations} 
\begin{eqnarray}
\hspace{-0.95in}&&  \quad \quad  \quad   \quad  \quad  \quad     
\label{prevsymmetry}
{\cal A}(x) \cdot \,   _2F_1 \Bigl( [a,b],[c],x \Bigr) 
\,   \,   = \,  \,    \,   _2F_1 \Bigl( [a,b],[c],y(x) \Bigr), 
\end{eqnarray}
along with the existence of $\, _nF_{n-1}$ analogues of 
the previous relation. 
Such remarkable identities correspond to {\em modular forms} 
that emerged in the analysis of multiple integrals related to
the square Ising model~\cite{IsingCalabi,IsingCalabi2,Christol,Diffalg}
or in other enumerative combinatorics context~\cite{perimeter}. 
They can be seen as a simple occurence of 
{\em infinite order}\footnote[9]{We have for instance in mind 
to provide exact representations of the renormalization 
group~\cite{Hindawi,broglie,Heegner}.}
covariance symmetries 
in physics~\cite{Hindawi} or enumerative combinatorics. 

The current paper generalizes these previous results 
 beyond hypergeometric 
functions\footnote[5]{Or even Heun functions, see~\cite{Aziz}.},
analyzing the conditions for  order-$N$ linear differential 
operators with an arbitrary number of 
singularities\footnote[1]{Far beyond operators with
hypergeometric solutions, or pullbacked  hypergeometric solutions.}  
to be {\em pullback invariant up to conjugations}: 
\begin{eqnarray}
\label{vLNgenergg}
\hspace{-0.95in}&&  \quad \quad  \quad   \quad   \quad   \quad  
 {{1} \over {v(x)}} \cdot \,  L_N  \cdot \, v(x) 
\, \,\,  = \, \, \,  \, pullback\Bigl(L_N, \,  y(x)  \Bigr).
\end{eqnarray}
One finds that the pullbacks $\, y(x)$ are
{\em differentially algebraic}~\cite{Selected,IsTheFull}, 
being {\em necessarily solutions of 
the same Schwarzian equations} as in~\cite{Aziz}
\begin{eqnarray}
\label{SchwarNconcl}
\hspace{-0.95in}&& \quad \quad   \quad \quad  \quad \quad  
W(x) 
\, \, \,  \,-W(y(x)) \cdot  \, y'(x)^2
\, \, \,  \,+ \,  \{ y(x), \, x\} 
\, \,\, \, = \,\, \, \,  \, 0, 
\end{eqnarray}
where the function $\, W(x)$ encoding the Schwarzian equation (\ref{SchwarNconcl})
is a simple expression of the first two coefficients of the linear 
differential operator (see (\ref{wherecondN})). For order-two linear differential 
operators this Schwarzian condition turns out to be sufficient. In the case 
of linear differential operators with
selected differential Galois groups however, we showed, for orders three 
and four, that the  ``Calabi-Yau'' conditions
(see sections \ref{CalabiYaucondi})
are rigid enough to force the pullbacked-invariant (up to conjugation) 
operators (see (\ref{vLNgenergg})) to reduce to symmetric powers of an 
order-two linear differential operator.  

The reduction of the solutions 
of this Schwarzian differential equation to {\em modular correspondences}
was an open question in~\cite{Aziz}. Modular correspondences require 
the existence, for any integer $\, n$, of solutions of the Schwarzian 
equation (\ref{SchwarNconcl})
of the form $ \, y_n (x)=  \, a_n \cdot \,  x^n \,  + \cdots \, \,$
such that, for any integer $\, m$ and $\, n$, the following 
``pre-modular'' condition is satisfied:
\begin{eqnarray}
\label{premodular}
\hspace{-0.95in}&& \quad \quad   \quad \quad  \quad \quad  \,  \,  \, 
y_n(a_n, \, \, y_m(a_m,  \,\, x)) \, \,   = \, \,   \,  y_{n m}(a_n a_m^n, \, \, x).
\end{eqnarray}
We derived in this paper a necessary and sufficient condition to obtain 
such ``pre-modular'' 
solutions for the ``Schwarzian condition''  (\ref{SchwarNconcl}).
This condition turns out to be a simple condition on the Laurent series 
of $ \,W(x)$ encoding the Schwarzian condition:
\begin{eqnarray}
\label{adB}
\hspace{-0.95in}&& \quad \quad \quad \quad \quad  \quad \quad 
W(x) \,  \, = \, \,   \, 
 -\frac{1}{2\cdot x^2} \, \,  \, + \frac{b}{x}
  \,  \, \, + \sum_{m=0}^{\infty} \, a_m \cdot x^m.
\end{eqnarray}

In light of what we have discussed so far, the current paper 
generates more questions than answers 
that give directions for further research. We have seen 
for example that (\ref{adB})
is a necessary and sufficient condition for 
obtaining ``pre-modular'' solutions for the ``Schwarzian condition'', 
corresponding, in general, to a {\em transcendental\footnote[2]{The series 
$\, y_n(x)$ (see (\ref{premodular}))  are {\em differentially algebraic}, 
but, {\em not necessarily algebraic functions}.} 
declination of modular correspondences}.
To have modular correspondences one needs the existence of
{\em selected values}  
of the parameters such that the solution series 
$ \, y_n(x)= \,  a_n \cdot  \, x^n \,  + \cdots \, $ (see (\ref{curious5})) 
actually {\em reduce to algebraic functions}. Is it only in the case of 
modular correspondences that such algebraic reductions for 
selected values take place ? 

Then we showed that an order-two linear differential operator 
emerging in the context of avoiding 
permutations counting~\cite{Conway,Bona}, provides a good illustration 
of a generalization of the pullback-covariance (\ref{prevsymmetry})
or of the pullback invariance up to conjugation (\ref{vLNgenergg}):
the $ \, _2F_1 \Bigl( [-1/4,3/4],[1],x \Bigr)$ 
that comes up in the context of avoiding 
permutations counting~\cite{Conway,Bona}, verify
a relation (see (\ref{followingidentity}), (\ref{followingidentity2})), 
whose general form is given by   
\begin{eqnarray}
\label{adB2F1}
\hspace{-0.95in}&&
\quad \quad \quad \quad \quad \quad 
\Phi(y(x)) \, \,\,  \,
\, \,\, =  \, \, \,
\Bigl(   {\cal A}(x) \cdot \, {{d} \over { dx}}
\, \, + {\cal B}(x)   \Bigr)
\, \cdot \,  \Phi(x),
\end{eqnarray}
giving a non-trivial explicit example of a 
{\em pullback invariance of an operator up to operator homomorphisms} 
(see (\ref{onepullbackhomo}))
\begin{eqnarray}
\label{onepullbackhomoconcl}
\hspace{-0.95in}&&  \quad   \quad   \quad   \quad   \quad  \quad \quad 
 M_{N-1} \cdot \,  L_N \, \,\,  = \, \, \,  \, 
pullback\Bigl(L_N, \,  y(x)  \Bigr) \cdot L_{N-1}.
\end{eqnarray}
Equation (\ref{Automat}) providing an exact representation 
of the {\em Landen} transformation 
(generator of the renormalization group) on $\, \tilde{\chi}^{(2)}$,
together with the explicit calculations of section \ref{pullbackhomomorphism},
make quite clear that conditions like (\ref{adB2F1}) 
provide a natural and interesting generalization 
of {\em modular forms}, going beyond the Schwarzian 
equation (\ref{SchwarNconcl}).

At last, we examined the equivalence of two different linear differential 
operators, under pullback and conjugation, yielding again some Schwarzian 
condition relating these two linear differential operators 
(see relation (\ref{SchwarUUhatAlmkvistgener})), and we discussed 
the consequence of such equivalence on the corresponding 
Yukawa couplings. These results revisiting and complementing 
the results of~\cite{Duco}, provide 
powerful tools to analyze various 
symmetry and classification problems of selected linear 
differential operators, in particular  linear differential 
operators of the Calabi-Yau type~\cite{Tables} 
({\em not necessarily of order four}~\cite{Christol}). 

When dealing with {\em linear} differential operators, 
we have seen the emergence of Schwarzian derivatives, 
consequence of the fact that
{\em the Schwarzian derivative is appropriate for the composition of functions}~\cite{What} 
(see the chain rule of the  Schwarzian derivative
of the composition of function).
Do {\em higher order Schwarzian derivatives}~\cite{Matone,Matone2,HigherSchw,HigherSchw2} 
occur for pullback-symmetries of {\em non-linear} ODE's, 
or, more generally, for {\em functional} equations? 

Restraining oneself to the univariate {\em linear} differential 
operators case, let us remark that 
if condition (\ref{prevsymmetry}), or (\ref{vLNgenergg}),
describe effectively all the modular forms that 
often occur in physics~\cite{IsingCalabi,IsingCalabi2,Diffalg},  
or enumerative combinatorics~\cite{perimeter}, a
pullback symmetry up to conjugation constraint like (\ref{vLNgenergg}) 
could be restrictive in some sense since it seems to 
yield systematic reduction\footnote[5]{At least in the case where the
operators verify Calabi-Yau conditions and thus 
have selected differential Galois groups.} to order-two linear differential 
operators. In contrast the simple hypergeometric 
example of section \ref{pullbackhomomorphism} seems to provide a natural 
generalization of {\em modular forms}:
 the {\em pullback invariance of an operator up to operator homomorphisms} 
condition (\ref{onepullbackhomoconcl}) promises to cover a larger ensemble of
exact representations of 
symmetries in physics or enumerative combinatorics. In particular 
the emergence of conditions like (\ref{adB2F1})
of higher order, namely generalized covariance 
(\ref{onepullbackhomoconcl}) for the representation of the Landen 
transformation\footnote[1]{And more generally the modular correspondences 
providing exact representations of the generators of the renormalization 
group~\cite{Hindawi,Heegner}.} 
on the other $\, n$-fold $\, \tilde{\chi}^{(n)}$'s of the Ising susceptibility 
(see~\cite{High,bo-gu-ha-je-ma-ni-ze-08,ze-bo-ha-ma-05b}), together with 
their corresponding large order {\em reducible} linear differential operators, 
is a challenging open problem.

\vskip .3cm
\vskip .3cm

{\bf Acknowledgments:} 
We would like to thank
 S. Boukraa,  M. van Hoeij and J-A. Weil for very fruitful 
discussions   on differential systems. We thank A.J. Guttmann for
providing an interesting pullbacked hypergeometric example.
This work has been performed 
without any ERC, ANR or MAE financial support.

\vskip .5cm
\vskip .5cm

\appendix

\section{A simple reducible linear differential operator of order four} 
\label{reducibleapp}

Let us consider an order-four linear differential operator
which is the square of an order-two linear differential operator: 
$\, L_4 \, = \, L_2 \cdot \, L_2$, 
where $\, L_2 \, = \, D_x^2 \, + p(x) \cdot \, D_x \, +q(x)$. 
This reducible order-four linear differential operator $\, L_4$ is of the form 
$\, D_x^4 \, + \, p_r(x) \cdot \, D_x^3 \, + \,  q_r(x) \cdot \, D_x^2 \, + \, \cdots \, \, $ 
where the two coefficients $\, p_r(x)$ and  $\, q_r(x)$ read respectively:
\begin{eqnarray}
\label{p4q4}
\hspace{-0.95in}&& \quad \quad \quad   
p_r(x)\, \, = \, \, \, 2 \cdot \, p(x), \quad \quad \quad 
q_r(x)\, \,  = \, \, \, 
p(x)^2 \, \, +  2 \cdot \, q(x) \,\,  +  2 \cdot \, {{ d p(x)} \over { dx}}. 
\end{eqnarray}
The coefficients of the order-four operator $\, L_4$ 
{\em verify the Calabi-Yau condition}\footnote[2]{
The {\em exterior square} 
of that an order-four operator $\, L_4 \, = \, L_2^2 \, \, $  is of order 
{\em five} instead of order six. This is a general result: the order of the  
symmetric squares of operators  $\, L_{2\,n} \, = \, L_2^n \, \, $ is 
less than $\, 2\, n \, (2\, n\, -1)/2$. Such $\, n$-th powers verify 
higher order Calabi-Yau conditions.} (\ref{Calabi}). 
We even have the identity\footnote[9]{More generally, the order-one 
linear differential operator $\, D_x \, +p(x) \, \, $ 
rightdivises the exterior square of the $\, n$-th power 
of $\, L_2$, for any integer $\, n$.} that the exterior square 
of $\, L_4 \, = \, L_2^2$ is the product of 
an order-one operator (having the wronskian of $\, L_4$ as solution), 
an order-three operator which is the symmetric square of $\, L_2$ 
and again the same order-one operator:
\begin{eqnarray}
\label{identp4q4}
\hspace{-0.95in}&& \quad \quad \quad   \quad  
Ext^2(L_2 \cdot \, L_2) \, \, \, = \, \, \,  \,
(D_x \, +p(x)) \cdot \, Sym^2(L_2) \cdot \, (D_x \, +p(x)). 
\end{eqnarray}
For this reducible order-four linear differential operator $\, L_4\, = \, L_2^2$ 
the first steps of the  $\, L_4^{(p)} \, = \, \, L_4^{(c)}$ calculations give
a function $\, W(x)$ given by (\ref{wherecond4}), namely 
$\, W_r(x) \, = \, 3/10 \cdot \,  p'_r(x)\, + 3/40 \cdot \,  p_r(x)^2\, - \, q_r(x)/5$. 
Using (\ref{p4q4}) one can rewrite $\, W_r(x)$ in terms of  $\, p(x)$ and $\, q(x)$.
One gets an expression similar to (\ref{wherecond}) but {\em different}, 
namely $\, W_r(x) \, = \,(p'(x)+p(x)^2/2 \, -2 \, q(x))/5$, which 
is exactly (\ref{wherecond}) {\em but divided by} $\, 5$. Therefore 
the pullback condition on this square 
operator $\, L_4 \, = \, L_2^2 \,\, $ {\em does not reduce} to the 
pullback condition on the (underlying) $\, L_2$. 

The change of variable $\, x \, \rightarrow \, y(x)$ on a 
linear differential operator which is the product of two  operators, 
is the product of these two linear differential operators 
on which this change of variable has been performed. More precisely 
{\em with our normalization of the pullback} of a linear differential operator 
a condition $\, L_4^{(p)} \, = \, \, L_4^{(c)}$ would give the relation 
\begin{eqnarray}
\label{identp4q4}
\hspace{-0.95in}&&  \quad  
 {{1} \over {y'(x)^4}} \cdot \, pullback(L_2^2, y(x)) \,  \,  \, = 
 \nonumber \\
 \hspace{-0.95in}&& \quad  \quad   
 \, = \, \,  \,  \Bigl({{1} \over {y'(x)^2}} \cdot \, pullback(L_2, y(x))\Bigr)
 \cdot \, \Bigl({{1} \over {y'(x)^2}} \cdot \, pullback(L_2, y(x))\Bigr)
\nonumber \\
\hspace{-0.95in}&& \quad  \quad  
 \, = \, \,  \, \Bigl({{1} \over {y'(x)^2}} \cdot \,
  {{1} \over {v(x)}} \cdot \, L_2 \cdot \, v(x) \Bigr)
 \cdot \, \Bigl( {{1} \over {y'(x)^2}} \cdot \,
 {{1} \over {v(x)}} \cdot \, L_2 \cdot \, v(x) \Bigr)
 \\
\hspace{-0.95in}&& \quad \quad   
 \, = \, \,   
{{1} \over {y'(x)^4}} \cdot \,   
\Bigl({{1} \over {v(x)}} \cdot \, M_2 \cdot \,  L_2 \cdot \, v(x)  \Bigr)
\quad \quad  \, \,   \hbox{where:} \quad  \, \,   
 M_2 \, = \, \,   y'(x)^2  \cdot \, L_2 \cdot \,  {{1} \over {y'(x)^2}}. 
\nonumber 
\end{eqnarray}
In other words the pullback of $\, L_4 \, = \, L_2^2 \,$ corresponds 
to a conjugate of {\em another} order-four linear differential
operator $\, M_4 \, = \,  M_2 \cdot \,  L_2$, which is not $\, L_4$ but 
 is also reducible into two different order-two linear differential 
operators. Note that the order-two linear differential operator 
$\, M_2$ depends on the change of variable $\, \, x \, \rightarrow \, \, y(x)$.

\vskip .1cm 

\section{Order-five linear differential operators} 
\label{order-five}

Let us consider an  {\em irreducible} order-five linear 
differential operator
\begin{eqnarray}
\label{L_5}
\hspace{-0.95in}&& \, \,  
L_5 \, \, \, = \, \, \, \, 
 D_x^5  \, \, + \, p(x) \cdot \, D_x^4 \, \, + \, \, q(x)\cdot \, D_x^3
 \, \, + \,  \, r(x) \cdot \, D_x^2 \,  \, + \, \, s(x)\cdot \, D_x \, + \, \, t(x),
\end{eqnarray}
and let us also introduce two other  linear differential operator of order five,
the operator $\, L_5^{(c)}$ conjugated of (\ref{L_5}) by a function $\, v(x)$,
namely $\, L_5^{(c)} \, = \, \, 1/v(x) \cdot \, L_5 \cdot \, v(x)$, and the 
pullbacked operator $\, L_5^{(p)}$ which amounts to changing 
$\, x \, \rightarrow \, \, y(x)$ in $\, L_5$.  
Imposing a generalized (symmetric) Calabi-Yau condition amounts to imposing that the
symetric square  of (\ref{L_5})  is of order less than (the generic order) $\, 15$.
Using this (symmetric) Calabi-Yau condition to perform any calculation is 
a very difficult task
since this condition corresponds to a huge polynomial in the coefficients and their 
derivatives. However, similarly to what we did in section \ref{symmCalabiYau} we 
can introduce a parametrization, similar to (\ref{U1U3param}) of this huge (symmetric) 
Calabi-Yau condition. We saw in~\cite{Canonical} that the (symmetric) Calabi-Yau 
condition for an order-five linear differential operator $\, L_5$ 
(which amounts to saying that the 
symmetric square of $\, L_5$ is of order less than $\, 15$), amounts to saying that
$\, L_5$ has the following decomposition 
\begin{eqnarray}
\label{L4symm5}
\hspace{-0.95in}&& \quad \quad \quad \quad  \quad \quad  \quad 
L_5 \, \,\, = \, \, \,\, 
(U_1 \cdot \,  V_1 \cdot \,  U_3 \, + \, \, U_1 \, + \,  U_3) \cdot \, e(x), 
\end{eqnarray}
where $\, U_1$ and $\, U_3$ are order-one,  order-one, and  
order-three {\em self-adjoint} linear differential operators of the 
form previously given with (\ref{U1U3}) and  (\ref{U1U3U1}), and $\, V_1$ 
is another order-one self-adjoint operator:
\begin{eqnarray}
\label{V1}
\hspace{-0.95in}&& \quad \quad \quad \quad \quad  \quad \quad 
V_1 \,\, \, = \, \, \, \,\, e(x) \cdot \, D_x
\,\, + \,  \, \, {{1} \over {2}} \cdot \,  {{d \, e(x)} \over {dx}},  
\end{eqnarray}
It is straightforward to get the coefficients of the order-five operator (\ref{L_5}):
 \begin{eqnarray}
\label{U1V1U3param}
\hspace{-0.95in}&& \quad \quad 
p(x) \, \, = \, \, \, \,   {{7} \over { 2}} \cdot \, {{a'(x)} \over {a(x)}}
\,\, \,  + \,  {{1} \over { 2}} \cdot \, {{c'(x)} \over {c(x)}}
\,\,   \, + \,   4 \cdot \, {{d'(x)} \over {d(x)}}
\, \,  \, + \,    {{3} \over { 2}} \cdot \, {{e'(x)} \over {e(x)}}, 
\quad \quad \cdots 
\end{eqnarray}
This gives a parametrization of the (symmetric) Calabi-Yau condition 
and thus a way to perform calculations for an order-five operator that
verifies this huge (symmetric) Calabi-Yau condition. 
Again, one finds that just imposing this (symmetric) Calabi-Yau condition 
is not sufficient to have $\, L_5^{(p)} \, \, = \, \, \, L_5^{(c)}$.

There is one subcase of that huge polynomial condition that can be written 
explicitly (in a similar manner we wrote 
the Calabi-Yau  (\ref{Calabi}) and  symmetric  Calabi-Yau (\ref{CalabiSymm}) 
conditions, see  (\ref{converseA}), (\ref{converseB}), (\ref{converseC}) below). 

Let us consider an order-two linear differential operator 
$\, L_2 = \,  D_x^2 \, +A(x) \cdot \, D_x$
$ \, +B(x)$,
and the symmetric fourth power of $\, L_2$, the coefficients of that  
order-five operator read:
\begin{eqnarray}
\label{symm4}
\hspace{-0.95in}&&   \quad 
p(x) \,  \, = \, \, \, 10 \cdot A(x), \quad \quad \quad  \, \,\,  
q(x) \,  \, = \, \, \,\,  
35 \cdot \, A(x)^2 \, \, \,+20 \cdot \,B(x)
 \, \,\, +10 \cdot \, {{d A(x)} \over {dx }},
\nonumber \\
\hspace{-0.95in}&&  \quad 
r(x) \,  \, = \, \, \,\, 50 \cdot \, A(x)^3 \,\, +120\cdot \, B(x) \cdot \, A(x)
 \,\, +45 \cdot \, A(x) \cdot \, {{d A(x)} \over {dx }} 
\nonumber \\
\hspace{-0.95in}&&  \quad  \quad \quad \quad \quad 
 \, +30 \cdot \,  {{d B(x)} \over {dx }} 
\, \, +5  \cdot \,  {{d^2 A(x)} \over {dx^2 }},
 \\
\hspace{-0.95in}&&  
s(x) \,  \, = \, \, \, \,
24  \cdot \, A(x)^4 \,\,  +208 \cdot \, A(x)^2 \cdot \, B(x)
 \, \, +46 \cdot \, A(x)^2  \cdot \,  {{d A(x)} \over {dx }} 
\nonumber \\
\hspace{-0.95in}&& \quad  \quad  \quad   
 \, +120 \cdot \,  {{d B(x)} \over {dx }} \cdot \, A(x) \,  \,
+11 \cdot \, A(x) \cdot \, {{d^2 A(x)} \over {dx^2 }}
 \,  \,+64 \cdot \, B(x)^2 
\nonumber \\
\hspace{-0.95in}&& \quad  \quad  \quad  
 \, \, +56 \cdot \, B(x) \cdot \,  {{d A(x)} \over {dx }}
\, +7 \cdot \, \Bigl( {{d A(x)} \over {dx }}\Bigr)^2 
\, \, +18 \cdot \,  {{d^2 B(x)} \over {dx^2 }} \,  \,
+ {{d^3 A(x)} \over {dx^3 }}, 
\nonumber \\
\hspace{-0.95in}&&  \quad  
t(x) \,  \, = \, \, \, \, 96 \cdot \, A(x)^3 \cdot \,B(x)
 \, \, +104 \cdot \,A(x)^2 \cdot \, {{d B(x)} \over {dx }} 
\, \, +128 \cdot \,A(x) \cdot \,B(x)^2 \, 
\nonumber \\
\hspace{-0.95in}&&  \quad  \quad  \quad 
+80 \cdot \,A(x) \cdot \,B(x) \cdot \, {{d A(x)} \over {dx }}  \,  \,
+36 \cdot \,  {{d^2 B(x)} \over {dx^2 }}  \cdot \,A(x) \,  \,
+64 \cdot \,B(x) \cdot \, {{d B(x)} \over {dx }}  
\nonumber \\
\hspace{-0.95in}&&  \quad  \quad  \quad 
 +8 \cdot \,B(x) \cdot \,  {{d^2 A(x)} \over {dx^2 }}\,  \,
 +28 \cdot \,  {{d B(x)} \over {dx }} \cdot \,  {{d A(x)} \over {dx }}
 \,  \, +4 \cdot \, {{d^3 B(x)} \over {dx^3 }}.
\nonumber 
\end{eqnarray}
Conversely this means $\, A(x) \, = \, p(x)/10$ and
\begin{eqnarray}
\label{converseA}
\hspace{-0.95in}&&  \quad   \quad 
r(x) \, \, = \, \, \, \, - {{4} \over {25}} \cdot \, p(x)^3 
\, \, \, - {{6} \over {5}} \cdot \, p(x)  \cdot \, {{d p(x) } \over { dx}}
\, \, \, + {{3} \over {5}} \cdot \, p(x) \cdot \, q(x)
 \nonumber \\
\hspace{-0.95in}&&  \quad \quad \quad  \quad   \quad  \quad 
 \, \, - {{d^2 p(x) } \over { dx^2}}
\, \, \, +{{3} \over {2}} \cdot \, {{d q(x) } \over { dx}}, 
 \\
\label{converseB}
\hspace{-0.95in}&&  \quad  \quad   
s(x) \, \, = \, \, \,  \, - {{9} \over {625}} \cdot \, p(x)^4 \,\, \,
 -{{58} \over {125}} \cdot \,  p(x)^2  \cdot \, {{d p(x) } \over { dx}} 
\, \, \, -{{1} \over {125}} \cdot \,  p(x)^2  \cdot \, q(x)
\nonumber \\
\hspace{-0.95in}&&  \quad \quad   \quad  \quad \quad 
 -{{28} \over {25}} \cdot \, p(x) \cdot \, {{d^2 p(x) } \over { dx^2}}
 \,  \, +{{3} \over {5}} \cdot \,  p(x)\cdot \, {{d q(x) } \over { dx}}  
\, \, \, 
- {{17} \over {25}} \cdot \, \Bigl(  {{d p(x) } \over { dx}}  \Bigr)^2
\nonumber \\
\hspace{-0.95in}&&  \quad \quad   \quad  \quad \quad  
-\, {{1} \over {25}} \cdot \, {{d p(x) } \over { dx}}  \cdot \, q(x) 
\, \, + {{4} \over {25}} \cdot \, q(x)^2  
\, \, -{{4} \over {5}} \cdot \,  {{d^3 p(x) } \over { dx^3}}
 \, \, + {{9} \over {10}} \cdot \,   {{d^2 q(x) } \over { dx^2}},  
 \\
\label{converseC}
\hspace{-0.95in}&& \quad  \quad  
t(x) \, \, = \, \, \,
 - {{11} \over {25}} \cdot \, 
 {{d p(x) } \over { dx}}  \cdot \,  {{d^2 p(x) } \over { dx^2}} \,
- {{8} \over {25}} \cdot \, p(x)  \cdot \,  {{d^3 p(x) } \over { dx^3}}
 \,  +{{4} \over {625}} \cdot \, p(x)^3 \cdot \,  {{d p(x) } \over { dx}}
 \nonumber \\
\hspace{-0.95in}&&  \quad \quad \quad  \quad  
 -{{11} \over {625}} \cdot \, p(x)^3 \cdot \, q(x) \,   -{{17} \over {125}} 
\cdot \, p(x)^2 \cdot \, {{d^2 p(x) } \over { dx^2}} 
 -{{1} \over {250}} \cdot \,  p(x)^2 \cdot \, {{d q(x) } \over { dx}} \,  
\nonumber \\
\hspace{-0.95in}&&  \quad \quad\quad  \quad  
 -{{3} \over {25}} \cdot \,  p(x) \cdot \, \Bigl(  {{d p(x) } \over { dx}}  \Bigr)^2
\,  \, +{{4} \over {125}} \cdot \, p(x) \cdot \, q(x)^2 \, +{{9} \over {50}}
 \cdot \, p(x) \cdot \,  {{d^2 q(x) } \over { dx^2}} 
 \nonumber \\
\hspace{-0.95in}&&  \quad \quad \quad  \quad  
 -{{1} \over {50}} \cdot \,  {{d p(x) } \over { dx}}  \cdot \,  {{d q(x) } \over { dx}} 
\,  \, -{{3} \over {25}} \cdot \,  q(x) \cdot \,  {{d^2 p(x) } \over { dx^2}} 
 \, \, + {{4} \over {25}} \cdot \,  q(x) \cdot \,  {{d q(x) } \over { dx}} 
  \\
\hspace{-0.95in}&&  \quad \quad \quad  \quad  
 - {{17} \over {125}} \cdot \,   p(x)
  \cdot \,   q(x)  \cdot \,  {{d p(x) } \over { dx}}  \, \,
- {{1} \over {5}} \cdot \,   {{d^4 p(x) } \over { dx^4}} \, \,
 +{{1} \over {5}} \cdot \, {{d^3 q(x) } \over { dx^3}} 
\, +{{7} \over {3125}} \cdot \, p(x)^5.
\nonumber
\end{eqnarray}
When the three conditions (\ref{converseA}), (\ref{converseB}), 
(\ref{converseC}) are verified, the symmetric square of the 
order-five linear differential operator
 $\, L_5$ is of order $\, 9$ instead of $\, 15$ (and thus its differential 
Galois group is $\, SO(5, \,  \mathbb{C})$).  The three conditions  
(\ref{converseA}), (\ref{converseB}), (\ref{converseC}) are necessary for
$\, L_5$ to be reducible to the symmetric cube of an underlying 
order-two linear differential operator.
If one imposes the three conditions  
(\ref{converseA}), (\ref{converseB}), (\ref{converseC}), the order-five 
linear differential operator is simply conjugated to its adjoint:
\begin{eqnarray}
\label{simply5}
\hspace{-0.95in}&& \quad   \quad \quad  \quad \quad  \quad \quad   \, \,
 L_5 \cdot \, w(x)^{2/5}  \,\, \, = \,\, \, \, w(x)^{2/5}  \cdot \, adjoint(L_5),
\end{eqnarray}
where $\, w(x)$ denotes the wronskian of $\, L_5$.
Recalling that an order-two linear differential operator 
$\, L_2 \, = \, \, D_x^2 \, +A(x)\cdot \, D_x \, +B(x)$,
having a wronskian $\, w_2(x)$ is such that 
$\, L_2 \cdot \, w_2(x)  \,\, \, = \,\, \, \, w_2(x) \cdot \, adjoint(L_2)$, the identity 
(\ref{simply5}) is a simple consequence of the fact that the order-five operator 
reduces to the symmetric fourth power of an order-two linear differential operator.

{\em Note that by imposing the two conditions}\footnote[1]{We have the same 
result imposing the two conditions (\ref{converseA}) and (\ref{converseC}), or 
(\ref{converseB}) and (\ref{converseC})}
(\ref{converseA}), (\ref{converseB}), the symmetric square of the order-five 
operator $\, L_5$ {\em becomes of the generic order} $\, 15$, yet the symmetric 
square of $\, L_5$ does not have a rational solution (the operator 
 $\, L_5$ and its adjoint are not homomorphic: the differential Galois group 
of  $\, L_5$ is not  equal, or included, in the orthogonal group 
$\, SO(5, \,  \mathbb{C})$).

\vskip .1cm

The identification of these two order-four linear differential 
operators $\,L_5^{(p)}$ and $\,L_5^{(c)}$
gives four conditions $\, {\cal C}_n$, 
$\, n= \, 4,  \,  3,  \, 2, \, 1, \, 0$, corresponding respectively to 
identification of the $\, D_x^n$ 
coefficients of $\, L_5^{(p)}$ and $\, L_5^{(c)}$. 

Performing the same pullback-compatibility calculations we did for 
order-three, and order-four operators for $\, L_5$ 
is a tremendously difficult task in a general framework.

The first calculation steps 
can be performed, giving the exact expression 
of the conjugation function $\, v(x)$ from $\, {\cal C}_4$ as:
\begin{eqnarray}
\label{v4for5}
\hspace{-0.95in}&& \quad \quad  \quad \quad \quad  \quad \quad  \quad \quad 
v(x) \, \, \, = \, \, \, \,  
 y'(x)^{ -2} \cdot \,   \Bigl({{w(x) } \over {w(y(x)) }}\Bigr)^{1/5}. 
\end{eqnarray}
and, eliminating the log-derivative $\, v'(x)/v(x)$ between  
$\, {\cal C}_4$ and  $\, {\cal C}_5$, giving the Schwarzian equation
\begin{eqnarray}
\label{condition5}
\hspace{-0.95in}&& \quad   \quad \quad  \quad \quad  \quad 
 W(x) 
\, \, \,  \,-W(y(x)) \cdot  \, y'(x)^2
\, \, \,  \,+ \,  \{ y(x), \, x\} 
\, \,\, \, = \,\, \, \,  \, 0, 
\end{eqnarray}
where, this time:
\begin{eqnarray}
\label{wherecond5}
\hspace{-0.95in}&& \quad   \quad \quad  \quad \quad  \quad 
W(x)  \, \, = \, \,  \, \, \,{{1} \over {5}} \cdot \,  {{ d p(x)} \over { dx}} 
 \, \, \,  + \, \, 
 {{1} \over {25}} \cdot \,   p(x)^2  \, \,  \, \,  - \, {{q(x)} \over {10}}.
\end{eqnarray}
Again one finds that the expression of $\, W(x)$ given by (\ref{wherecond5})
gives back the expression (\ref{wherecond})
when $\, p(x)$ and $\, q(x)$ are deduced from (\ref{symm4})
($\,  p(x)$ and $\, q(x)$ becoming $\, A(x)$ and $\, B(x)$).

\vskip .1cm

The condition that we called in~\cite{Diffalg}
 {\em symmetric Calabi-Yau condition}
for the operator $\, L_5$ (corresponding to impose that its 
symmetric square is of order less than $\, 15$) is a huge polynomial 
condition on the coefficients of $\, L_5$ and its derivative. Seeing 
if these pullback-compatibility calculations  yield  necessarily 
 the huge (symmetric Calabi-Yau\footnote[2]{Meaning that 
the order-five operator has a $\, SO(5, \,  \mathbb{C})$ differential Galois group.})
condition and the {\em  three} conditions (\ref{converseA}), (\ref{converseB}), 
(\ref{converseC}), or, in other words, that the order-five linear differential 
operator necessarily 
reduces again to (a symmetric fourth power of) an underlying 
order-two linear differential operator, remains an open question. 

\vskip .2cm 

\section{Reduction of the order-three ODE (\ref{equaF}) to the  order-two ODE (\ref{FepsAR})
in the rank-two case (\ref{WAR}).} 
\label{Reduction32}

The order-three linear differential 
equation (\ref{equaF}) on $\, F(x)$ should reduce 
to the  order-two linear ODE (\ref{FepsAR}) in the 
rank-two subcase (\ref{WAR}). When
 $\, A_R(x) \, = \, \, -w'(x)/w(x)$, the order-three linear 
differential operator $\, {\cal L}_3$ (see (\ref{equaFop}))
 has three solutions: 
\begin{eqnarray}
\label{three}
\hspace{-0.95in}&& \quad  \quad \quad  \quad  \quad  \quad \quad 
{{1} \over {w(x)}}, \,  \quad \quad \quad   {\cal S}_F, \,
  \quad \quad \quad \hbox{and:} 
\quad  \quad \quad  w(x) \cdot \, {\cal S}_F^2.
\end{eqnarray}
This can be seen as a consequence of the fact that 
 the order-two linear differential operator $\, {\cal L}_F$
rightdivides the order-three operator $\, {\cal L}_3$: 
\begin{eqnarray}
\label{equaFoprevisit}
\hspace{-0.95in}&& \quad  \quad 
 {\cal L}_3  
 \,  \,   \,  = \,  \,  \, \,\, 
\Bigl(D_x \, + A_R(x)\Bigr) \cdot \, D_x   \cdot \,
 \Bigl(D_x \, - A_R(x)\Bigr)\,  = \,  \,  \, \, 
\Bigl(D_x \, + A_R(x)\Bigr) \cdot \, {\cal L}_F.
\end{eqnarray}
In this rank-two subcase (\ref{WAR}), the function $\, F(x)$ 
is $\, {\cal S}_F$ {\em and not the third solution} 
$\, w(x) \cdot \, {\cal S}_F^2 \, $
 {\em which prevails in the general Schwarzian case} 
(see (\ref{firstbister})). 
The form of the last solution $\, w(x) \cdot \, {\cal S}_F^2 \, $ 
can be deduced from the fact that order-three  linear differential
operator $\, {\cal L}_3$ is the symmetric square of an order-two 
{\em self-adjoint} operator $\, {\cal L}_2$ (see (\ref{equaFop})) 
which is simply conjugated to the order-two operator 
$\, {\cal L}_F$ given by (\ref{FepsARop}):
\begin{eqnarray}
\label{FactoM2first}
\hspace{-0.95in}&& \quad \quad   \quad  \quad  \, 
 \, \,   \, \, 
 {\cal L}_2\,\,\,   \,   = \, \,  \,\,   D_x^2 \, -{{W(x)} \over {2}}
\,\,  \,  = \,  \,  \, \, 
\Bigl(D_x \, + {{A_R(x)} \over {2}}\Bigr) \cdot \,
 \Bigl(D_x \, - {{A_R(x)} \over {2}}\Bigr)
 \nonumber \\
\hspace{-0.95in}&& \quad \quad \quad \quad \quad \quad  \quad   \quad  
\,\, = \,  \,  \, \, w(x)^{1/2} \cdot \, {\cal L}_F \cdot \,w(x)^{-1/2}
\end{eqnarray}
which has clearly the solution $\,  w(x)^{1/2} \cdot \, {\cal S}_F$
as well as the solution $\, w(x)^{1/2} \cdot \, w(x)^{-1}$
$ \, = \, w(x)^{-1/2}$, deduced from the solutions of 
$\, {\cal L}_F$ (see (\ref{FepsARop})).  The three solutions 
(\ref{three}) correspond to all the products of these two solutions
namely the square of $\, w(x)^{-1/2}$ and  
$\,  w(x)^{1/2} \cdot \, {\cal S}_F$, and their product. Note that the 
factorization (\ref{FactoM2first}) {\em requires condition} 
(\ref{WAR}) {\em to be satisfied}. 

\vskip .1cm

{\bf Remark :} Recalling (\ref{equaFop}), (\ref{FepsARop}),
  (\ref{FactoM2first}), one can see that the rightdivision
(\ref{equaFoprevisit}) can be seen as a consequence of 
the identity
\begin{eqnarray}
\label{infact}
\hspace{-0.95in}&& \quad \quad \quad \quad   \, \,  \, \,  \, 
Sym^2\Bigl( {\cal L}_2\Bigr)  \,  \,  = \,  \,  \,
 Sym^2\Bigl( \Bigl(D_x \, + {{A_R(x)} \over {2}}\Bigr) \cdot \,
 \Bigl(D_x \, - {{A_R(x)} \over {2}}\Bigr) \Bigr) 
\nonumber \\ 
\hspace{-0.95in}&& \quad \quad  \quad  
 \quad \quad  \quad \quad \quad \quad  \quad  \, 
\,  \,  = \,  \,  \,  \Bigl(D_x \, + A_R(x)\Bigr) \cdot \, D_x   \cdot \,
 \Bigl(D_x \, - A_R(x)\Bigr).
\end{eqnarray}

\vskip .2cm 

\section{Mirror maps for $\, _2F_1([1/12,\, 5/12], \, [1], \, x)$.} 
\label{Mirror}

\vskip .1cm

The modular correspondences
$\, x \, \rightarrow \, \, y(x) \, \, $ are {\em infinite order} 
 algebraic transformations such that
\begin{eqnarray}
\label{modularform2explicitapp}
\hspace{-0.95in}&& \quad \quad  \quad  \quad  \quad 
 _2F_1\Bigl([{{1} \over {12}}, \, {{5} \over {12}}], \, [1], \, y(x)  \Bigr)
\, \,  = \, \, \, \, 
 {\cal A}(x) \cdot \,
 _2F_1\Bigl([{{1} \over {12}}, \, {{5} \over {12}}], \, [1], \,  x  \Bigr),
\end{eqnarray}
where $\, {\cal A}(x)$ is an algebraic function.
The modular correspondences $\, y(x)$ are solutions of the Schwarzian 
condition (\ref{condition1ter}), where $\, W(x)$ simply related to the 
function $\, F(x)$ (see (\ref{equaF})) are given by 
equations (\ref{firstbister}). These  modular correspondences
have series expansion at $\, x \, = \, 0$ of the form 
\begin{eqnarray}
\label{Qseriesnapp}
\hspace{-0.95in}&&    \quad    
y_n(x) \, \,\, = \, \, \, P(Q^n(x))  \,\, = \, \, \,
  1728 \cdot \, \Bigl({{ x} \over {1728}}\Bigr)^n  \, \, + \,   \, \cdots 
\quad \quad \quad 
 n \, = \, \, 2, \, 3, \, 4, \,\,  \cdots 
\end{eqnarray}
where $\, P(x)$ and $\, Q(x)$ are such that $\, P(Q(x)) \, = \, Q(P(x)) \, = \, x$,  
corresponding to the ``simplest'' examples of 
{\em mirror maps}~\cite{Aziz}. More precisely,
the well-known ``mirror maps''~\cite{LianYau1,LianYau,Doran,Doran2}
are often described as series with {\em integer coefficients}~\cite{Kratten,Kratten2}. 
These series correspond to a rescaling of $\, P(x)$ and $\, Q(x)$ 
by $\, 1728$, namely~\cite{Aziz}:
\begin{eqnarray}
\label{Qseries1728}
\hspace{-0.95in}&&    
{{Q(1728 \cdot \, x)} \over {1728}} \, \, = \, \, \, 
 x \, \, +744\,{x}^{2}\,\,  +750420\,{x}^{3}\, +872769632\,{x}^{4}
\, +1102652742882\,{x}^{5} \, + \, \, \cdots
 \nonumber
\end{eqnarray}
and:
\begin{eqnarray}
\label{Qseries1728}
\hspace{-0.95in}&&    
{{P(1728 \cdot \, x)} \over {1728}} \, \, = \, \, \,
 x \, \, -744\,{x}^{2} \,\,  +356652\,{x}^{3} 
\, -140361152\,{x}^{4} \, +49336682190\,{x}^{5} \, \, + \, \, \, \cdots 
\nonumber
\end{eqnarray}

The two functions $\,P(x)$ and $\, Q(x)$ are 
{\em differentially algebraic}~\cite{Selected,IsTheFull},
but  {\em not holonomic} functions.
Introducing 
the function $\, Q(x) \, = \, \, \exp(\Theta(x))$, equation (\ref{equaFthus}) 
with $\, \lambda \, = \, \, 0$ yields the following Schwarzian 
relations on $\, Q(x)$
\begin{eqnarray}
\label{identityQbister}
\hspace{-0.95in}&&  \, \,  \quad    \quad        \quad         
 W(x) \,\, + \,  \{Q(x), \, x\} \, 
+ \, {{1} \over {2}} \cdot \,  \Bigl({{ Q'(x)} \over {Q(x)}}\Bigr)^2
\, \, = \, \, \, 0,  \quad    \quad  \quad     \quad   \hbox{or:}  
\\
\hspace{-0.95in}&&  \, \,   \quad   \quad   \quad             
 W(x) \, + \,  \{\ln(Q(x)), \, x \} \, = \, 0 
\quad \quad  \, \,  \, \, \,  \hbox{where:}  \quad \, \,  \quad  \, \,     
{{Q'(x)} \over {Q(x)}} \, \, = \, \, \, \, {{1} \over {F(x)}},
\end{eqnarray}
when $\, P(x)$ the (composition) inverse of $\, Q(x)$ verifies 
the functional equation and Schwarzian equation:
\begin{eqnarray}
\label{Pfunc}
\hspace{-0.95in}&&  \quad  \quad   
x \cdot \, {{ d P(x)} \over { dx}}  \, \, = \, \, \, F(P(x)),
 \quad   \quad     \{P(x), \, x\} \,\,\, \,
 -\,  {{ 1} \over {2 \cdot \, x^2}} \,\,\, - \, W(P(x))
\, \,\,\,  = \, \,\, \, 0. 
\end{eqnarray}
Note that the one-parameter commuting 
family (\ref{yinfini}) solution of the Schwarzian 
equation (\ref{condition1ter}), can be 
expressed using these two functions $\, P(x)$ and $\, Q(x)$
as $\, y_1(a_1, \, x)  =  P(a_1 \cdot \, Q(x))\,\, $
where $\,\,\, a_1 \, = \, \, \exp(\epsilon)$.  

\vskip .2cm 

\section{Selected subcase: Heun function examples.} 
\label{subcaseHeun}

\vskip .1cm

Since the classification of Heun function is an interesting non trivial
problem, let us use the condition (\ref{WAR}), 
$\, W(x) \, = \, A'_R(x) \, + \, A_R(x)^2/2$, 
to find the Heun functions corresponding to such factorizations (like the example 
analysed in detail in~\cite{Aziz}). The Heun function 
$\, {\it HeunG} \left( a, \,q, \, \alpha,\, \beta,\, \gamma,\, \delta, \, x \right)$ 
is solution of a linear differential operator of order two 
$\, L_2 \, = \, \, D_x^2 \,\, +A(x) \cdot \, D_x \, +B(x)\,$ 
where $\, \,A(x)$ and  $\, B(x)$
read:
\begin{eqnarray}
\label{Heun2F1with}
\hspace{-0.96in}&&   \quad    
A(x)  \,\,\, = \, \, \,  \,  \, 
{\frac { (\alpha+\beta \, +1) \cdot \,  {x}^{2} \, 
- \, ( ( \delta +\gamma) \cdot \,  a 
\, +\alpha \, -\delta\,  +\beta\, +1) \cdot \,  x 
\,\, +\gamma \cdot \,a}{ x \cdot \, (x-1)  \cdot \, (x \, -a) }},
\\
\label{Heun2F1withB}
\hspace{-0.96in}&& 
\quad 
B(x) \, \, \,= \, \, \, \,  \, 
 {\frac {\alpha \, \,\beta \cdot \,x \, \, -q}{
x \cdot \, (x-1)  \cdot \, (x \, -a) }}.
\end{eqnarray}
One thus simply deduces the corresponding function $\, W(x)$ 
function from the formula (\ref{wherecond}), namely 
$\, W(x) \, = \, A'(x) \, +A^2(x)/2 \, -2 \, B(x)$. 
At first sight we exclude the values $\, a \, = \, 0$ and $\, a\,= \,1$
in order to have Heun functions with four singularities 
$\, 0, \, 1, \, a, \, \infty$ to avoid trivial subcases where 
the Heun functions could reduce to $\, _2F_1$ hypergeometric 
functions. If one imposes that the function $\, W(x)$ is of the form 
(\ref{WAR}), the rational function $\, A_R(x)$ must be of the form:
\begin{eqnarray}
\label{mustbe}
\hspace{-0.96in}&&   \quad   \quad
   \quad   \quad    \quad    \quad  \, \,
A_R(x)  \, \,\, = \, \, \, \, \,
{{u} \over {x\, -a}} \, \,\, +\, \, {{v} \over {x}}
 \, \,\, + \,  \, {{w} \over {x\, -1}}. 
\end{eqnarray}
The identification of $\, W(x)$ given by (\ref{WAR}) with $\, A_R$ 
of the form (\ref{mustbe}), with
$\, W(x) \, = \, A'(x)  \, +A^2(x)/2 \, -2 \, B(x)$
where $\, A(x)$ and $\, B(x)$ are given by (\ref{Heun2F1with}) 
and (\ref{Heun2F1withB}), gives a set of five equations in the
parameters of the Heun function and in the three coefficients 
$\, u, \, v, \, w$ in (\ref{mustbe}), the simplest one being 
\begin{eqnarray}
\label{mustbe}
\hspace{-0.96in}&&   \quad   \quad   \quad 
  \quad    \quad    \quad    \quad  
 a^2 \, \cdot \, (\gamma \, -v) \cdot \,  (\gamma \, -2 \, +v) 
\,  \, \, \,  = \,  \, \,  \, \, 0. 
\end{eqnarray}
The example analysed in~\cite{Aziz} corresponding to the
factorization condition (\ref{WAR}) corresponds to the 
following values of these parameters:
\begin{eqnarray}
\label{parameters}
\hspace{-0.96in}&&   \quad   \, \,  
 a \,  = \,  \, M, \,  \,  \,  \,  \,  \,  q \,  = \,  \, (M+1)/4,  
\,  \, \,   \,  \, \, \alpha= \, 1/2, \,  \,  \,   \,  \,\,  
 \beta=  \, 1,  \,   \, \,   \,  \, \,\gamma =  \, 3/2,
  \, \,\,    \,  \,   \, \delta=  \, 1/2,
\\ 
\hspace{-0.96in}&&   \quad   \, \quad \, \, \hbox{with:} 
\quad \quad   \quad   \quad    \, \, 
u \, = \, \, 1/2, \, \,  \,  \, \,   \,  \,  \, \,   v \, = \, \, 1/2,
 \, \, \,  \,   \,  \,  \, \,   \,  w  \, = \, \,   1/2, 
\end{eqnarray}
which corresponds to the $\, \gamma \, -2 \, +v \, = \, 0$ branch of (\ref{mustbe}).
The analysis of these five equations gives four solutions that we have excluded 
at first sight because they corresponds to $\, a \, = \, 1$ and yield reduction 
to $\, _2F_1$ hypergeometric functions\footnote[1]{Like for instance 
$\, a \,  = \,  \, 1, \, \,  \,  q \,  = \,  \, \beta \cdot \, \gamma$, 
with $\,  \,  \, u \, = \, \, \alpha \, -\beta \, -\delta \, -\gamma \, +1,\,  \, 
 \, v \, = \, \, \gamma, \, \, \,  w  \, = \, \,   \delta$.  }, except when 
a bunch of conditions occur
\begin{eqnarray}
\label{parameters1}
\hspace{-0.96in}&&   \quad    \quad   \quad
\alpha \, -\gamma \,+1\,\, = \,\,\,0, \, \, \, \,    \, \, \, \, 
\beta \, -\delta \, -1\, \, \, = \,\,\,0, \,\,   \, \, \, \, \,  \, 
\alpha \, -\delta \, -\gamma \, +2\,\, = \,\,\,0, 
\\
\hspace{-0.96in}&&   \, \,    
\hbox{and}  \quad  \quad   \quad 
\alpha \, -\gamma \, -1 \, = \, \, 0, \, \,  \, \, \, \,\,    \, 
\alpha \, -\delta \, -\gamma \, = \, \, 0, \, \, \, \, \, \,   \,  \, 
\beta \, -\delta \, +1 \, = \, \, 0, \, \, 
\quad \, \,    \cdots  \quad 
\end{eqnarray}
The example analyzed in~\cite{Aziz} corresponding to (\ref{parameters}) 
is equivalent  to 
\begin{eqnarray}
\label{parameters2}
\hspace{-0.96in}&&   \quad   \quad   \quad  \quad   \quad \quad   \quad \quad   \quad
\alpha \, \,  -\gamma \,\,  +1\,\,\,  = \, \,\,\,0. 
\end{eqnarray}

\vskip .2cm 

\section{Pullback invariance up to operator homomorphisms: a simple hypergeometric example.} 
\label{Asimpler}

 Let us consider the order-two linear differential operator 
\begin{eqnarray}
\label{moinsunquart}
\hspace{-0.95in}&&  \quad   \quad  \quad   \quad \quad  
{\cal L}_2 \, \, = \, \, \, \,  \, 
D_x^2 \,  \,  \, + \, {{3\, x \, -2} \over { 2 \cdot \, x  \cdot \, (x-1)}} \cdot  \, D_x 
  \,   \, \,   -\,  {{3} \over { 16 \cdot \, x  \cdot \, (x-1)}},
\end{eqnarray}
which has the hypergeometric function solution 
$\, \,  _2F_1([-1/4, 3/4],[1],x)$.
We have the following homomorphism of the type (\ref{twopullbacks}) 
between $\, {\cal L}_2$ pullbacked by two simple different 
rational functions $\, p_1(x)$ and $\, p_2(x)$:
\begin{eqnarray}
\label{twopullbacksexample}
\hspace{-0.95in}&&  \quad \, \, \,   
  pullback\Bigl({\cal L}_2 , \,  \, p_1(x)  \Bigr) 
 \cdot \, L_1  \cdot \, \alpha(x)   
\, \,\,\,    = \, \, \,  \,  \alpha(x) \cdot \,  M_1 \cdot  \, 
 pullback\Bigl({\cal L}_2 ,  \,  \, p_2(x)  \Bigr),  
 \\
\hspace{-0.95in}&& 
\label{p1p2}  
\hbox{where:} \, \, \,    \quad  \, \,
p_1(x) \, = \, \,  {{- \, 64 \, x } \over { (1-x) \cdot \, (1\, -9\,x)^3 }}, 
\quad  \quad
 p_2(x) \, = \, \, {{- \, 64 \, x^3 } \over { (1-x)^3 \cdot \, (1\, -9\,x) }}, 
 \\
\label{twopullbacksexamplewhere}
\hspace{-0.95in}&&   
 \alpha(x) \, \, = \, \, \,
 x^3 \cdot \Bigl( {{ 1\, -x} \over {1\, -9 \, x}} \Bigr)^{1/2}, 
\quad \, \, \,  \,   
M_1 \, \, = \, \,\, 
 8 \cdot \,{\frac { (1 -9\,x) }{ (1\, -x) \cdot \,  {x}^{2}}} \cdot \, D_x 
 \, \, +{\frac {171\,{x}^{2}-142\,x+19}{ (1 \, -x)^{2} \cdot \, {x}^{3}}}, 
\nonumber \\
\hspace{-0.95in}&&    \hbox{and:} \quad   \quad   \quad    \quad   
L_1 \,\,   \,   = \, \,\, \,  
 8 \cdot \,{\frac { (1 -9\,x) }{ (1 \, -x) \cdot \, {x}^{2}}} \cdot \, D_x 
 \,  \,\,   -{\frac {189\,{x}^{2}-226\,x+21}{ (1\, -x)^{2} \cdot \, {x}^{3}}}. 
\end{eqnarray}
Denoting $\, A$ and $\, B$ the two rational pullbacks $\, p_1(x)$ 
and $\, p_2(x)$ in (\ref{twopullbacksexample}) one finds 
that they are related by the following rational algebraic curve:
\begin{eqnarray}
\label{modularcurve}
\hspace{-0.95in}&&  \quad  \quad    \Gamma_3(A, \, B) \, = \, \,     \, 
4096  \cdot \, A \, B \cdot \,  ({A}^{2}{B}^{2} \, + 1)\,\,  
 -4608  \cdot \, A \, B \cdot \, ({A}{B}\, +1) \cdot \, (A \,+B) \,
 \nonumber \\
\hspace{-0.95in}&&  \quad  \quad    \quad      \quad        \quad    \, 
 \,  -({A}^{4}\, -900\,{A}^{3}B+28422\,{A}^{2}{B}^{2}\, -900\,A{B}^{3}+{B}^{4})
 \, \,\, =  \, \,\,   \,\, 0. 
\end{eqnarray}
The two {\em Hauptmoduls} parametrizing the 
{\em modular equation}\footnote[2]{See equation (108) in subsection 5.1 
of~\cite{Aziz}.} 
corresponding to the representation of 
$\, \tau \, \rightarrow \, 3 \, \tau$, are given as follows:
\begin{eqnarray}
\label{3tau}
\hspace{-0.95in}&&  \quad \quad 
P_1(x) \, \, = \, \, \, \,{\frac {1728 \, \, x}{ (x+27)  \cdot \, (x+3)^{3}}}, \quad \quad 
P_2(x) \, \, = \, \, \, \,{\frac {1728 \, \, {x}^{3}}{ (x+27)  \cdot \, (x+243)^{3}}}.
\end{eqnarray}
Note that we have the following relations between $\, p_1(x)$ and $\, p_2(x)$, 
and the two Hauptmoduls  $\, P_1(x)$ and $\, P_2(x)$: 
\begin{eqnarray}
\label{notethat}
\hspace{-0.95in}&& \quad \quad \quad  \quad \quad  \quad  
p_1(x) \,\, = \, \, \, P_1(-27 \, x),   \, \, \quad \quad \quad
p_2(x)  \,\, = \, \, \, P_2(- 243 \, x),   \, \, 
\end{eqnarray}
which explain the compatibility between the two relations:
\begin{eqnarray}
\label{notethat}
\hspace{-0.95in}&& \quad \quad  \quad  \quad \quad    \quad  
p_2(x) \, \, = \, \, p_1\Bigl({{1} \over {9 \, x}}\Bigr),  \, \,  \quad \quad \quad 
P_2(x) \, \, = \, \, P_1\Bigl({{729} \over { x}}\Bigr).
\end{eqnarray}
Relation (\ref{twopullbacksexample}) yields the following identity 
on the $\, _2F_1$ hypergeometric function
\begin{eqnarray}
\label{followingidentity}
\hspace{-0.95in}&&  \quad  \quad  \quad 
_2F_1\Bigl([-{{1} \over {4}}, \, {{3} \over {4}}], \, [1], \, p_1(x)\Bigr)
  \, \,\, =  \, \, \,\, 
 {\cal L}_1 \Bigl( \, _2F_1\Bigl([-{{1} \over {4}}, \, {{3} \over {4}}], \, [1], \, p_2(x)\Bigr) \Bigr), 
\\
\label{calL1}
\hspace{-0.95in}&&    \hbox{where:} \quad   \, \, \,  
{\cal L}_1  \, \,\, =  \, \, \,
   {{ 8 \cdot \,(1-9\,x)^{1/2} } \over {
 3 \cdot \, (1-\,x)^{1/2} }} \cdot \, x  \cdot \, {{d} \over { dx}} 
\, \, +{\frac { 1 \,-3\,x \,-45\,{x}^{2}\,-81\,{x}^{3} }{ (1-x)^{3/2} \cdot \, (1-9\,x)^{3/2}}}, 
\end{eqnarray}
\begin{eqnarray}
\label{followingidentity2}
\hspace{-0.95in}&&   \quad  \quad  \quad 
_2F_1\Bigl([-{{1} \over {4}}, \, {{3} \over {4}}], \, [1], \, p_2(x)\Bigr) 
 \, \,\, =  \, \, \,\, 
 {\cal L}_2 \Bigl( \,  _2F_1\Bigl([-{{1} \over {4}}, \, {{3} \over {4}}], \, [1], \, p_1(x)\Bigr)\Bigr), 
\\
\label{calL2}
\hspace{-0.95in}&&    \hbox{where:} \quad   \, \, \,   
{\cal L}_2 
\, \,\, =  \, \, \,  
-\, {{ 8 \cdot \,(1-\,x)^{1/2} } \over { 3 \cdot \, (1-\,9\, x)^{1/2} }} \cdot \, x  \cdot \, {{d} \over { dx}} 
\, \, +{\frac { 1 \,+5\,x \,+3\,{x}^{2}\,-9\,{x}^{3} }{ (1-x)^{3/2} \cdot \, (1-9\,x)^{3/2}}}.
\end{eqnarray}
Introducing  the order-two linear differential operator $\, H_1$ annihilating the 
pullbacked hypergeometric function 
$\,  _2F_1([-1/4, \, 3/4], \, [1], \, p_1(x))$:
\begin{eqnarray}
\label{H2}
\hspace{-0.95in}&&   \,  \,  \, \, \,  
H_1 \, \, = \, \,\, \, D_x^2 \, \,  \, +  {{ (1\, -3\, x)^2 } \over {
x \cdot \, (1\, -x) \cdot \, (1\, -9 \, x) }} \cdot \, D_x 
\, \,  + \, {{ 12} \over {x \cdot \, (1\, -x)^2 \cdot \, (1\, -9 \, x)^2 }}, 
\end{eqnarray}
the compatibility between relation (\ref{followingidentity}) 
and  (\ref{followingidentity2}) is a consequence of the identity
\begin{eqnarray}
\label{compatibility}
\hspace{-0.95in}&&   \quad  \quad  \quad \quad \quad \quad 
{\cal L}_1 \cdot \, {\cal L}_2 \, \, \, \, =  \, \, \, \,
  1 \, \, \, - \, {{64 \, x^2} \over {9}}  \,  \cdot \, H_1, 
\end{eqnarray}
namely that the product $\,{\cal L}_1 \cdot \, {\cal L}_2 \, $ is equal 
to $\, 1$ modulo $\, H_1$. Of course  introducing  the order-two linear 
differential operator $\, H_2$ annihilating the 
pullbacked hypergeometric function $\,  _2F_1([-1/4, \, 3/4], \, [1], \, p_2(x))$
one also has a very similar identity:
\begin{eqnarray}
\label{compatibility2}
\hspace{-0.95in}&&   \quad  \quad  \quad \quad \quad \quad 
{\cal L}_2 \cdot \, {\cal L}_1 \, \, \, \, =  \, \, \, \,
  1 \, \, \, - \, {{64 \, x^2} \over {9}}  \,  \cdot \, H_2, 
\end{eqnarray}
which means that the product $\,{\cal L}_2 \cdot \, {\cal L}_1$ is equal 
to $\, 1$ modulo $\, H_2$. 

\vskip .1cm 

One can get rid of the unpleasant square roots 
in (\ref{calL1}), (\ref{calL2}) introducing instead of the 
pullbacked hypergeometric functions 
$\,  _2F_1([-1/4, \, 3/4], \, [1], \, p_2(x))$ 
and $\,  _2F_1([-1/4, \, 3/4], \, [1], \, p_1(x))$,
the functions 
\begin{eqnarray}
\label{twofunctions}
\hspace{-0.95in}&&   
\Xi_2(x) \,\,\, = \, \, \, \,\,
x \cdot \, (1\, -x)^{3/4} \cdot (1\, -9\, x)^{1/4}  
\cdot \, _2F_1\Bigl([-{{1} \over {4}}, \, {{3} \over {4}}], \, [1], \, p_2(x)\Bigr), 
\\
\hspace{-0.95in}&&  
\Xi_1(x) \,\, = \,\,  \,\, 3^{7/2} \cdot \, \Xi_2 \Bigl({{1} \over {9 \, x}} \Bigr)
\,\, = \, \, \, {{(1\, -x)^{1/4} \cdot (1\, -9\, x)^{3/4} } \over {x^2}} 
\cdot \, _2F_1\Bigl([-{{1} \over {4}}, \, {{3} \over {4}}], \, [1], \, p_1(x)\Bigr).
\nonumber 
\end{eqnarray}
$\Xi_2(x)\, $ is a series with integer coefficients
\begin{eqnarray}
\label{Xi2series}
\hspace{-0.95in}&& \quad 
\Xi_2(x) \,\,\, = \, \, \, \, \, x \,\, \, -3\,{x}^{2}
 \, \, \,  -6\,{x}^{3} \, \,-22\,{x}^{4} \, \,-108\,{x}^{5} \, \,
-612\,{x}^{6}  \,\,-3786\,{x}^{7} \, \,-24858\,{x}^{8} 
\nonumber \\
\hspace{-0.95in}&& \quad \quad \quad \quad  \,  \,  \, 
\,-170406\,{x}^{9} \, \,-1207014\,{x}^{10} \, \, -8771850\,{x}^{11} 
\,\, \,\, + \,\, \cdots 
\nonumber 
\end{eqnarray}
when $\, \Xi_1(x) \, $ is a Laurent series with integer coefficients.
These two functions are simply related as follows: 
\begin{eqnarray}
\label{twofunctionssimply}
\hspace{-0.96in}&& \quad \quad  \quad  \quad  
\Xi_1(x) \,\,\, = \, \, \, \,  {\cal M}_1\Bigl(  \Xi_2(x)  \Bigr)
  \quad  \quad  \quad  \quad \quad \quad \hbox{where:} 
\\
\label{U1def}
\hspace{-0.96in}&& \quad  \quad  \quad  \quad 
 {\cal M}_1\,\, = \, \, \,  \, {{8} \over {3}} \cdot \, 
 {{ (1\, -9 \, x) } \over { x^2 \cdot \, (1\, -x) }} \cdot \, D_x 
\,\, \,  + {{117 \, x \, -5 } \over {3 \cdot \, x^3 \cdot \, (1\, -x)  }}. 
\end{eqnarray}
In fact the function $\, \Xi_2(x)$ is solution of the order-two linear 
differential operator $\, \Omega_2$
\begin{eqnarray}
\label{opXi2}
\hspace{-0.95in}&&   \quad  \quad  \quad 
\Omega_2 \, \,= \, \, \,  \,  D_x^2 \,  \, \,  
-{\frac {(1 \, -3\,x)}{x \cdot \, (1 \, -x) }} \cdot \, D_x \, \,  \, 
+{\frac {1 \, -9 \, x \, +36\,{x}^{2}  }{{x}^{2} 
\cdot \, (1 \, -9\,x)  \cdot \, (1 \, -x) }},  
\end{eqnarray}
with a remarkable duality property. 
{\em It is homomorphic to its pullback by} $\, x \, \rightarrow \, 1/9/x$:
\begin{eqnarray}
\label{opXi2}
\hspace{-0.95in}&&   \quad  \quad  \quad \quad
pullback\Bigl(\Omega_2, \, {{1} \over { 9\, x}}\Bigr) \cdot \, {\cal M}_1 
 \, \, \,= \, \, \, \,  {\cal N}_1 \cdot \,  \Omega_2
\\
\hspace{-0.95in}&&   \quad  \hbox{where:}  \quad \quad \quad \quad 
{\cal N}_1  \, \,= \, \,  \,
 {{8 \cdot \, (1\, -9 \, x) } \over { 3 \cdot \, (1\, -x) \cdot \, x^2 }} \cdot \, D_x 
\,\,\, - \, {{27\, x \, -11 } \over { 3 \cdot \, (1\, -x) \cdot \, x^3  }}. 
\end{eqnarray}
The simple relation (\ref{twofunctionssimply}), which is a rewriting 
of (\ref{followingidentity}) with the order-one operator
$\, {\cal L}_1$ being replaced by the order-one operator $\, {\cal M}_1$,   
 is an obvious consequence 
of the homomorphism (\ref{opXi2}). Of course we also have 
the (mirror) relation\footnote[2]{Consequence 
of the (mirror) homomorphism relation: 
$\,\,\, {\cal N}_2 \cdot \, pullback((\Omega_2, \, 1/9/x) 
 \, \, \,= \, \, \, \,   \Omega_2 \cdot \, {\cal M}_2 $.}, compatible 
with (\ref{twofunctionssimply}), 
which is a rewriting of (\ref{followingidentity2}) with the order-one operator
$\, {\cal L}_2$ being replaced by  the order-one operator $\, {\cal M}_2$   
\begin{eqnarray}
\label{twofunctionssimplybis}
\hspace{-0.96in}&& \quad  \quad  \quad \quad  \, \,  \,
\Xi_2(x) \,\,\, = \, \, \, \,  {\cal M}_2\Bigl(  \Xi_1(x)  \Bigr)
  \quad  \quad  \quad  \quad \quad \quad \hbox{where:} 
\\
\label{U2def}
\hspace{-0.96in}&& \quad  \quad  \quad  \quad \, \,  \,
 {\cal M}_2\,\, = \, \, \,  
  -\, {{ 8 \cdot \,(1\, - \, x) \cdot \, x^4 } \over {
3 \cdot \, (1\, -9 \, x) }} \cdot \, D_x 
\,\, + {{(5 \, x \, -13) \cdot \, x^3 } \over {3 \cdot \, (1\, -9 \, x)  }}. 
\end{eqnarray}
Note that $\, {\cal M}_1$ and $\, {\cal M}_2$ given by (\ref{U1def}) 
and (\ref{U2def}) are related  by the
involutive change of variable $\, x \, \rightarrow \, 1/9/x$:
\begin{eqnarray}
\label{involutivechange}
\hspace{-0.96in}&& \, \,  
{\cal M}_1 \, \, = \, \, \, 
6561 \cdot \, pullback\Bigl({\cal M}_2, \, {{1} \over {9 \, x}}\Bigr),
   \, \,  \, \,  \, \, \,  
6561 \cdot \, {\cal M}_2 \, \, = \, \, \, 
pullback\Bigl({\cal M}_1, \, {{1} \over {9 \, x}}\Bigr).
\end{eqnarray}
Denoting $\, \Omega_1$ the order-two operator annihilating $\, \Xi_1$,
the compatibility between the relations (\ref{twofunctionssimply}) 
and (\ref{twofunctionssimplybis}) corresponds to the relations:
\begin{eqnarray}
\label{compatibilitybis}
\hspace{-0.96in}&& \quad \quad  \,  \,  \,  
{\cal M}_1 \cdot \, {\cal M}_2 \, = \, \, 
1 \, \, - \, {{64 \, x^2 } \over {9 }} \cdot \, \Omega_1,  
\quad \quad  \, \,  
{\cal M}_2 \cdot \, {\cal M}_1 \, = \, \, 
1 \, \, - \, {{64 \, x^2 } \over {9 }} \cdot \, \Omega_2,
\quad  
\end{eqnarray}
which should be compared with (\ref{compatibility2}) and (\ref{compatibility}).   

\vskip .2cm 

Relations (\ref{followingidentity2}), or\footnote[1]{Or relations 
(\ref{followingidentity}) or (\ref{twofunctionssimply}), 
but in that case the series corresponding to $\, y(x)$ are Puiseux series : 
$\, y(x) \, = \, x^{1/3} \, + \cdots  $} (\ref{twofunctionssimplybis}),   
can be seen as a particular 
case of a generalized pullback symmetry condition of the form
\begin{eqnarray}
\label{followingidentitygenera}
\hspace{-0.95in}&&  
\quad \, \, \,  \, \,  
_2F_1\Bigl([\alpha, \, \beta], \, [\gamma], \, y(x)\Bigr) 
\, \,\, =  \, \, \,\,
\Bigl(   {\cal A}(x) \cdot \, {{d} \over { dx}} 
\, \, + {\cal B}(x)   \Bigr)
\, \cdot \,   _2F_1\Bigl([\alpha, \, \beta], \, [\gamma], \, x \Bigr), 
\end{eqnarray}
where $\,   {\cal A}(x)$ and $\, {\cal B}(x)$ are algebraic functions. Identities 
like (\ref{followingidentity}) can be seen as generalizations of the identities 
$\,_2F_1([\alpha, \, \beta], \, [\gamma], \, y(x)) $
$= {\cal A}(x) \cdot \, _2F_1([\alpha, \, \beta], \, [\gamma], \, x)$
 analysed in~\cite{Aziz}.

\vskip .2cm

\vskip .2cm

\subsection{Representation of the composition of the algebraic transformations $\, x \, \rightarrow \, y(x)$.} 
\label{pullbackhomomorphismapp}

\vskip .1cm

We want to see the algebraic transformations $\, x \, \rightarrow \, y(x) \, $ as symmetries. 
In particular we want to have a representation of the composition of 
these algebraic transformations, like:
\begin{eqnarray}
\label{followingidentitygenera2}
\hspace{-0.95in}&&     \, 
_2F_1\Bigl([\alpha, \, \beta], \, [\gamma], \, y(y(x))\Bigr)  
\, \,\, =  \, \, \,\,
\Bigl(   {\cal A}_2(x) \cdot \, {{d} \over { dx}} 
\, \, + {\cal B}_2(x)   \Bigr)
\, \cdot \,   _2F_1\Bigl([\alpha, \, \beta], \, [\gamma], \, x \Bigr). 
\end{eqnarray}
Let us show here that by building on the previous example we can actually provide
identities of the type (\ref{followingidentitygenera2}). Introducing 
\begin{eqnarray}
\label{followingidentity9intro}
\hspace{-0.95in}&& 
q_1(x) \,\, = \, \, \, \,
{\frac { -1728 \cdot \, x \cdot \, (1\, -81\,x \, + 2187 \,{x}^{2}) }{
 (1\, -81\,x)^{9} \cdot \, (1\, -27\,x)  \cdot \, (1\, + 2187\,x^2) }}, 
\\
\label{followingidentity9intro2}
\hspace{-0.95in}&&  
q_2(x)  \,\, = \, \, \, \, \,   q_1\Bigl({{1 } \over { 2187 \, x }}  \Bigr) \,\, = \, \, \, \, \,
{\frac {  -1728 \cdot \, 3^{24} \cdot \, {x}^{9} \cdot \, ( 1 \, -81\,x \, +2187\,{x}^{2}) }{
 (1\, +2187\,{x}^{2})  \cdot \, (1\, - 27\,x)^{9} \cdot \, (1\, -81\,x) }}.
\end{eqnarray}
one has the new pullback symmetry relation similar to (\ref{followingidentity}):
\begin{eqnarray}
\label{followingidentity9}
\hspace{-0.95in}&&  
\quad \quad \quad \, \,
_2F_1\Bigl([-{{1} \over {4}}, \, {{3} \over {4}}], \, [1], \, q_1(x)\Bigr) 
 \, \,\, =  \, \, \, \,
 {\hat L}_1 \Bigl( \,  _2F_1\Bigl([-{{1} \over {4}}, \, {{3} \over {4}}], \, [1], \, q_2(x)\Bigr)\Bigr), 
\end{eqnarray}
where:
\begin{eqnarray}
\label{followingidentity9where}
\hspace{-0.95in}&&   \quad   \quad    \quad  \,\,
{\hat L}_1 \, \,\, =  \, \, \, \, {{32} \over {9}} \cdot \, 
{\frac {x \cdot \, (1 \, -81\,x \, +2187\,{x}^{2})  \cdot \, U_1(x) }{
 (1 \, -81\,x)  \cdot (1 \, -27\,x)^{5}}} \cdot \, D_x
\nonumber \\
\hspace{-0.95in}&&   \quad  \quad   \quad   \quad   \quad   \quad   \quad  
 + \, {\frac {V_1(x)}{ 
(1 \, -108\,x \, +2187\,{x}^{2})  \cdot \, (1\, -81\,x)  \cdot \, (1 \, - 27\,x)^{5}}},
\\
\hspace{-0.95in}&&     \quad   \, \,  
U_1(x) \, \, = \,\,\, \,  
1\, \, \,-81\,x\, \,\, +4374\,{x}^{2}\,\, \, -177147\,{x}^{3}\, \, \, +4782969\,{x}^{4}, 
\\
\hspace{-0.95in}&&     \quad   \, \,   
V_1(x) \, \, = \,\,\,  \,  \; 
1\, \,\, -26244\,{x}^{2}\,\, \, +3779136\,{x}^{3}\,\, \, -277412202\,{x}^{4} \,\, \, +12397455648\,{x}^{5}
\nonumber \\
\hspace{-0.95in}&&    \quad   \quad   \quad  \quad   \quad  \,
-311486073156\,{x}^{6}\,  \, +3012581722464\,{x}^{7}\, \,  +22876792454961\,{x}^{8}.
\end{eqnarray}
One also has the new pullback symmetry relation similar to (\ref{followingidentity2}) 
\begin{eqnarray}
\label{followingidentity9bis}
\hspace{-0.95in}&&  
\quad \quad \quad 
_2F_1\Bigl([-{{1} \over {4}}, \, {{3} \over {4}}], \, [1], \, q_2(x)\Bigr) 
 \, \,\, =  \, \, \, \,
 {\hat L}_2 \Bigl( \,  _2F_1\Bigl([-{{1} \over {4}}, \, {{3} \over {4}}], \, [1], \, q_1(x)\Bigr)\Bigr), 
\end{eqnarray}
\begin{eqnarray}
\label{followingidentity9wherebis}
\hspace{-0.95in}&&   \quad   \quad    \quad  
{\hat L}_2 \, \,\, =  \, \, \,
 -\, {{32} \over {9}} \cdot \, 
{\frac {x \cdot \, (1 \, -81\,x \, +2187\,{x}^{2})  \cdot \, U_2(x) }{
 (1\, -81\,x)^{5} \cdot \, (1\, - 27\,x) }}
\cdot \, D_x
\nonumber \\
\hspace{-0.95in}&&    \quad  \quad   \quad    \quad   \quad   \quad   \quad  
+{\frac {V_2(x) }{ 
(1 \, -108\,x \, + 2187\,{x}^{2})  \cdot \, (1\, -81\,x)^{5} \cdot \, (1\, - 27\,x) }}, 
\\
\hspace{-0.95in}&&    \, \,    \, \,   
U_2(x) \, \, = \,\, \,
1\,\,\, -81\,x\,\,\, +4374\,{x}^{2}\,\, -177147\,{x}^{3}\,\, +4782969\,{x}^{4}, 
\\
\hspace{-0.95in}&&    \, \,  \, \,   
V_2(x) \, \, = \,\,\,
 1\,\,\, +288\,x\,\,\, -65124\,{x}^{2}\,\, +5668704\,{x}^{3}\,\, -277412202\,{x}^{4}\, \, +8264970432\,{x}^{5}
\nonumber \\
\hspace{-0.95in}&&    \quad   \quad   \quad  \quad   \quad 
-125524238436\,{x}^{6}\,+22876792454961\,{x}^{8}.
\end{eqnarray}
Let us introduce  the order-two linear differential operator $\, {\hat H}_1$ 
annihilating the pullbacked hypergeometric function 
$\,  _2F_1([-1/4, \, 3/4], \, [1], \, q_1(x))$:
\begin{eqnarray}
\label{hatH2}
\hspace{-0.95in}&&  
{\hat H}_1 \, \, = \, \,\, \, D_x^2 \,\,\, 
+  {\frac {\alpha_1(x)}{ ( 1 \, -81\,x)  \cdot \, (1\, - 27\,x)  
\cdot \, (1\, + 2187\,{x}^{2}) 
 \cdot \, (1 \, -81\,x +2187\,{x}^{2}) \cdot \,  x}} \cdot \, D_x 
\nonumber \\
\hspace{-0.95in}&&   \quad  \quad 
- \,{\frac {324}{x \cdot \, (1 \, -81\,x +2187\,{x}^{2})  \cdot \, 
 (1\, +2187\,{x}^{2})^{2} \cdot \, (1-81\,x)^{2} \cdot \, (1 \, - 27\,x) ^{2}}}, 
\end{eqnarray}
where
\begin{eqnarray}
\label{hatH2where}
\hspace{-0.95in}&& 
\alpha_1(x) \, = \, \,  \, 
1 \, \,  +2187\,{x}^{2} \, -354294\,{x}^{3}\,+23914845\,{x}^{4} 
\, -774840978\,{x}^{5}\,+10460353203\,{x}^{6}.
\nonumber
\end{eqnarray}
The compatibility between relation (\ref{followingidentity}) and  
(\ref{followingidentity2}) is a consequence of the identity:
\begin{eqnarray}
\label{compatibility}
\hspace{-0.95in}&&   \quad  \quad 
{\hat L}_1  \cdot \, {\hat L}_2  
\, \, \, \, =  \, \, \, \,\, 1 \, \, \, + R_{1,2}(x)  \,  \cdot \, {\hat H}_1, 
\quad \quad  \quad  \quad  \quad \hbox{where:} 
\\
\hspace{-0.95in}&&   \quad   \quad 
R_{1,2}(x) \, \, =  \, \, \,
- \, {{1024} \over {81}} \cdot \,
 {\frac { {x}^{2} \cdot \, (1 \, -81\,x \, +2187\,{x}^{2})^{4} \cdot \, 
(1\, +2187\,{x}^{2})^{2}}{ (1\, -81\,x)^{6} \cdot \, (1\, - 27\,x)^{6}}}. 
\end{eqnarray}
Of course  introducing  the order-two linear differential operator $\, {\hat H}_2$ 
annihilating the pullbacked hypergeometric function 
$\,  _2F_1([-1/4, \, 3/4], \, [1], \, q_2(x))$,
one also has a similar identity {\em with the same rational function} $\, R_{1,2}(x)$:  
\begin{eqnarray}
\label{compatibility}
\hspace{-0.95in}&&   \quad \quad  \quad  \quad  \quad  \quad  \quad  \quad 
{\hat L}_2  \cdot \, {\hat L}_1  \, \, \, \, =  \, \, \, \, \,
1 \, \, \,\, + R_{1,2}(x)  \,  \cdot \, {\hat H}_2.
\end{eqnarray}
Again we have that $\, {\hat L}_1$ and $\,  {\hat L}_2$ are 
obtained from each other by the (involutive) change of variable 
$\, x \, \longleftrightarrow \, \, \, 1/2187/x$:
\begin{eqnarray}
\label{involutivechange}
\hspace{-0.96in}&& \, 
-9 \, \cdot \,{\hat L}_1 \, \, = \, \, \, 
 pullback\Bigl({\hat L}_2, \, {{1} \over {2187 \, x}}\Bigr),
   \quad   \,  \, 
{\hat L}_2 \, \, = \, \, \, 
-9 \, \cdot \, pullback\Bigl({\hat L}_1, \, {{1} \over {2187 \, x}}\Bigr).
\end{eqnarray}
Note that the two pullbacks $\, q_1(x)$ and $\, q_2(x)$ 
(see (\ref{followingidentity9intro}), (\ref{followingidentity9intro2}))
are related to the two 
previous pullbacks $\, p_1(x)$ and $\, p_2(x)$ (see (\ref{p1p2})):
\begin{eqnarray}
\label{related9}
\hspace{-0.95in}&&  \quad  \quad     \quad    
q_1(x) \, \, = \, \, \,
 p_1\Bigl( 27 \cdot \,x \cdot \,  (1 \, -81\,x \, +  2187\,{x}^{2})  \Bigr),
\\
\hspace{-0.95in}&&    
 \quad \quad        \quad    
q_2(x) \, \, = \, \, \,  p_2\Bigl( {\frac { 19683 \cdot \, {x}^{3}}{ 
1 \, -81\,x \, +  2187\,{x}^{2} }}  \Bigr) \, \, = \, \, \, 
p_1 \Bigl( {{1 \, -81\,x \, +  2187\,{x}^{2} } \over { 177147 \cdot \, x^3}} \Bigr).
\end{eqnarray}
Recalling
$\, \, \Phi(x)  = \,  _2F_1([-1/4, \, 3/4], \, [1], \, p_1(x))\, $
the new identities (\ref{followingidentity9}) and  (\ref{followingidentity9bis}) read 
\begin{eqnarray}
\label{followingidentity9ter}
\hspace{-0.95in}&&  
\quad \quad 
\Phi \Bigl( 27 \cdot \,x \cdot \,  (1 \, -81\,x \, +  2187\,{x}^{2})  \Bigr)
  \, \,\, =  \, \, \, \,
 {\hat L}_1  \Bigr(\Phi\Bigl( {{1 \, -81\,x \, +  2187\,{x}^{2} } \over {
    177147 \cdot \, x^3}} \Bigr) \Bigr), 
\\
\label{followingidentity9ter2}
\hspace{-0.95in}&&  
\quad \quad 
 \Phi\Bigl(  {{1 \, -81\,x \, +  2187\,{x}^{2} } \over { 177147 \cdot \, x^3}}\Bigr) 
\, \,\, =  \, \, \,  \,
{\hat L}_2 \Bigl(\Phi\Bigl( 27 \cdot \,x \cdot \,  (1 \, -81\,x \, +  2187\,{x}^{2}) \Bigr) \Bigr),  
\end{eqnarray}
or, introducing $\, \, \, \Psi(x) \, = \,  _2F_1([-1/4, \, 3/4], \, [1], \, q_1(x))$:
\begin{eqnarray}
\label{followingidentity9quat}
\hspace{-0.95in}&&  
\quad \quad  \quad  \, \, \,  
  \Psi(x)  \, \,\, =  \, \, \,
 {\hat L}_1  \Bigr(\Psi\Bigl( {{1 } \over {2187 \cdot \, x}} \Bigr)\Bigr), 
\quad \quad 
 \Psi\Bigl({{1 } \over {2187 \cdot \, x}}  \Bigr)
  \, \,\, =  \, \, \, {\hat L}_2 \Bigl( \Psi(x) \Bigr).  
\end{eqnarray}
Denoting $\, A$ and $\, B$ the two pullbacks 
in (\ref{followingidentity9ter}),  (\ref{followingidentity9ter2}),
\begin{eqnarray}
\label{AandB}
\hspace{-0.95in}&&  
\quad   \, \, \, \,
\, A  \, =   \, \, 
27 \cdot \,x \cdot \,  (1 \, -81\,x \, +  2187\,{x}^{2}), 
\quad   \quad 
\, B  \,=  \,  \,
  {{ 1 \, -81\,x \, +  2187\,{x}^{2}} \over {  177147 \cdot \,x^3}}, 
\end{eqnarray}
one sees that they 
are related by the simple $A, \, B$ symmetric algebraic curve:
\begin{eqnarray}
\label{algcurve}
\hspace{-0.95in}&&  \quad   \quad   
9 \,{A}^{3}{B}^{3} \, \, \, -30 \, {A}^{2}{B}^{2}  \, \,\,  +12\,AB \cdot \, (A \, +B) 
\, \, \, -{A}^{2} \,-AB \,-{B}^{2} \, \, = \, \, \, 0.
\end{eqnarray}
Let us consider the algebraic equation (\ref{modularcurve}), 
that we denote $\, \Gamma_3(A, \, B) \, = \, \, 0$ because it is 
so closely related to the modular equation representing 
$\, \tau \, \rightarrow \, 3 \, \tau$ (see their close relation with the Hauptmoduls 
(\ref{3tau}) and (\ref{notethat})). 
Performing the resultant in $\, B$ of  the polynomial $\, \Gamma_3(A, \, B)$
with the same one  $\, \Gamma_3(B, \, C)$ one gets a new algebraic equation
 $\, \Gamma_9(A, \, C) \, \, = \, \, \, 0$.  The two pullbacks 
$\, q_1(x)$ and $\, q_2(x)$ {\em are actually a rational parametrization of that
new algebraic equation} $\, \Gamma_9(A, \, C) \, \, = \, \, \, 0$. In 
other words, if we think identity (\ref{followingidentity2}) as a
symmetry transformation identity of the type (\ref{followingidentitygenera}), 
the new identity (\ref{followingidentity9}) must be seen as the identity for
the iteration of that transformation:
\begin{eqnarray}
\label{followingidentitygenera2bis}
\hspace{-0.95in}&&  
\,  \, \,   
_2F_1\Bigl([\alpha, \, \beta], \, [\gamma], \, y(y(x))\Bigr) 
\, \,\, =  \, \, \,\,
\Bigl(   {\cal A}_2(x) \cdot \, {{d} \over { dx}} 
\, \, + {\cal B}_2(x)   \Bigr)
\, \cdot \,   _2F_1\Bigl([\alpha, \, \beta], \, [\gamma], \, x \Bigr). 
\end{eqnarray}
We are very close to a modular form, the previous algebraic curve (\ref{modularcurve}) 
playing the role of the {\em modular equation}\footnote[2]{Given by equation (108) 
in subsection 5.1.1 in~\cite{Aziz}.} 
(see (\ref{notethat})), and the algebraic curve 
$\, \Gamma_9(A, \, C) \, \, = \, \, \, 0$ playing the role of the 
modular equation corresponding to $\, \tau \, \rightarrow \, 9 \cdot \, \tau$.

 Note that if one calculates the function 
$\, W(x) \, = \, A'(x) +A(x)^2/2 \, -2 \, B(x)$ corresponding to the order-two 
operator  $\, {\cal L}_2$,  one gets 
\begin{eqnarray}
\label{WcalL2}
\hspace{-0.95in}&&  \quad 
W(x) \,\,    = \, \,  \, 
{\frac {x\, -4}{ 8 \cdot \, (x-1) \cdot \,  x}}
 \,  \, = \, \,\, \,  -{\frac{1}{2 \, x^2}} \,\, \,   -{\frac{7}{8 \, x}}
\,   \,\,  -{\frac{5}{4}} \,\, \,   -{\frac{13}{8}} \, x \,\,   \, -2 \, x^2 \,\,  
\, \, + \, \, \cdots \, \, 
\end{eqnarray}
which is of the form 
$\,\, \, W(x) \, = \, \, -1/2/x^2 \, \, + \,  \cdots$
(in contrast with the result for $\, \tilde{\chi}^{(2)}$, see (\ref{Laurentdonotmatch})).

\vskip .2cm

\section{Schwarzian conditions for different Calabi-Yau operators with related Yukawa couplings} 
\label{CalabiSchwa}

\vskip .1cm

\subsection{Revisiting a Calabi-Yau operator in~\cite{Duco}} 
\label{CalabiSchwa1}

Following Almkvist, van Straten and Zudilin~\cite{Duco}, 
let us consider the order-four linear differential operator $\, L_4$ 
such that its exterior square  
annihilates\footnote[5]{See also~\cite{Candelas}.} 
\begin{eqnarray}
\label{5F4}
\hspace{-0.95in}&& \,  \quad  \quad  \quad  \quad  \quad   \quad   \quad  \, \,  \, 
_5F_4\Bigl([{{1} \over {2}}, \, a, \, 1-a, \, b, \, 1-b], [1, \, 1, \, 1, \, 1], \, x\Bigr). 
\end{eqnarray}
This  order-four linear differential operator such that its exterior square 
is order-five (it verifies the Calabi-Yau condition (\ref{Calabi})) reads 
\begin{eqnarray}
\label{L45F4}
\hspace{-0.95in}&& \quad  \quad \quad    \quad 
 L_4\, \, \, = \, \,  \, \, \,
 D_x^4 \, \, \, \, + \, P(x) \cdot  D_x^3 \,  \, \, \, 
+ \, Q(x) \cdot  D_x^2 \, \, \,  \,+ \, R(x) \cdot  D_x  \, \, \, + \, S(x), 
\end{eqnarray}
where $\, P(x)$ and $\, Q(x)$ read:
\begin{eqnarray}
\label{whereM4}
\hspace{-0.95in}&&  \quad  \quad  \quad  \quad   
 P(x)\, \, = \, \,  \, \, {\frac {4\, -5\,x}{x \cdot \, (1\, -x) }}, \quad \, \,    
\nonumber \\
\hspace{-0.95in}&&   \,  \quad  \quad  \quad  \quad  
 Q(x) \, \, = \, \,  \, \,  \,
\,{\frac { (3  \,x \, - 2) \cdot \, (11 \, x \, - 10) }{ 8 \cdot \, {x}^{2} \cdot \, (x-1)^{2}}} 
\, \,\,  \, +\,{\frac { a \cdot \,(1 -a)  \, +b \cdot \, (1-b) }{2 \cdot \, x \cdot \, (x-1) }}.
\end{eqnarray}
The other rational functions $\, R(x)$ and $\, S(x)$ are more involved rational 
functions that will not be given here. The operator $\, L_4$ can be seen as the 
``exterior (or antisymmetric) square root\footnote[1]{See the concept of 
Yifan Yang pullback introduced in~\cite{Yifan}.}'' 
of the order-five linear differential operator that annihilates 
the $\,_5F_4$ hypergeometric function (\ref{5F4}).  

\vskip .1cm 

{\bf Remark:} In~\cite{Duco} the authors introduce a proxy of the exact 
``exterior square root'' $\, L_4$ namely the so-called Yifan Yang pullback,
given in general by the equations in the section ``Definition'' page 10 
of~\cite{Yifan}\footnote[2]{The author of~\cite{Yifan} has benefited 
from an unpublished result by Yifan Yang. Note that there is a misprint 
in~\cite{Yifan} in the ``Definition'' of  Yifan Yang pullback: on top 
of page 11, the term $\, b_3\, b_4/25$ {\em should be replaced by}  
$\, b_3\, b_4'/25$. With this correction the exact `exterior square root'' 
$\, L_4$ and the Yifan Yang pullback $\, M_4$ are related by a simple 
conjugation $\, M_4 \cdot \, u(x) \, = \, \, u(x) \cdot \, L_4$, where 
$\, 3/10 \cdot \, b_4 \, = \, -u'(x)/u(x)$.}
and, in this example, by equations (3.11), page 278 in~\cite{Duco}, which 
reads
\begin{eqnarray}
\label{L45F4}
\hspace{-0.95in}&&   
 M_4\, \, \, = \, \,  \, \, \,
 D_x^4 \, \, \, \, +   P_{YY}(x) \cdot  D_x^3 \,  \, \, \, 
+  Q_{YY}(x) \cdot  D_x^2 \, \, \,  \,+ R_{YY}(x) \cdot  D_x  \, \, \, +  S_{YY}(x), 
\end{eqnarray}
where $\, P_{YY}(x)$ and $\, Q_{YY}(x)$ read:
\begin{eqnarray}
\label{whereM4}
\hspace{-0.95in}&&  \quad  \quad  \quad  \,   \,  \,     
 P_{YY}(x)\, \, = \, \,  \, \,  \,  
{\frac {2 \cdot \, (3\, -5\,x)}{x \cdot \, (1\, -x) }},
\nonumber \\
\hspace{-0.95in}&&   \,  \quad  \quad  \quad  \,  \,    \,   
 Q_{YY}(x) \, \, = \, \,  \, \,  
\,{\frac { 99\, x^2 \, -122 \, x \, +28 }{ 4 \cdot \, {x}^{2} \cdot \, (x-1)^{2}}} 
\, \,\,  
+\,{\frac { a \cdot \,(1 -a)  \, +b \cdot \, (1-b) }{2 \cdot \, x \cdot \, (x-1) }},
\end{eqnarray}
the other rational functions $\, R_{YY}(x)$ and $\, S_{YY}(x)$ being more involved 
rational functions that will not be given here. The  ``Yifan Yang pullback'' 
$\, M_4$ is related to the exact ``exterior square root'' $\, L_4$  by a simple 
conjugation $\, M_4 \cdot \, u(x) \, = \, \, u(x) \cdot \, L_4$, with 
$\, u(x) \, = \, \, x^{-1/2} \cdot \, (1-x)^{-3/4}$. In general one may prefer
to introduce the Yifan Yang pullback defined page 10 and 11 of~\cite{Yifan}
instead of the exact ``exterior square root'', because the corresponding formulae 
are simpler. It does not make any difference however since 
the two operators are simply conjugated.

\vskip .1cm 

Let us consider the order-four linear differential operator $\, {\cal L}_4$ 
given on page 284 of~\cite{Duco} which annihilates the Hadamard product of two 
simple $\, _2F_1$ hypergeometric functions:
\begin{eqnarray}
\label{Hadamard}
\hspace{-0.95in}&& \,  \quad  \quad  \quad 
\Bigl({{1} \over {1\, -x}} \cdot \, _2F_1([a, \, 1-a], [1], \, x)\Bigr)
 \, \star   \,
\Bigl({{1} \over {1\, -x}} \cdot \, _2F_1([b, \, 1-b], [1], \, x)\Bigr).   
\end{eqnarray}
This  order-four operator $\, {\cal L}_2$ reads 
\begin{eqnarray}
\label{calL2hatP}
\hspace{-0.95in}&& \quad  \quad \quad   
  {\cal L}_4 \, \, \, \, = \, \,  \, \,\,
 D_x^4 \, \,\, \, + \, \hat{P}(x) \cdot  D_x^3
 \, \,\, \, + \, \hat{Q}(x) \cdot  D_x^2 
\,  \, \,+ \, \hat{R}(x) \cdot  D_x   \, \, \, + \, \hat{S}(x), 
\end{eqnarray}
where:
\begin{eqnarray}
\label{wherehatP4}
\hspace{-0.96in}&&   \,  
\hat{P}(x) \, \, = \, \,  \, \,
  2 \,{\frac {5\,{x}^{2}+4\,x-3}{x \cdot \, (x+1)  \, (x-1) }}, 
\nonumber \\
\hspace{-0.96in}&&   \, 
\hat{Q}(x)  \, \, = \, \,  \, \, 
2 \cdot \,{\frac { a \cdot \,(1 -a)  \, +b \cdot \, (1-b) }{x \cdot \, (x-1)^{2}}}
\,\, \,  
+{\frac {25\,{x}^{4}+40\,{x}^{3}-16\,{x}^{2}-32\,x+7}{
{x}^{2} \cdot \, (x+1)^{2} \, (x-1)^{2}}}.
\end{eqnarray}
Introducing the pullback $\, y(x)$ and the function $\, v(x)$
\begin{eqnarray}
\label{pullter}
\hspace{-0.95in}&& \quad  \quad  \quad   \quad   \quad   \quad  
 y(x) \, \, = \, \, \,  {\frac { -4 \cdot \, x}{ (1 \, -x)^{2}}},  
\quad  \quad  \, \, \,   
 v(x)  \, \, = \, \, \, \Bigl({\frac {x \cdot \, (1 \, +x) }{1 \, -x}} \Bigr)^{1/2}, 
\end{eqnarray}
one has the relation
\begin{eqnarray}
\label{vL4}
\hspace{-0.95in}&&  \quad \quad  \quad   \quad   \quad   \quad  
 v(x) \cdot \,  {\cal L}_4   \cdot \, {{1} \over {v(x)}}
\, \,\,  = \, \, \,  \, 
pullback\Bigl(L_4, \,  {\frac {-4 \, x}{ (1 \, -\,x)^{2}}}  \Bigr).
\end{eqnarray}
and one verifies that a  Schwarzian equation 
(\ref{SchwarUUhatAlmkvist}) is actually verified 
for (\ref{whereM4}) and (\ref{wherehatP4})
\begin{eqnarray}
\label{SchwarUUhatAlmkvist}
\hspace{-0.95in}&& \quad  \quad \quad  \quad  \quad 
\hat{U}_R(x)  \,\, \,  \, - \, U_M(y(x)) \cdot \, y'(x)^2 
\, \,\,  + \,  \{y(x), \, x\} 
 \,\,\, \, = \, \,\, \,   \, 0, 
\end{eqnarray}
with:
\begin{eqnarray}
\label{U}
\hspace{-0.95in}&& \quad  \quad \quad  \quad   \quad  \, 
U_M(x) \, \, \, = \, \, \, \,
 -{{Q(x)} \over {5}} \,\,   + \, {{3} \over {40}} \cdot \,  P(x)^2  \,
 \, + \, {{3} \over {10}} \cdot \, {{ d P(x)} \over {dx}},  
\\
\label{Uhat}
\hspace{-0.95in}&& \quad  \quad \quad \quad   \quad    \, 
\hat{U}_R(x) \, \, \, = \, \, \, \,
 -{{\hat{Q}(x)} \over {5}} \, \,  + \, {{3} \over {40}} \cdot \,  \hat{P}(x)^2  \,
 \, + \, {{3} \over {10}} \cdot \, {{ d \hat{P}(x)} \over {dx}}.  
\end{eqnarray}
This Schwarzian equation (\ref{SchwarUUhatAlmkvist}), together with 
the definitions (\ref{U}) and  (\ref{Uhat}), are exactly the 
Schwarzian equation (6.5) together with definition (6.4),
page 290 of~\cite{Duco}. 

\vskip .1cm

\subsubsection{Schwarzian conditions for Calabi-Yau operators and Yukawa couplings.\\} 
\label{Yukawa}

Let us calculate the series expansion of the nome and {\em Yukawa couplings}~\cite{Christol}
of  $\, L_4$ and $\,  {\cal L}_2$. In order to perform the calculations for 
arbitrary values of $\, a$ and $\, b$, let us introduce the same 
variables $\, s$ and $\, p$ as the one introduced by~\cite{Duco}:
\begin{eqnarray}
\label{ps}
\hspace{-0.95in}&&  \quad \quad \quad 
s \, \, = \, \, \,  a \cdot \, (1-a) \,  +b \cdot \, (1-b), \quad \quad \quad 
p \, \, = \, \, \, a \cdot \, b \cdot \,(1-a) \cdot \,(1-b).
\end{eqnarray}
Considering the subcase $\, a\, = \, \, 3$ and $\, b \, = \, \, 5$, 
the nome of $\, L_4$ reads
\begin{eqnarray}
\label{nomeL4}
\hspace{-0.95in}&&  \quad  \quad  \quad 
q_x(L_4) \, \, = \, \, \,  \, 
x \, \, \,  + \, (2\,p \, -s \,+1) \cdot \,  {{x^{2}} \over {2}}  \, \, 
\nonumber \\ 
\hspace{-0.95in}&& \quad  \quad \quad  \quad \,\,
+ \, (93\,{p}^{2}-98\,ps+26\,{s}^{2}+112\,p-60\,s+40)
 \cdot \,  {{x^{3}} \over {128}} 
 \\ 
\hspace{-0.95in}&& \quad  \quad \quad  \quad \,\,
+ \, (27748\,{p}^{3} -45289\,{p}^{2}s +24798\,p{s}^{2} -4554\,{s}^{3} +55759\,{p}^{2}
\nonumber \\ 
\hspace{-0.95in}&& \quad  \quad \quad \quad \quad  \quad \,\,
-61734\,ps+17190\,{s}^{2}+43848\,p-24516\,s+13608)
 \cdot {{ x^{4}} \over {62208}}    \, \,  \,  \, \, + \, \,  \, \cdots,  
\nonumber 
\end{eqnarray}
while the nome of $\,  {\cal L}_4 $ reads:
\begin{eqnarray}
\label{nomeL4}
\hspace{-0.95in}&& \quad    \quad   \, \,    
q_x({\cal L}_4) \, \, = \, \, \,   
 -\, {{1} \over {4}} \cdot \,
 q_x(L_4)\Bigl(  {\frac { -4 \cdot \, x}{ (1 \, -x)^{2}}}  \Bigr) 
 \, \,\, = \, \, \,\, \,\, 
x \,\, \, \,-2 \cdot \, (2\,p \, -s) \cdot \, {x}^{2}
\nonumber \\ 
\hspace{-0.95in}&& \quad \quad \quad  \quad          \,  \,
\, + \, \Bigl((93\,{p}^{2}\, -98\,ps\, +26\,{s}^{2}\, -16\,p\, +4\,s\Bigr)
 \cdot \,  {{ x^{3}} \over {8 }}
\nonumber \\ 
\hspace{-0.95in}&& \quad \quad  \quad   \quad    \,  \,
\, - \, \Bigl(27748\,{p}^{3}\, -45289\,{p}^{2}s\, +24798\,p{s}^{2}
\,-4554\,{s}^{3}\, +9708\,ps \, -12038\,{p}^{2}
\nonumber \\ 
\hspace{-0.95in}&& \quad \quad \quad  \quad  \quad  \quad      \,  \,
-1764\,{s}^{2}\,+1080\,p\,-216\,s\Bigr) \cdot {{ x^{4}} \over {972 }}
\, \, \, \,  \, + \, \,  \, \cdots 
\end{eqnarray}
The respective Yukawa couplings of $\, L_4$ and $\, {\cal L}_4$ read:
\begin{eqnarray}
\label{YukL4}
\hspace{-0.95in}&&  
K_x(L_4) \, \, = \, \, \,\, 
1 \,  \, \, - \left( 5\,p+1-2\,s \right) \cdot  \, x \, \,\,
 \,   + \,  \Bigr(825\,{p}^{2}\,-638\,ps\,+120\,{s}^{2}\, +244\,p\, -80\,s \Bigl)
 \cdot \, {{ {x}^{2}} \over {64}}
\nonumber \\ 
\hspace{-0.95in}&& \quad  \quad   \quad    \,  \, 
\, - \, \Bigl(119240\,{p}^{3}\,-133883\,{p}^{2}s\,+48642\,p{s}^{2}\,
-5688\,{s}^{3} \,-20346 \,ps\,+35609\,{p}^{2}
\nonumber \\ 
\hspace{-0.95in}&& \quad  \quad  \quad \quad     \quad  \quad  \,  \, 
+2448\,{s}^{2}\,-3420\,p\,+1728\,s \Bigr) \cdot {{x^3} \over { 5184}}
\, \,\,\, + \, \, \, \cdots 
\end{eqnarray}
\begin{eqnarray}
\label{YukcalL2}
\hspace{-0.95in}&&    \quad \,  \,  \, 
K_x({\cal L}_4) \, \, = \,\, \, 
K_x(L_4)\Bigl(  {\frac { -4 \cdot \, x}{ (1 \, -x)^{2}}} \Bigr)
\, \,\, = \, \, \, \,\, \,
1 \, \,\,\, +4 \cdot \, (5\,p \, -2\,s \, +1) \cdot \, x
\nonumber \\
\hspace{-0.95in}&& \quad \quad    \quad \,  \,  \, \,
\, + \, \Bigl(825\,{p}^{2}\,-638\,ps\,+120\,{s}^{2}
\,+404\,p\,-144\,s\,+32\Bigr)  \cdot \,  {{x^{2}} \over {4}}
\nonumber \\
\hspace{-0.95in}&& \quad \quad  \quad \,  \,  \, \, 
\, + \, \Bigl(119240\,{p}^{3}\,-133883\,{p}^{2}s\,+48642\,p{s}^{2}
\,-5688\,{s}^{3}\, -72024\,ps\, +102434\,{p}^{2}
\nonumber \\
\hspace{-0.95in}&& \quad \quad  \quad \quad  \quad \, \,   \,  \, \, 
+12168\,{s}^{2}\,+21204\,p\,-6696\,s\, +972\Bigr)
 \cdot \,  {\frac {{x}^{3}}{81}}
\, \, \, \, \, \, + \, \, \cdots 
\end{eqnarray}
In terms of the nome the Yukawa couplings read:
\begin{eqnarray}
\label{YukcalL2nome}
\hspace{-0.95in}&&    \quad \quad 
K_q(L_4) \, \, = \, \, \,  
1  \, \,  \,  - \, (5\,p \,-2\,s \,+1)  \cdot \, q
\nonumber \\
\hspace{-0.95in}&&    \quad \quad   \quad   \quad  \,
\, + \, \Bigl(1145\,{p}^{2}\,  -926\,ps\,  +184\,{s}^{2}
\,  +468\,p\,  -176\,s\,  +32\Bigr)  \cdot \, {{q^{2}} \over {64}}
\\
\hspace{-0.95in}&&    \quad  \quad  \quad    \quad  \,
\, - \Bigl(571795\,{p}^{3}\,  -698524\,{p}^{2}s\,  +280506\,p{s}^{2}
\,  -36972\,{s}^{3} \,  +355447\,{p}^{2}
\nonumber \\
\hspace{-0.95in}&&    \, \,  \quad  \quad    \quad  \quad  \quad \quad 
 -273162\,ps\,  +51390\,{s}^{2}\,  +54072\,p\,  -18900\,s\,  +1944\Bigr)
 \cdot \, {{q^{3}} \over {10368}} 
 \, \,\,\, \,  \,  +  \, \, \, \cdots
 \nonumber
\end{eqnarray} 
and 
\begin{eqnarray}
\label{YukcalL2nome}
\hspace{-0.95in}&&    \quad  \quad   \quad
K_q({\cal L}_4) \, \, \, = \, \, \, K_q(L_4)(-4 \cdot \, q) 
\, \, \, = \, \, \, \, \,   1  \, \,  \, 
+ 4 \cdot \, (5\,p \,-2\,s \,+1)  \cdot \, q
\nonumber \\
\hspace{-0.95in}&&    \quad    \quad \quad   \quad  \,\, \, \, 
+\, \Bigl(1145\,{p}^{2}\, -926\,ps\, +184\,{s}^{2}\, +468\,p\, -176\,s\, +32\Bigr)
  \cdot \, {{q^{2}} \over {4}}
\\
\hspace{-0.95in}&&    \quad    \quad \quad    \quad  \,\, \,
\, + \Bigl(571795\,{p}^{3}\, -698524\,{p}^{2}s\, +280506\,p{s}^{2}\, 
-36972\,{s}^{3}\,  +355447\,{p}^{2}
\nonumber \\
\hspace{-0.95in}&&    \, \, \, \,  \quad    \quad  \quad  \quad  \quad \quad 
 -273162\,ps\, +51390\,{s}^{2}\, +54072\,p\, -18900\,s\, +1944\Bigr)
 \cdot \, {{q^{3}} \over {162}} 
 \, \,\,\, \,  +  \, \, \, \cdots
 \nonumber
\end{eqnarray} 

\vskip .1cm

On this example we see that the nome and Yukawa couplings expressed 
in terms of the $\, x$ variable, are simply related 
(see (\ref{nomeL4}), (\ref{YukcalL2})) by the pullback 
transformation. The Yukawa couplings expressed in term of the nome 
of the two linear differential operators are related  in an even 
more simple and ``universal'' way:
 $\, K_q({\cal L}_4) \, \, = \, \, \, K_q(L_4)(-4 \cdot \, q)$.
This is a general result (see Appendix E of~\cite{Christol}). For 
a pullback $\, y(x)$ with a series expansion of the form 
\begin{eqnarray}
\label{pullform}
\hspace{-0.95in}&&    \quad \quad 
\quad \quad \quad \quad \quad  \quad \quad 
y(x) \,\, \,\, = \, \, \, \,\,\lambda \cdot x^n \,\,\, \, + \, \, \cdots,  
\end{eqnarray} 
the nome and Yukawa couplings expressed in terms of the $\, x$ variable
of two order-four linear differential  operators such that 
\begin{eqnarray}
\label{vL4suchthat}
\hspace{-0.95in}&&  \quad \quad  \quad \quad  \quad   \quad   \quad   \quad  
 v(x) \cdot \,  {\cal L}_4   \cdot \, {{1} \over {v(x)}}
\, \,\,  = \, \, \,  \, pullback\Bigl(L_4, \,  y(x) \Bigr),
\end{eqnarray}
are simply related as follows:
\begin{eqnarray}
\label{YukcalL2gen}
\hspace{-0.95in}&&    \quad  \quad  \quad  
 q_x({\cal L}_4)^n \, \, = \, \, \,   
 \, {{1} \over {\lambda}} \cdot \, q_x(L_4)\Bigl( y(x) \Bigr), 
 \quad  \quad \,
K_x({\cal L}_4) \, \, = \, \, \,K_x(L_4)\Bigl( y(x) \Bigr). 
\end{eqnarray}
Their Yukawa couplings, expressed in terms of the nome, 
{\em are related in an even simpler ``universal'' way}: 
\begin{eqnarray}
\label{YukcalL2nomegener}
\hspace{-0.95in}&&    \quad  \quad  \quad 
 \quad    \quad  \quad  \quad \quad  \quad  \quad 
K_q({\cal L}_4) \, \,\, = \, \, \, K_q(L_4)(\lambda  \cdot \, q^n).
\end{eqnarray} 
The previous example corresponded to the case $\, n\, = \, 1$ and $\, \lambda \, = -4$.
In the case  $\, n\, = \, 1$ and $\, \lambda \, = \, 1$, the pullback 
is a deformation of the identity $\, y(x) \, = \, \, x \, + \, \, \cdots \,$ 
and the  Yukawa couplings expressed in terms of the nome of the two operators
are equal. One thus recovers Proposition (6.2) of~\cite{Duco}
where the Yukawa couplings coincide.

\vskip .2cm

\subsection{Schwarzian conditions for Calabi-Yau operators related by pullback and conjugation.} 
\label{CalabiSchwa2}

In fact the Schwarzian condition (\ref{SchwarUUhatAlmkvist}) can be obtained 
in a {\em totally general framework} where two order-four  linear differential 
operators are equal up 
to pullback and conjugation. Let us consider two  order-four operators 
$\, L_4$ and $\, M_4$ such that 
\begin{eqnarray}
\label{vL4gener}
\hspace{-0.95in}&&  \quad  \quad  \quad \quad  \quad   \quad   \quad   \quad  
 v(x) \cdot \,  M_4   \cdot \, {{1} \over {v(x)}}
\, \,\,  = \, \, \, \,  \, pullback\Bigl(L_4, \,  y(x)  \Bigr).
\end{eqnarray}
A straightforward calculation similar to the one performed 
in section \ref{order-four} yields the Schwarzian relation\footnote[2]{This 
result is the same as the one in~\cite{Duco}.}
\begin{eqnarray}
\label{SchwarUUhatAlmkvistgener}
\hspace{-0.95in}&& \quad    \quad  \quad \quad  \quad 
W(M_4, \, x)  \,\,\, -\, \, W(L_4, \, y(x)) \cdot \, y'(x)^2 \, \,\, + \, \{y(x), \, x\}
 \, \,\,  = \, \,\, \, 0,
\end{eqnarray}
where the $\, W(M_4, \, x)$ and $\, W(L_4, \, x)$ are given by (\ref{wherecond4}), 
the $\, p(x)$ and $\, q(x)$ being the ones of the corresponding operators
$\, M_4$ and $\, L_4$:
\begin{eqnarray}
\label{wherecondM4L4genera}
\hspace{-0.95in}&& \quad   \quad
W(M_4, \,x)  \, \, = \, \,  \, \, \,
{{3} \over {10}} \cdot \,  {{ d p(M_4, \, x)} \over { dx}}  \, \, \,  + \, \, 
 {{3} \over {40}} \cdot \,   p(M_4, \,x)^2  \, \,  \, \, 
 - \, {{q(M_4, \,x)} \over {5}},
\\
\hspace{-0.95in}&& \quad   \quad
W(L_4, \,x)  \, \,\, = \, \,  \, \, \,
{{3} \over {10}} \cdot \,  {{ d p(L_4, \, x)} \over { dx}}  \, \, \,  + \, \,  \, 
 {{3} \over {40}} \cdot \,   p(L_4, \,x)^2  \, \,  \, \,  \, 
 - \, {{q(L_4, \,x)} \over {5}}.
\end{eqnarray}

\vskip .2cm

{\bf Remark 1:} There is nothing specific with order-four linear differential 
operators, one has the same result for two operators of {\em arbitrary orders} 
$\, N$ equal up to pullback and conjugation (see (\ref{vL4gener})): the 
expressions of $\, W(M_N, \, x)$ and $\, W(L_N, \, x)$ being the ones given 
in (\ref{wherecondN}), (\ref{wherecondNmore}). One also has 
\begin{eqnarray}
\label{SchwarUUhatAlmkvistgenerN}
\hspace{-0.95in}&& \quad   \quad  \quad \quad  \quad 
W(M_N, \, x)  \,\,\,
 -\, \, W(L_N, \, y(x)) \cdot \, y'(x)^2 \, \,\, + \, \{y(x), \, x\}
 \, \, = \, \,\, \, 0.
\end{eqnarray}

\vskip .1cm

{\bf Remark 2:} The expressions of $\, W(M_N, \, x)$ and $\, W(L_N, \, x)$ are 
related by (\ref{SchwarUUhatAlmkvistgenerN}). Let us assume that  $\, W(L_N, \, x)$
is compatible with the modular correspondences structures (existence of solutions 
of the Schwarzian equations of the form 
$\, y(x) \, = \, a_n \cdot \, x^n \, + \, \, \cdots \, $ with (\ref{curious5})).
One thus has $\, W(L_N, \, x)\, = \, \, -1/2/x^2\, + \, \, \cdots  \, \, $ Is this 
condition automatically satisfied for $\, W(M_N, \, x)$ as a consequence of 
 (\ref{SchwarUUhatAlmkvistgenerN}) ? For pullbacks of the form 
$\, y(x) \, = \, a_n \cdot \, x^n \, + \, \, \cdots, \, $ 
the function  $\, W(M_N, \, x)$ deduced from 
 (\ref{SchwarUUhatAlmkvistgenerN}), reads:
\begin{eqnarray}
\label{SchwarUUhatAlmkvistgenerNnn}
\hspace{-0.95in}&& \quad \quad   \quad     \,    
W(M_N, \, x) \, \,\,  = \,\,\,\, 
  W(L_N, \, y(x)) \cdot \, y'(x)^2 \, \,  - \, \{y(x), \, x\}
\nonumber \\
\hspace{-0.95in}&& \quad \quad \,  \quad  \quad \quad   \quad   
\, = \, \, \Bigl( -{{n^2} \over { 2 \, x^2}} \, + \, \cdots    \Bigr)
 \,\, 
+ \, \Bigl(  {{n^2 \, -1} \over { 2 \, x^2}} \, + \, \cdots   \Bigr)
 \, \,  \, = \, \, \, - {{1} \over { 2 \, x^2}}  \, \,\, + \, \cdots 
\end{eqnarray}
The condition (\ref{Laurent})  
{\em for the modular correspondences structures is thus preserved by pullbacks}. 

\vskip .2cm

\subsection{More general framework} 
\label{moregeneral}

For arbitrary orders we observed that the functions $\, W(x)$ 
that occur in the Schwarzian conditions are left invariant
under conjugations of the operators (\ref{accordingxtext}) 
and (\ref{accordingx2text}). 
More generally, one can consider operators that are not conjugated 
by a function $\, \rho(x)$, yet homomorphic, in the sense of the 
equivalence of operators\footnote[1]{Two linear differential operators 
$\, L_N$ and $\, {\tilde L}_N$ of order $\,N$ are 
homomorphic~\cite{Diffalg,Canonical} when there exists operators 
(intertwiners) of order at most $\, N-1$,  
such that $\,\,\, M_{N-1} \, L_N  -  {\tilde L}_N \, {\tilde M}_{N-1}$
$  =  0$. }. For a given operator $\, L_N$ of order-$\, N$, one can
easily obtain operators $\, {\tilde L}_N$ homomorphic to $\, L_N$. For 
instance, for an order-two  linear differential operator 
$\, L_2  = \,D_x^2 +A(x)\, D_x +B(x)$,
introducing the order-one operator  
$\,\, L_1 \, = \, \eta(x) \,D_x +\rho(x)$, 
an  order-two operator $\, {\tilde L}_2$ homomorphic to $\, L_2$
is easily obtained performing\footnote[9]{In Maple 
just to rightdivision(LCLM($L_2, \, L_1$),$ \, L_1$).} the 
rightdivision by $\, L_1$ of the LCLM of $\, L_2$ and $\, L_1$. 
If one now compares the functions $\, W(x)$ corresponding 
respectively to  $\, L_2$ and $\, {\tilde L}_2$, one sees 
that they are {\em quite different}, except when 
$\, \eta(x) \, = \, 0$, in which case  one reduces the operator 
equivalence to a conjugation by a function $\rho(x)$.
The analysis of the conditions for two order-$N$ operators
$\, L_N$ and $\, M_N$ to be homorphic up 
to pullback
\begin{eqnarray}
\label{vLNgener}
\hspace{-0.95in}&&  \quad \quad  \quad   \quad   \quad   \quad  
 M_{N-1} \cdot \,  M_N   
\, \,\,  = \, \, \,  \,
 pullback\Bigl(L_N, \,  y(x)  \Bigr) \cdot L_{N-1}, 
\end{eqnarray}
is a much more general problem corresponding to massive calculations
even if one restricts to operators that are homomorphic to their adjoint
(thus corresponding to selected, orthogonal or symplectic, 
differential Galois groups\footnote[5]{In that general framework 
(\ref{vLNgener}), we do not have the Calabi-Yau, or symmetric Calabi-Yau, 
equations that help us to perform our calculations.}). Performing 
such calculations will require new tools and ideas.
This cannot be performed in general (like we did
in the first section of this paper) but could be considered on 
particular problems emerging from physics or enumerative 
combinatorics, where the operators will be of some ``selected'' form.

\end{document}